\documentclass{emulateapj}

\usepackage{graphicx} 
\usepackage{longtable} 

\usepackage{array} 
\usepackage{amssymb,amsmath} 

\newcommand{\numax}{\mbox{$\nu_{\rm max}$}}
\newcommand{\Dnu}{\mbox{$\Delta \nu$}}

\newcommand{\muHz}{\mbox{$\mu$Hz}}
\newcommand{\kep}{\mbox{\textit{Kepler}}}
\newcommand{\teff}{\mbox{$T_{\rm eff}$}}
\newcommand{\logg}{\mbox{$\log g$}}
\newcommand{\feh}{\mbox{$\rm{[Fe/H]}$}}
\newcommand{\msun}{\mbox{$M_{\sun}$}}
\newcommand{\rsun}{\mbox{$R_{\sun}$}}
\newcommand{\ntot}{\mbox{1797}}
\newcommand{\nkoi}{\mbox{77}}
\newcommand{\npl}{\mbox{107}}

\newcommand{\rearth}{\mbox{$R_{\ensuremath{\oplus}}$}}

\def\apjl{ApJ}%
\def\apjs{ApJS}%
\def\ao{Appl.~Opt.}%
\def\aap{A\&A}%
\slugcomment{accepted for publication in ApJ}

\shorttitle{Asteroseismology of Kepler Planet-Candidate Host Stars}
\shortauthors{D. Huber et al.}

\begin{document}

\title{Fundamental Properties of Kepler 
Planet-Candidate Host Stars using Asteroseismology}

\author{
Daniel Huber\altaffilmark{1,2}, 
William J. Chaplin\altaffilmark{3,4}, 
J{\o}rgen Christensen-Dalsgaard\altaffilmark{4}, 
Ronald L. Gilliland\altaffilmark{5}, 
Hans Kjeldsen\altaffilmark{4}, 
Lars A. Buchhave\altaffilmark{6,7}, 
Debra A. Fischer\altaffilmark{8}, 
Jack  J. Lissauer\altaffilmark{1}, 
Jason F. Rowe\altaffilmark{1}, 
Roberto Sanchis-Ojeda\altaffilmark{9}, 
Sarbani Basu\altaffilmark{8}, 
Rasmus Handberg\altaffilmark{4}, 
Saskia Hekker\altaffilmark{10}, 
Andrew W. Howard\altaffilmark{11}, 
Howard Isaacson\altaffilmark{12}, 
Christoffer Karoff\altaffilmark{4}, 
David W. Latham\altaffilmark{13}, 
Mikkel N. Lund\altaffilmark{4}, 
Mia Lundkvist\altaffilmark{4}, 
Geoffrey  W. Marcy\altaffilmark{12}, 
Andrea Miglio\altaffilmark{3}, 
Victor Silva Aguirre\altaffilmark{4}, 
Dennis Stello\altaffilmark{14,4}, 
Torben Arentoft\altaffilmark{4}, 
Thomas Barclay\altaffilmark{1}, 
Timothy  R. Bedding\altaffilmark{14,4}, 
Christopher J. Burke\altaffilmark{1}, 
Jessie L. Christiansen\altaffilmark{1}, 
Yvonne P. Elsworth\altaffilmark{3}, 
Michael R. Haas\altaffilmark{1}, 
Steven D. Kawaler\altaffilmark{15}, 
Travis S. Metcalfe\altaffilmark{16}, 
Fergal Mullally\altaffilmark{1}, and 
Susan E. Thompson\altaffilmark{1} 
}
\altaffiltext{1}{NASA Ames Research Center, Moffett Field, CA 94035, USA}
\altaffiltext{2}{NASA Postdoctoral Program Fellow; \mbox{daniel.huber@nasa.gov}}
\altaffiltext{3}{School of Physics and Astronomy, University of Birmingham, Birmingham B15 2TT, UK}
\altaffiltext{4}{Stellar Astrophysics Centre, Department of Physics and Astronomy, Aarhus University, Ny Munkegade 120, DK-8000 Aarhus C, Denmark}
\altaffiltext{5}{Center for Exoplanets and Habitable Worlds, The Pennsylvania State University, 525 Davey Lab, University Park, PA 16802}
\altaffiltext{6}{Niels Bohr Institute, University of Copenhagen, DK-2100 Copenhagen, Denmark}
\altaffiltext{7}{Centre for Star and Planet Formation, Natural History Museum of Denmark, University of Copenhagen, DK-1350 Copenhagen, Denmark}
\altaffiltext{8}{Department of Astronomy, Yale University, New Haven, CT 06511, USA}
\altaffiltext{9}{Department of Physics, Massachusetts Institute of Technology, 77 Massachusetts Ave., Cambridge, MA 02139, USA}
\altaffiltext{10}{Astronomical Institute 'Anton Pannekoek', University of Amsterdam, Science Park 904, 1098 XH Amsterdam, The Netherlands}
\altaffiltext{11}{Institute for Astronomy, University of Hawaii, 2680 Woodlawn Drive, Honolulu, HI 96822, USA}
\altaffiltext{12}{Department of Astronomy, University of California, Berkeley, California 94720, USA}
\altaffiltext{13}{Harvard-Smithsonian Center for Astrophysics, 60 Garden Street, Cambridge, Massachusetts 02138, USA}
\altaffiltext{14}{Sydney Institute for Astronomy (SIfA), School of Physics, University of Sydney, NSW 2006, Australia}
\altaffiltext{15}{Department of Physics and Astronomy, Iowa State University, Ames, IA 50011 USA}
\altaffiltext{16}{Space Science Institute, Boulder, CO 80301, USA}

\begin{abstract}
We have used asteroseismology to determine fundamental properties for 66 
\kep\ planet-candidate host stars,
with typical uncertainties of 3\% and 7\% in radius and mass, respectively. 
The results include new 
asteroseismic solutions for four host stars with confirmed planets  
(Kepler-4, Kepler-14, Kepler-23 and Kepler-25) and increase the total number of \kep\ 
host stars with asteroseismic solutions to \nkoi.
A comparison with stellar properties in the planet-candidate 
catalog by Batalha et al.\ shows that
radii for subgiants and giants obtained from spectroscopic 
follow-up are systematically too low by up to a factor of 1.5, while the properties for 
unevolved stars are in good agreement. 
We furthermore apply asteroseismology to confirm that a large majority of cool main-sequence 
hosts are indeed dwarfs and not misclassified giants.
Using the revised stellar properties, we recalculate the 
radii for \npl\ planet candidates in our sample, and 
comment on candidates for which the radii change from a
previously giant-planet/brown-dwarf/stellar regime to a sub-Jupiter size, or vice versa.
A comparison of stellar densities from asteroseismology with densities derived from  
transit models in Batalha et al.\ assuming circular orbits shows significant 
disagreement for more than half of the sample 
due to systematics in the modeled impact parameters, 
or due to planet candidates which may be in eccentric orbits.
Finally, we investigate tentative correlations between host-star masses and planet-candidate 
radii, orbital periods, and multiplicity,  
but caution that these results may be influenced by the small sample size and detection biases.
\end{abstract}


\section{Introduction}
Nearly 700 confirmed 
planetary systems outside our solar system have been discovered in the past two decades. 
The vast majority of these planets have been detected using 
indirect techniques such as transit photometry or Doppler velocities, 
which yield properties of the planet only as a function of the properties of the 
host star. The accurate knowledge of the fundamental properties of host stars, particularly 
radii and masses, is therefore of great 
importance for the study of extrasolar planets.

The \kep\ Mission \citep{borucki10,koch10b} has revolutionized exoplanet science in the 
last few years, yielding thousands of new exoplanet candidates 
\citep{borucki11b,borucki11,batalha12}.
A serious problem in the interpretation of \kep\ planet detections, 
occurrence rates, and ultimately the determination of the frequency of habitable planets,
is the accuracy of stellar parameters. Almost all \kep\ planet-candidate hosts 
(also designated as Kepler Objects of Interest, or KOIs) 
are too faint to have measured parallaxes, and hence 
stellar parameters mostly rely on the combination of broadband photometry, stellar model 
atmospheres and evolutionary tracks, as done for the Kepler Input Catalog \citep[KIC,][]{brown11}. 
Biases in the KIC have been shown to reach up to 50\% in radius and 0.2\,dex in $\log g$ 
\citep{verner11b}, introducing serious uncertainties in the derived planetary properties. 
A more favorable case are hosts for which high-resolution spectroscopy is available, 
which yields strong constraints on the evolutionary state. Recent spectroscopic
efforts on planet-candidate hosts have concentrated on cool M-dwarfs
\citep{johnson12,muirhead12,muirhead12b} as well as 
F-K dwarfs \citep{buchhave12}.
Nevertheless, spectroscopic analyses are often affected by 
degeneracies between $T_{\rm eff}$, $\log g$ and [Fe/H] \citep{torres12}, and only yield 
strongly model-dependent constraints on stellar radius and mass.

An excellent alternative to derive accurate stellar radii and masses of host stars
is asteroseismology, the study of stellar oscillations 
\citep[see, e.g.,][]{brown94, CD04, aerts10, gilliland10}. Prior to the \kep\ mission, 
asteroseismology of exoplanet hosts was restricted to a few stars with detections 
from ground-based Doppler observations 
\citep{bouchy05,vauclair08,wright11} or the Hubble space telescope \citep{gilliland11,nutzman11}. 
This situation has dramatically changed 
with the launch of the \kep\ space telescope, which provides
photometric data 
suitable for both transit searches and asteroseismology. First results for 
previously-known transiting planet hosts in the \kep\ field were
presented by \citet{cd10}, and several \kep\ planet discoveries have since benefitted 
from asteroseismic constraints of host-star properties 
\citep{batalha11,howell12,barclay12,borucki12,carter12,chaplin12,barclay12b,gilliland13}.
In this paper, we present the first 
systematic study of \kep\ planet-candidate host stars using asteroseismology.

\section{Determination of Fundamental Stellar Properties}

\subsection{Background}

Solar-like oscillations are acoustic standing waves excited by near-surface convection 
\citep[see, e.g.,][]{houdek99,houdek06,samadi07}. 
The oscillation modes are characterized by
a spherical degree $l$ (the total number 
of surface nodal lines), a radial order $n$ (the number of nodes from the surface to 
the center of the star)
and an azimuthal order $m$ (the number of surface nodal lines that cross the equator). The 
azimuthal order $m$ is generally only important if the $(2l+1)$ degeneracy of frequencies 
of degree $l$ is lifted 
by rotation. 

Solar-like oscillations 
involve modes of low spherical degree $l$ and high radial order $n$, and hence the
frequencies can be asymptotically described by a series of 
characteristic separations \citep{vandakurov68,tassoul80,gough86}. 
The large frequency separation \Dnu\ is the spacing between modes with the same spherical 
degree $l$ and consecutive radial order $n$, and probes the 
sound travel time across the stellar diameter. 
This means that \Dnu\ is related to the mean stellar density and is expected to scale as 
follows \citep{ulrich86}:

\begin{equation}
\Delta\nu = \frac{({M/M_{\sun}})^{1/2}}{(R/R_{\sun})^{3/2}} \Delta\nu_{\sun} \: .
\label{equ:dnu}
\end{equation}

\noindent
Another fundamental observable is the frequency at 
which the oscillations have maximum power (\numax).
As first argued by \citet{brown91}, \numax\ for sun-like stars is expected to scale with the 
acoustic cut-off frequency and can therefore be related to fundamental 
stellar properties, as follows \citep{KB95}:

\begin{equation}
\nu_{\rm max} = \frac{ M/M_{\sun}}{(R/R_{\sun})^{2}\sqrt{T_{\rm eff}/T_{\rm eff,\sun}}} \nu_{\rm max,\sun} \: .
\label{equ:nmax}
\end{equation}

\noindent
Equation (\ref{equ:nmax}) shows that \numax\ is mainly dependent on surface gravity, and 
hence is a good indicator of the evolutionary state.
Typical oscillation frequencies range from $\sim$ 3000\,\muHz\ for 
main sequence stars like our Sun down to $\sim$ 300\,\muHz\ for low-luminosity red giants, 
and a few \muHz\ for high-luminosity giants. 
We note that while individual oscillation frequencies provide more detailed 
constraints on properties such as stellar ages 
\citep[see, e.g.,][]{dogan10,metcalfe10,dimauro11,mathur12,metcalfe12}, 
the extraction of frequencies is generally only possible for high S/N detections. 
To extend our study to a large ensemble of planet-candidate hosts, we therefore concentrate solely  
on determining the global oscillation properties \Dnu\ and 
\numax\ in this paper.

It is important to note that Equations (\ref{equ:dnu}) and (\ref{equ:nmax}) 
are approximate relations which require calibration. For comprehensive reviews 
of theoretical and empirical tests of both relations 
we refer the reader to \citet{belkacem12} and \citet{miglio13}, but we 
provide a brief summary here.
It is well known that Equation (\ref{equ:nmax}) is on less firm ground than Equation 
(\ref{equ:dnu}) due to uncertainties in modeling convection which 
drives the oscillations, although recent progress on the theoretical understanding 
of Equation (\ref{equ:nmax}) has been made \citep{belkacem11}.
\citet{stello09} showed that both relations agree with models 
to a few percent, which was supported 
by investigating the relation between \Dnu\ and \numax\ for a large ensemble 
of \kep\ and CoRoT stars, and by comparing derived radii and masses 
to evolutionary tracks and synthetic stellar 
populations \citep{hekker09,miglio09,kallinger10,huber10,mosser10,huber11b,silvaaguirre11,miglio12c}. 
More recently, \citet{white11} showed that 
that \Dnu\ calculated from individual model frequencies can systematically 
differ from \Dnu\ calculated using Equation (\ref{equ:dnu}) by 
up to 2\% for giants and for dwarfs with $M/\msun\gtrsim1.2$. 
Similar results were found by \citet{mosser13}, who investigated the influence of 
correcting the observed \Dnu\ to the value expected in the high-frequency asymptotic limit.
In general, however, comparisons with individual frequency modeling have 
shown agreement within 
2\% and 5\% in radius and mass, respectively,
both for dwarfs \citep{mathur12} and for giants \citep{dimauro11,jiang11}.

Empirical tests
have been performed using independently determined fundamental properties from 
Hipparcos parallaxes, eclipsing binaries, cluster stars, and optical long-baseline interferometry 
\citep[see, e.g.,][]{stello08,bedding11b,brogaard12,miglio11,miglio12b,huber12b,silva12}. 
For unevolved stars ($\logg \gtrsim 3.8$) no evidence for 
systematic deviations has yet been determined within the observational uncertainties, 
with upper limits of $\lesssim 4$\% in radius \citep{huber12b} and $\lesssim 10$\% 
in mass \citep{miglio11}. 
For giants and evolved subgiants ($\logg \lesssim 3.8$) similar results have been 
reported, although a systematic deviation of $\sim 3\%$ in \Dnu\ 
has recently been noted for He-core burning red giants 
\citep{miglio12b}.

In summary, for stars considered in this study, Equations (\ref{equ:dnu}) and 
(\ref{equ:nmax}) have been tested theoretically to 
$\sim 2\%$ and $\sim 5\%$, as well as empirically to $\lesssim 4\%$ and $\lesssim 10\%$ 
in radius and mass, respectively. While it should be kept in mind that future revisions of these 
relations based on more precise empirical data are possible, it is clear that these 
uncertainties are significantly smaller than for classical methods to determine radii and masses 
of field stars.

\subsection{Asteroseismic Analysis}

Our analysis is based on \kep\ short-cadence \citep{gilliland10b} and long-cadence 
\citep{jenkins10} data through Q11. We have used simple-aperture photometry 
(SAP) data for our analysis. We have analyzed all available data for the \ntot\
planet-candidate hosts listed in the cumulative catalog by \citet{batalha12}. Before 
searching for oscillations, transits need to be removed or 
corrected since the sharp structure in the time domain would cause significant power 
leakage from low frequencies into the oscillation frequency domain. This was done using a 
median filter with a length chosen according to the measured duration of the 
transit. In an alternative approach, all transits were phase-clipped from the 
time series using the periods and epochs listed in \citet{batalha12}. Note that for typical 
transit durations and periods, the induced gaps in the time series have little 
influence on the resulting power spectrum. Finally, all time series were high-pass 
filtered by applying a quadratic Savitzky-Golay filter \citep{savitzky64} 
to remove additional low-frequency power due to stellar activity and 
instrumental variability. For short-cadence data, the typical cut-off frequency was 
$\sim100$\,\muHz, while for long-cadence data a cut-off of $\sim1$\,\muHz\ was applied. 

\begin{figure}
\begin{center}
\resizebox{\hsize}{!}{\includegraphics{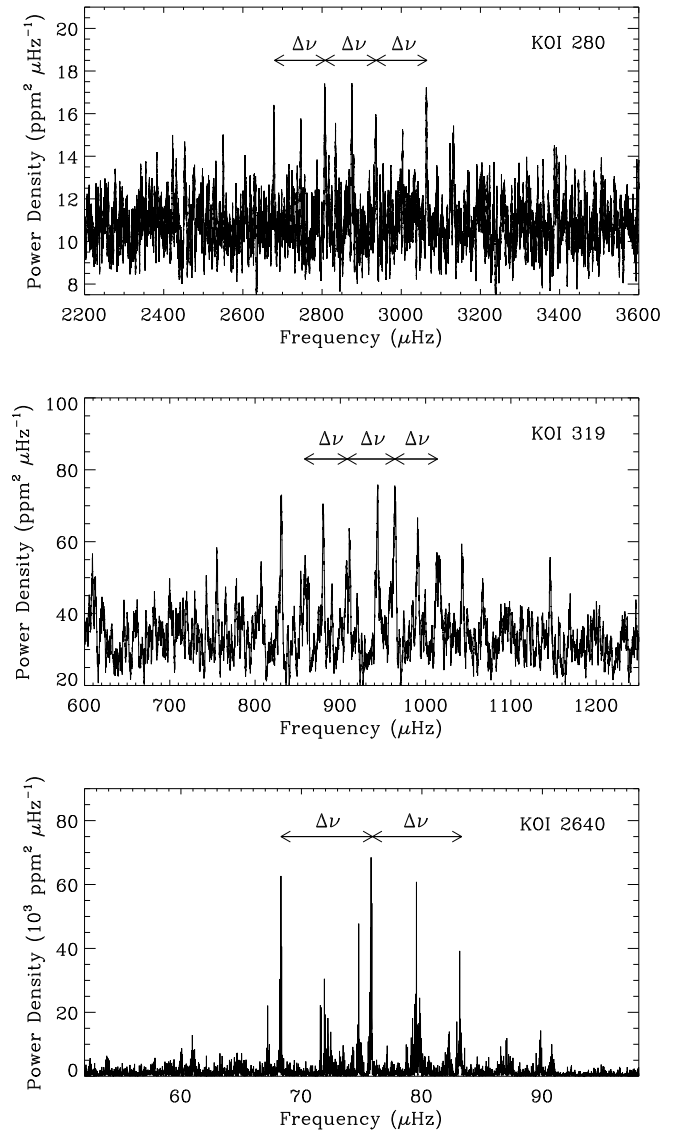}}
\caption{Power spectra for three \kep\ planet-candidate host stars
with detected solar-like oscillations.
The panels show three representative hosts in 
different evolutionary stages: a main-sequence star (top panel), a subgiant (middle panel) 
and a red giant (bottom panel). For the latter long-cadence data were 
used, while the former two have been calculated using short-cadence data. 
The large frequency separation \Dnu\ is indicated in each panel.
Note the increase 
in the y-axis scale from the top to bottom panel, illustrating the increase in 
oscillation amplitudes for evolved stars.}
\label{fig:examples}
\end{center}
\end{figure}

To detect oscillations and extract the global oscillation 
parameters \Dnu\ and \numax, we have used the analysis pipelines described by \citet{huber09}, 
\citet{hekker10c}, \citet{karoff10}, \citet{verner11c} and \citet{lund12}. 
Note that these methods have been extensively tested 
on \kep\ data and were shown to agree well with other 
methods \citep{hekker11,verner11,hekker12b}. 
We successfully detect oscillations in a total of 77 planet-candidate hosts (including 11 
stars for which asteroseismic solutions have been published in separate studies). 
For 69 host stars short-cadence 
data were used, while 8 of them showed oscillations with \numax\ values low 
enough to allow a detection using long-cadence data. 
The final values for \Dnu\ and \numax\ are listed in Table 1 and 
were adopted from the method of \citet{huber09}, with 
uncertainties calculated by adding in quadrature the formal uncertainty 
and the scatter of the values over all other methods. 
Note that in some cases the S/N was too low to reliably estimate \numax, and hence only 
\Dnu\ is listed. 
For one host (KOI-1054) \Dnu\ could not be reliably determined, and hence 
only \numax\ is listed.
The solar reference values, which were calculated using the same method, 
are $\Delta\nu_{\sun}=135.1\pm0.1\,\muHz$ 
and $\nu_{\rm max,\sun}=3090\pm30\,\muHz$ 
\citep{huber11b}.

We note that in the highest S/N cases, the observational uncertainties on 
\Dnu\ are comparable to or lower than the accuracy to which Equation (\ref{equ:dnu}) has 
been tested (see previous section). To account for systematic errors in 
Equation (\ref{equ:dnu}), we adopt a conservative approach by
adding to our uncertainties in quadrature the difference between the observed \Dnu\ and 
the corrected \Dnu\ using Equation (5) in \citet{white11}. 
To account for the fact that \Dnu\ can be measured more precisely than
\numax, the same fractional uncertainties were added in quadrature 
to the formal \numax\ uncertainties.  The final median uncertainties in 
\Dnu\ and \numax\ are 2\% and 4\%, respectively.

Figure \ref{fig:examples} shows 
examples of power spectra for three stars in the sample, 
illustrating a main-sequence star (top panel), a subgiant 
(middle panel) and a red-giant (bottom panel). Note that the power spectra 
illustrate typical intermediate S/N detections. 
Broadly speaking the detectability of oscillations depends 
on the brightness of the host star and the evolutionary 
state, because oscillation amplitudes scale proportionally to stellar 
luminosity \citep[see, e.g.,][]{KB95,chaplin11b}. Among host stars that are 
close to the main-sequence ($\log g > 4.2$), the faintest star with detected oscillations 
has a \kep\ magnitude of $12.4$\,mag.

\subsection{Spectroscopic Analysis}

In addition to asteroseismic constraints,  
effective temperatures and metallicities are 
required to derive a full set of fundamental properties. For all stars in our sample 
high-resolution optical spectra were obtained as part of the \kep\ follow-up 
program \citep{gautier10}. 
Spectroscopic observations were taken using four different instruments: 
the HIRES spectrograph \citep{vogt94} on the 10-m telescope at Keck Observatory
(Mauna Kea, Hawaii), the FIES spectrograph \citep{djupvik10} on the 2.5-m Nordic Optical 
Telescope at the Roque de los Muchachos Observatory
(La Palma, Spain), the TRES spectrograph \citep{furesz08} on the 1.5-m Tillinghast 
reflector at the 
F. L. Whipple Observatory (Mt. Hopkins, Arizona), and the Tull Coud\'e spectrograph on the 
2.7-m Harlan J. Smith Telescope at McDonald Observatory (Fort Davis, Texas). 
Typical resolutions of the spectra range from 40,000--70,000.
Atmospheric parameters were derived using 
either the Stellar Parameter Classification 
\citep[SPC, see][]{buchhave12} or Spectroscopy Made Easy \citep[SME, see][]{valenti96} 
pipelines. 
Both methods match the observed spectrum to synthetic model spectra in the optical 
wavelengths and yield estimates of \teff, 
\logg, metallicity and $v\sin(i)$. 
Note that in our analysis we have 
assumed that the metal abundance [m/H], as returned by SPC, is equivalent to 
the iron abundance [Fe/H].
For stars with multiple observations each spectrum was analyzed individually, 
and the final parameters were calculated as an average of the individual results
weighted by the cross-correlation function (CCF), which gives a measure 
of the quality of the fit compared to the spectral template.
To ensure a homogenous set of parameters, we adopt the spectroscopic values from SPC, 
which was used to analyze the entire sample of host stars.
Table 1 lists for each planet-candidate host the details of the SPC analysis such as 
the number of spectra used, the average S/N of the observations, the 
average CCF, and the instrument used to obtain the spectra.

As discussed by \citet{torres12}, spectroscopic methods such as SME and SPC suffer 
from degeneracies between \teff, $\log g$ and [Fe/H]. Given the weak dependency of 
\numax\ on \teff\ (see Equation (\ref{equ:nmax})), asteroseismology can be used 
to remove such degeneracies by fixing \logg\ in the spectroscopic analysis to the 
asteroseismic value \citep[see, e.g.,][]{bruntt12,morel12,thygesen12}. This is done 
by performing the asteroseismic analysis (see next section) using initial estimates 
of \teff\ and [Fe/H] from spectroscopy, and iterating both analyses until convergence is 
reached (usually after one iteration). We have applied this method to all host stars in our 
sample to derive asteroseismically constrained values of \teff\ and [Fe/H], which are 
listed in Table 2. Note that since SPC has been less tested for giants, we have 
adopted more conservative error bars of 80\,K in \teff\ and 0.15\,dex in [Fe/H] 
for all evolved giants with $\logg < 3$ \citep{thygesen12}.
For all stars in our sample, we have added 
contributions of 59\,K in \teff\ and 0.062\,dex in [Fe/H] in quadrature 
to the formal uncertainties to account for 
systematic differences between spectroscopic methods, as suggested by \citet{torres12}.

\begin{figure}
\begin{center}
\resizebox{\hsize}{!}{\includegraphics{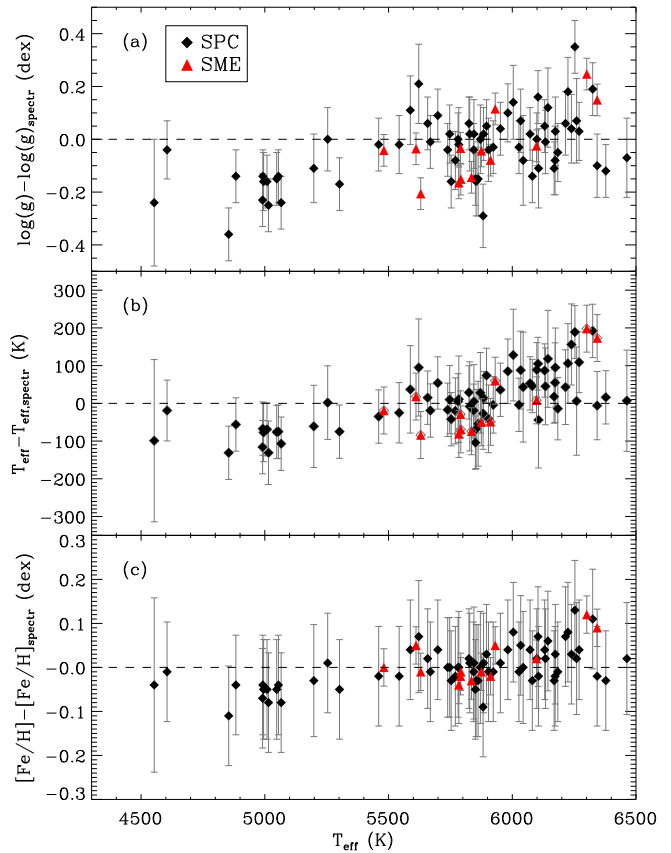}}
\caption{\textit{(a)} Comparison of \logg\ from a spectroscopic analysis with and 
without asteroseismic constraints on \logg. The difference is shown in the sense of constrained 
minus unconstrained analysis. Black diamonds show the sample analyzed 
with SPC, and red triangles stars analyzed with SME.
\textit{(b)} Same as panel (a) but for \teff.
\textit{(c)} Same as panel (a) but for \feh.}
\label{fig:partials}
\end{center}
\end{figure}

Our sample allows us to investigate the effects of fixing \logg\ on the determination of 
\teff\ and [Fe/H].
This is analogous to the work of \citet{torres12}, who 
used stellar densities derived from 
transits to independently determine \logg\ for a sample of main-sequence stars.
Figure \ref{fig:partials} shows the differences in \logg, \teff\ and [Fe/H] 
as a function of \teff. As in the \citet{torres12} 
sample, the unconstrained analysis tends to slightly underestimate  \logg\ (and 
hence \teff\ and 
[Fe/H]) for stars hotter than $\sim 6000$\,K. 
More serious systematics are found for stars with $\teff \lesssim 5400$\,K, which in our 
sample corresponds to subgiant and giant stars, for 
which \logg\ is systematically overestimated by up to 
0.2\,dex. The effect of these systematics on planet-candidate radii 
will be discussed in detail in Section 3.1.

Figure \ref{fig:partials} shows that changes in \logg\ are correlated 
with changes in \teff\ and [Fe/H]. We have investigated 
the partial derivatives $\Delta \teff/\Delta \logg$ and $\Delta \rm{[Fe/H]}/\Delta \logg$ 
and did not 
find a significant dependence on stellar properties such as effective temperature. 
The median derivatives for our sample are 
$\Delta \teff/\Delta \logg = 475 \pm 60$\,K/dex and 
$\Delta \rm{[Fe/H]}/\Delta \logg = 0.31 \pm 0.03$, respectively. Hence, a 
change of $\logg = 0.1$\,dex typically corresponds to a change of $\sim 50$\,K in 
\teff\ and 0.03\,dex in \feh.

The results in this study can also be used to test 
temperatures based on broadband photometry.
A comparison of 46 dwarfs which overlap with the sample of 
\citet{pinsonneault11} showed that the photometric temperatures 
(corrected for the spectroscopic metallicities in Table 2) are on average 
$190$\,K hotter than our spectroscopic estimates, with a scatter of $130$\,K. 
This offset is larger than previous comparisons based on a brighter comparison
sample \citep[see][]{pinsonneault11}, 
pointing to a potential problem with interstellar reddening. 
The results of our study, combined with other samples for which 
both asteroseismology and spectroscopy are available 
\citep{molenda11,bruntt12,thygesen12}, will be a valuable 
calibration sample to improve effective
temperatures in the \kep\ field based on photometric techniques
such as the infrared flux method \citep{casagrande10}.

\subsection{Grid-Modelling}

Given an estimate of 
the effective temperature, Equations (\ref{equ:dnu}) and 
(\ref{equ:nmax}) can be used to calculate the mass and radius of a star 
\citep[see, e.g.,][]{kallinger10c}. However, since both equations allow radius 
and temperature to vary freely for any given mass, a more refined method is to 
include knowledge from evolutionary theory to match the spectroscopic parameters with 
asteroseismic contraints.
This so-called grid-based method has been used extensively 
both for unevolved and evolved 
stars \citep{stello09,kallinger10,chaplin11a,creevey12,basu10,basu12}. 

\begin{figure}
\begin{center}
\resizebox{\hsize}{!}{\includegraphics{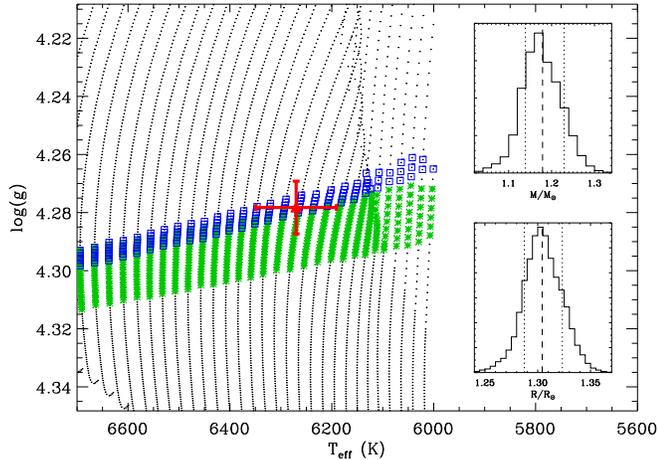}}
\caption{Surface gravity versus $T_{\rm eff}$ for BaSTI evolutionary tracks with 
a metallicity of $\rm{[Fe/H]}=+0.02$ in steps of 0.01\msun. Colored symbols 
show the $\pm 1\sigma$ constraints from 
\numax\ (green asterisks) and 
\Dnu\ (blue squares) for KOI-244 (Kepler-25). The determined position is shown as a red diamond.
The inset illustrates the distributions of Monte-Carlo 
simulations for stellar mass and radius, with dashed and dotted lines corresponding to 
the median and $\pm 1\sigma$ confidence limits, respectively.}
\label{fig:hrdexample}
\end{center}
\end{figure}

To apply the grid-based method, we have used different model tracks: 
the Aarhus Stellar Evolution Code \citep[ASTEC,][]{CD08a}, the
Bag of STellar Isochrones \citep[BaSTI,][]{basti}, 
the Dartmouth Stellar Evolution Database
\citep[DSEP,][]{dotter08}, the Padova stellar evolution code \citep{marigo08},
the Yonsei-Yale isochrones \citep[YY,][]{demarque04}, and the Yale Rotating Stellar 
Evoution Code \citep[YREC,][]{demarque08}. 
To derive masses and radii we have  
employed several different 
methods \citep{dasilva06,stello09,basu11,miglio12,silva13}. In brief, the methods calculate a 
likelihood function for a set of independent Gaussian observables 
$X$:

\begin{equation}
\mathcal{L}_{X} = \frac{1}{\sqrt{2\pi}\sigma_{X}} \exp{\left( \frac{-(X_{\rm obs}-X_{\rm model})^2}{2\sigma_{X}^2}\right)}
\end{equation}

\noindent
with $X=\{{T_{\rm eff}, \rm{[Fe/H]}, \numax, \Dnu}\}$. The combined likelihood is:

\begin{equation}
\mathcal{L} = \mathcal{L}_{T_{\rm eff}} \mathcal{L}_{\rm{Fe/H}} \mathcal{L}_{\numax} \mathcal{L}_{\Dnu} \: .
\label{equ:lh}
\end{equation}

\noindent
Note that for cases where only \Dnu\ could be measured, the \numax\ 
term in Equation (\ref{equ:lh}) was omitted.
The best-fitting model is then identified from the likelihood distribution 
for a given physical parameter, e.g. stellar mass and radius. Uncertainties are 
calculated, for example, 
by performing Monte-Carlo simulations using 
randomly drawn values for the observed values of $T_{\rm eff}$, [Fe/H], \numax\ and 
\Dnu. For an extensive comparison of these methods, including a discussion of 
potential systematics, we refer the reader to \citet{gai11}.

\begin{figure*}
\begin{center}
\resizebox{13cm}{!}{\includegraphics{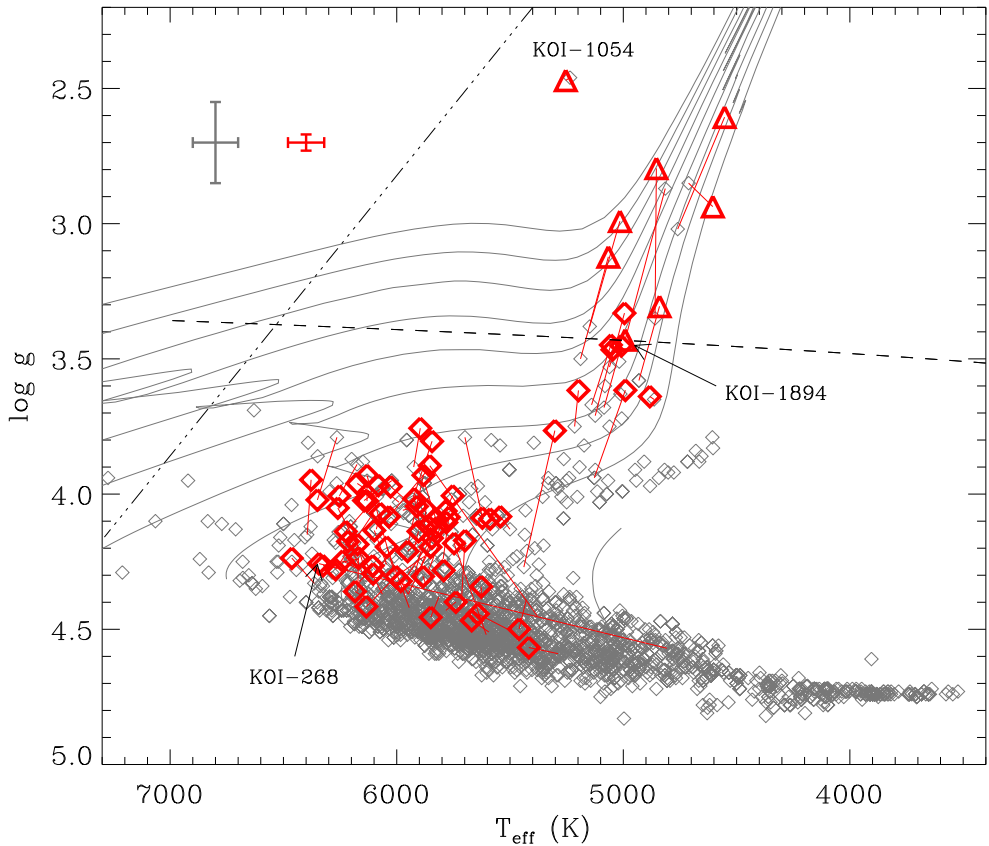}}
\caption{Surface gravity versus effective temperature for planet-candidate hosts 
in the \citet{batalha12} 
catalog (grey diamonds), together with solar metallicity Yonsei-Yale evolutionary tracks from 
0.8-2.6\,$M_{\sun}$ in 
steps of 0.2\,$M_{\sun}$ (grey lines). The dashed-dotted line marks the approximate location of 
the cool edge of the instability strip, and the dashed line marks the 
long-cadence Nyquist limit. 
Thick red symbols show the revised positions of 77 host stars with asteroseismic 
detections using long-cadence (triangles) and short-cadence (diamonds) data, 
respectively. Thin red lines 
connect the revised positions to the values in \citet{batalha12}. 
Typical error bars for stars with spectroscopic follow-up only (grey) and with 
asteroseismic constraints (red) are 
shown in the top left side of the plot.
A few host stars that 
are discussed in more detail in the text are annotated.}
\label{fig:hrd}
\end{center}
\end{figure*}

As an example, Figure \ref{fig:hrdexample} shows a diagram 
of $\log g$ versus $T_{\rm eff}$ for KOI-244 (Kepler-25), with evolutionary tracks taken 
from the BaSTI grid. The blue and green areas show models within 1-$\sigma$ of the 
observationally measured values of \Dnu\ and \numax, respectively. The insets show histograms 
of Monte-Carlo simulations for mass and radius. 
The best-fitting values were calculated as the median and 84.1 and 15.9 percentile 
(corresponding to the $1\,\sigma$ confidence limits) of the distributions. 

To account for systematics due to different model 
grids, the final parameters for each star were calculated as the median over all 
methods, with uncertainties estimated by adding 
in quadrature the median formal uncertainty and the scatter over all methods.
Table 2 lists the final parameters for all host stars in our sample. 
The median uncertainties in radius and mass are 3\,\% and 7\,\%, 
respectively, consistent with the limits discussed in 
Section 2.1. Note that for 11 hosts we have adopted the solutions published in separate 
papers. We also note that for some stars 
the mass and radius distributions are not symmetric due to the ``hook'' in 
evolutionary tracks just before Hydrogen exhaustion
in the core (see Figure \ref{fig:hrdexample}). 
In general, however, 
variations in the results of different model grids are larger than these asymmetries and 
hence it is valid to assume symmetric (Gaussian) error bars, as 
reported in Table 2. 
Table 2 also reports the 
stellar density directly derived from Equation (\ref{equ:dnu}).
Importantly, the stellar properties presented here are
independent of the properties of the planets themselves, 
and hence can be directly used to re-derive planetary parameters.
Table 3 lists revised radii and semi-major axes for the \npl\ planet candidates in our 
sample, calculated using the transit parameters in \citet{batalha12}.

To check the consistency of the stellar properties, we compared the final 
radius and mass estimates in Table 2 with those calculated by solving Equations 
(\ref{equ:dnu}) and (\ref{equ:nmax}) (the direct method) for stars which have both 
reliable \numax\ and \Dnu\ measurements. We found excellent agreement 
between both determinations, with no systematic deviations and a scatter 
consistent with the uncertainties from the direct method.

\section{Comparison with previous stellar parameters}

\subsection{Revised stellar parameters in \citet{batalha12}}

The planet-candidate catalog by \citet{batalha12} included a revision of stellar properties 
based on matching available constraints to Yonsei-Yale evolutionary tracks. 
This revision (hereafter referred to as YY values) was justified since KIC surface gravities 
for some stars, in particular for 
cool M-dwarfs and for G-type dwarfs, seemed unphysical compared to predictions from 
stellar evolutionary theory. 
When available, the starting values for this revision were spectroscopic solutions, 
while for the remaining stars KIC parameters were used as initial guesses.
Our derived stellar parameters allow us 
to test this revision based on our subset of planet-candidate hosts.

Figure \ref{fig:hrd} shows surface gravity versus effective 
temperature for all planet-candidate hosts in the \citet{batalha12} catalog. 
The nearly horizontal dashed line shows the long-cadence 
Nyquist limit, below which short-cadence data are needed to sufficiently sample the 
oscillations. Thick red symbols in the plot show all host stars for which we have detected 
oscillations using short-cadence data (diamonds) and long-cadence data (triangles). 
Note that the $\teff$ and $\log g$ values plotted for these stars were derived from 
the combination of asteroseismology and spectroscopy, as discussed in 
Section 2 
(see also Table 2). For each detection, a thin line connects the position 
of the star determined in this work to the values published in \citet{batalha12}.

\begin{figure}
\begin{center}
\resizebox{\hsize}{!}{\includegraphics{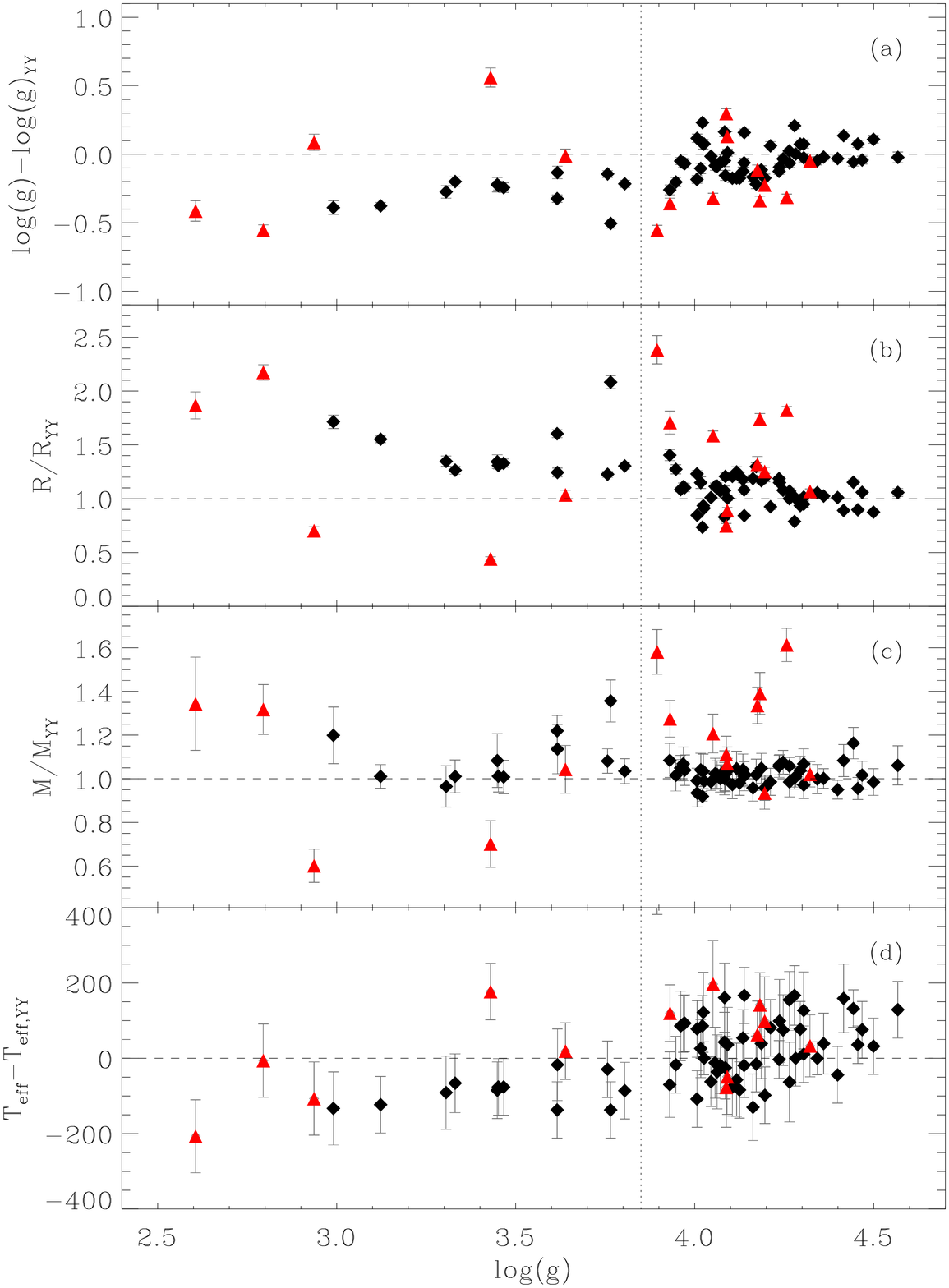}}
\caption{
\textit{(a)} Difference between $\log g$ determined from asteroseismology and 
$\log g$ given in 
\citet{batalha12} as a function of seismic $\log g$ for all host stars in our sample. Red triangles mark 
stars for which the revised parameters in \citet{batalha12} are based on 
KIC parameters, while black diamonds are stars for which spectroscopic solutions were 
used. The vertical dotted lines divides evolved stars ($\logg < 3.85$) from 
unevolved stars ($\logg > 3.85$). 
\textit{(b)} Same as panel (a) but for stellar radii.
\textit{(c)} Same as panel (a) but for stellar masses.
\textit{(d)} Same as panel (a) but for stellar effective temperatures. 
Note that KOI-1054 has been omitted from this figure 
since no full set of stellar properties was derived (see text).}
\label{fig:comp2}
\end{center}
\end{figure}

Figure \ref{fig:hrd} shows that our sample consists primarily of slightly 
evolved F to G-type
stars. This is due to the larger oscillation amplitudes in these stars 
compared to their unevolved counterparts. We observe 
no obvious systematic shift in $\log g$ for unevolved stars, while 
surface gravities for evolved giants and subgiants were generally overestimated compared 
to the asteroseismic values. To illustrate this further, Figure 
\ref{fig:comp2} shows the difference between fundamental properties (\logg, radius, 
mass and temperature) from the asteroseismic analysis and the values given by 
\citet{batalha12} as a function of surface gravity. Red triangles 
mark stars for which KIC parameters were used as initial values, while black diamonds 
are stars for which spectroscopic solutions were used. In the following we distinguish 
between unevolved and evolved stars using a cut at $\logg = 3.85$ (see 
dotted line in Figure \ref{fig:comp2}), which roughly divides our sample between 
stars before and after they reached the red-giant branch.

Table \ref{tab:diffs} summarizes the mean differences between values
derived in this work and the values given by \citet{batalha12}.
If we only use unevolved stars ($\log g > 3.85$), the differences for stars 
with spectroscopic follow-up (black diamonds in Figure \ref{fig:comp2}) 
are small, with an average difference of $-0.04\pm0.02$\,dex (scatter of 0.12\,dex) 
in \logg\ and $6\pm2$\% (scatter of 15\%) in 
radius. 
For stars based on KIC parameters (red triangles in Figure \ref{fig:comp2}), the 
mean differences are considerably larger with $-0.17\pm0.10$\,dex (scatter of 0.29 dex) 
in \logg\ and $41\pm17$\% (scatter of 52\%) in radius.
This confirms previous studies predicting overestimated KIC radii for \kep\ targets
\citep{verner11b,gaidos13} and shows that, as expected, spectroscopy yields a strong
improvement (both in reduced scatter and offset) compared to the KIC. This 
emphasizes the need for a systematic spectroscopic follow-up of all planet-candidate hosts.

The unevolved planet-candidate host with the largest change in stellar 
parameters from this study is 
KOI-268 (see annotation in Figure \ref{fig:hrd}), which has 
previously been classified as a late K-dwarf with $\teff \sim 4800$\,K, hosting a 1.6 
\rearth\ planet in a 110\,d orbit. Our asteroseismic analysis yields 
$\log g = 4.26\pm0.01$, which is clearly incompatible with a late-type dwarf. 
Follow-up spectroscopy revised \teff\ for this star to 6300\,K, correctly 
identifying it as a F-type dwarf. The revised radius for the planet candidate 
based on the stellar parameters presented here is $3.00\pm0.06$\,\rearth\ (compared 
to the previous estimate of 1.6\,\rearth).

We note that four unevolved hosts in our sample are confirmed planetary systems that until 
now had no available asteroseismic constraints: Kepler-4 \citep{borucki10b}, 
Kepler-14 \citep{buchhave11}, 
Kepler-23 \citep{ford12} and Kepler-25 \citep{steffen12}. 
The agreement with the 
host-star properties published in the discovery papers 
(based on spectroscopic constraints with evolutionary tracks) is good, and the
stellar parameters presented here will be valuable for future studies of these systems.
We have also compared the our results for Kepler-4 and Kepler-14 with 
stellar properties published by
\citet{southworth11} and \citet{southworth12}, 
and found good agreement within $1-\sigma$ for mass and radius.

\begin{table*}
\begin{small}
\begin{center}
\renewcommand\thetable{4}      
\caption{Mean differences between host-star properties in this study and as given 
in \citet{batalha12}.}
\begin{tabular}{l | c c | c c}        
\hline         
Parameter    & \multicolumn{2}{|c|}{$\logg > 3.85$} & \multicolumn{2}{|c}{$\logg < 3.85$} 	\\ 
    & Spectroscopy & KIC  & Spectroscopy & KIC													\\
\hline
$\Delta(\logg)$ (dex)	& $-0.04\pm0.02$ (0.12)	& $-0.17\pm0.10$ (0.29)	& $-0.27\pm0.03$ (0.11)	&	$-0.1\pm0.2$ (0.5)	 \\
$\Delta(R)$ (\%)		& $6\pm2$ (15)			& $41\pm17$ (52)		& $44\pm5$ (17)			&	$24\pm37$ (83) \\
$\Delta(M)$ (\%)		& $2\pm1$ (4)			& $21\pm7$ (21)			& $9\pm3$ (11)			&	$1\pm16$ (36) \\
$\Delta(\teff)$ (K)		& $31\pm12$ (84)		& $111\pm47$ (141)		& $-88\pm12$ (42)		&	$-25\pm70$ (155) \\					
\hline
\end{tabular} 
\label{tab:diffs} 
\end{center}
\flushleft Differences are given in the sense of values derived in this work minus the 
values given in \citet{batalha12}. Error bars are the standard error of the mean, 
and numbers in brackets are the standard deviation of the residuals.
\end{small}
\end{table*}

Turning to evolved hosts ($\log g < 3.85$), the differences between the asteroseismic 
and YY values for stars 
with spectroscopic follow-up (black diamonds in Figure \ref{fig:comp2}) 
are on average $-0.27\pm0.03$\,dex (scatter of 0.11\,dex) in $\log g$ and $44\pm5$\% 
(scatter of 17\%) in radius. 
For stars based on KIC parameters (red triangles in Figure \ref{fig:comp2}), the 
mean differences are $-0.1\pm0.2$\,dex (scatter of 0.5\,dex) in $\log g$ and $24\pm37$\% 
(scatter of 83\%) in radius.
Unlike for unevolved stars, the
parameters based on spectroscopy are systematically overestimated in $\log g$, and 
hence yield planet-candidate radii that are systematically underestimated 
by up to a factor of 1.5.
This bias is not present in the YY properties based on initial values in the 
KIC (although the scatter is high), and illustrates 
the importance of coupling asteroseismic constraints with spectroscopy,
particularly for evolved stars.

\begin{figure}
\begin{center}
\resizebox{\hsize}{!}{\includegraphics{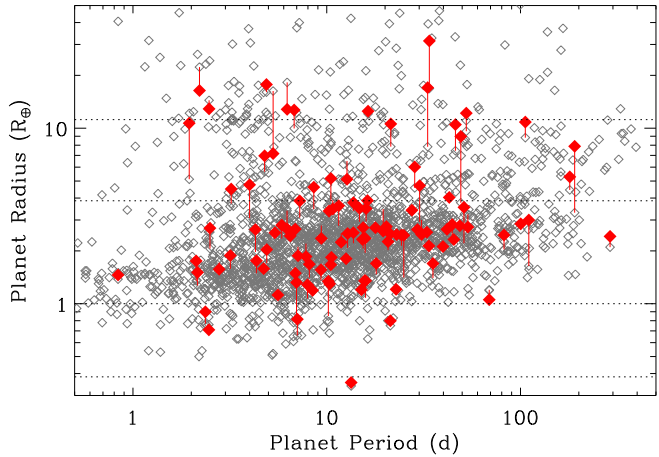}}
\caption{Planet radius versus orbital period for all candidates in the catalog by 
\citet{batalha12}. Thick red diamonds show the rederived radii for all planetary candidates 
included in our sample with revised radii $<50\rearth$, 
with thin red lines connecting the updated radii to those 
published in the \kep\ planet-candidate catalog \citep{batalha12}. 
Dotted lines mark the radii of Mercury, Earth, 
Neptune and Jupiter.}
\label{fig:comp_planet}
\end{center}
\end{figure}

Figure \ref{fig:comp_planet} shows planet-candidate radii versus 
orbital periods for the full planet-candidate catalog, highlighting 
all candidates in our sample with revised radii $<50\rearth$ in red.
As in Figure \ref{fig:hrd}, thin lines connect our rederived
radii to the values published by \citet{batalha12}.
For a few evolved host stars, the revised host radii
change the status of the candidates from planetary companions to objects that are 
more compatible with brown dwarfs or low-mass stars. The first planet-candidate host that was 
identified as a false-positive using asteroseismology was KOI-145.01, as discussed by 
\citet{gilliland10}. KOI-2640.01 is another example for a potential 
asteroseismically determined false-positive, with an increase of the companion radius from 
9\,\rearth\ to $17.0\pm0.8$\,\rearth. Additionally, the companions of KOI-1230 and KOI-2481 are now 
firmly placed in the stellar mass regime with revised radii of $64\pm2$\,\rearth\ ($0.58\pm0.02$\,\rsun) and 
$31\pm2$\,\rearth\ ($0.29\pm0.02$\,\rsun), respectively.
On the other hand, for KOI-1894 the companion radius becomes 
sufficiently small to qualify the companion as a sub-Jupiter size
planet candidate, with a decreased radius from 16.3\rearth\ to $7.2\pm0.4$\rearth.
KOI-1054 is a peculiar star in the sample with a very low metallicity 
($\rm{[Fe/H]}=-0.9$\,dex). Despite the lack of a reliable \Dnu\ measurement our 
study confirms that this star is a evolved giant with $\logg=2.47\pm0.01$\,dex, 
indicating that the potential companion with an orbital period of only 3.3 days is likely 
a false-positive.

\subsection{Identification of misclassified Giants}

A recent study by \citet{mann12} showed that 
$\sim 96$\% of all bright ($K_{\rm p}<14$) and cool ($T_{\rm eff} < 4500$\,K) stars in the 
\kep\ Input Catalog are giants. This raises considerable worry about a 
giant contamination among the cool planet-candidate host sample, and has implications 
for studies of planet detection completeness and 
the occurrence rates of planets with a given size.  
Asteroseismology provides an efficient tool to identify giants using 
\kep\ photometry alone, without the need for follow-up observations. 
Since oscillation amplitudes 
scale with stellar luminosity (see, e.g., the y-axis scale in Figure 
\ref{fig:examples}), giants show large amplitudes that are 
detectable in all typical \kep\ 
targets. In addition, oscillation timescales 
scale with stellar luminosity, with frequencies for 
red giants falling well below the long-cadence Nyquist limit 
\citep[see, e.g.,][]{bedding10b,hekker11c,mosser11c}. Hence, any cool
giant should show detectable oscillations using long-cadence data, 
which is readily available for all \kep\ targets. 

A few caveats to this method exist. First, there is a cut-off in 
temperature below 
which the variability is too slow to reliably measure oscillations. 
For data up to Q11 this limit is about $3700$\,K, 
corresponding to a $\numax\sim 1\muHz$
at solar metallicity. 
Second, there is evidence that tidal interactions from close stellar companions can suppress 
oscillations in giants, as first observed in the 
hierarchical triple system HD\,180891 \citep{derekas11,borkovits12,fuller13}. 
Hence, giant stars that are also in multiple systems with 
close companions could escape a detection with our method.

\begin{figure}
\begin{center}
\resizebox{\hsize}{!}{\includegraphics{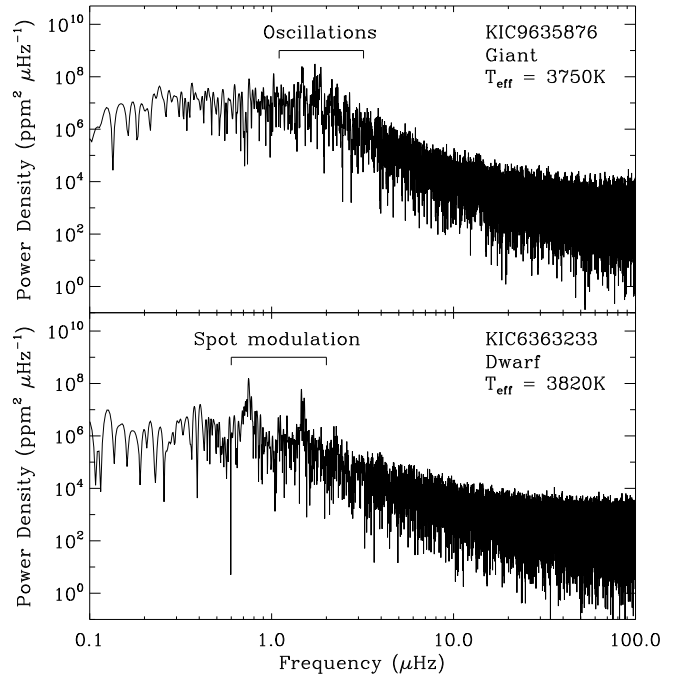}}
\caption{Comparison of long-cadence power spectra of a giant (top panel) and a 
dwarf (bottom panel) with similar effective temperatures in the sample by 
\citet{mann12}. Note that KIC9635876 has been classified as a dwarf in the KIC.}
\label{fig:comp1}
\end{center}
\end{figure}

To test the success rate of asteroseismic giant identifications, we 
analyzed 132 stars for which \citet{mann12} presented a spectroscopic luminosity class 
and for which several quarters of \kep\ data are available. 
As an example, Figure \ref{fig:comp1} compares the power spectra of the dwarf KIC6363233 
and the giant KIC9635876. 
The power spectrum of the dwarf
shows a strong peak at $\sim 0.8$\,\muHz\ accompanied by several harmonics, 
a typical signature of non-sinusoidal variability 
due to rotational spot modulation. The giant, on the other hand, shows 
clear power excess with regularly spaced peaks that are typical for solar-like oscillations, 
indicating the evolved nature of the object. 
Out of the 132 stars in the sample, 96 were 
identified as giants based on detection of oscillations, compared to 104 stars 
that were identified as giants by \citet{mann12}. All eight stars that were missed 
by the asteroseismic 
classification are cool ($\lesssim3700$\,K) giants, for which the 
oscillation timescales are likely too long to be resolved with the available amount of 
\kep\ data. Hence, this result implies a very high success rate when \teff\ restrictions 
are taken into account, and suggests 
that giants with suppressed oscillations by close-in stellar companions are rare.

As shown in Figure \ref{fig:hrd}, our analysis
did not yield a detection of oscillations compatible with giant stars in 
any of the cool main-sequence hosts. This confirms that the majority of these stars are 
indeed dwarfs. Additionally, we did not detect oscillations in stars near 
the 14\,Gyr isochrone of the YY models (which can be seen as a ``finger''
between $\teff=4700-5000$\,K and $\logg=3.8-4.0$ in Figure \ref{fig:hrd}). The 
non-detection of oscillations confirms 
that these stars must 
have $\logg \gtrsim 3.5$, and hence are either subgiants or cool dwarfs.

\subsection{Stellar Density from Transit Measurements}

The observation of transits allows a measurement of the 
semi-major axis as a function of the stellar radius ($a/R_{*}$), 
provided the eccentricity of the orbit is known. 
For the special case of circular orbits, $a/R_{*}$ is directly related to the mean 
density of the star \citep[see, e.g.,][]{seager03,winn10b}:

\begin{equation}
\langle\rho_{\star}\rangle \approx \frac{3\pi}{GP^{2}}\left(\frac{a}{R_{\star}}\right)^3 \: ,
\label{equ:rho}
\end{equation}

\noindent
where $G$ is the gravitational constant and $P$ is the orbital period. 
Equation (\ref{equ:rho}) can be used 
to infer host-star properties in systems with transiting exoplanets, for example to 
reduce degeneracies between spectroscopic parameters 
\citep[see, e.g.,][]{sozzetti07,torres12}. 

\begin{figure}
\begin{center}
\resizebox{\hsize}{!}{\includegraphics{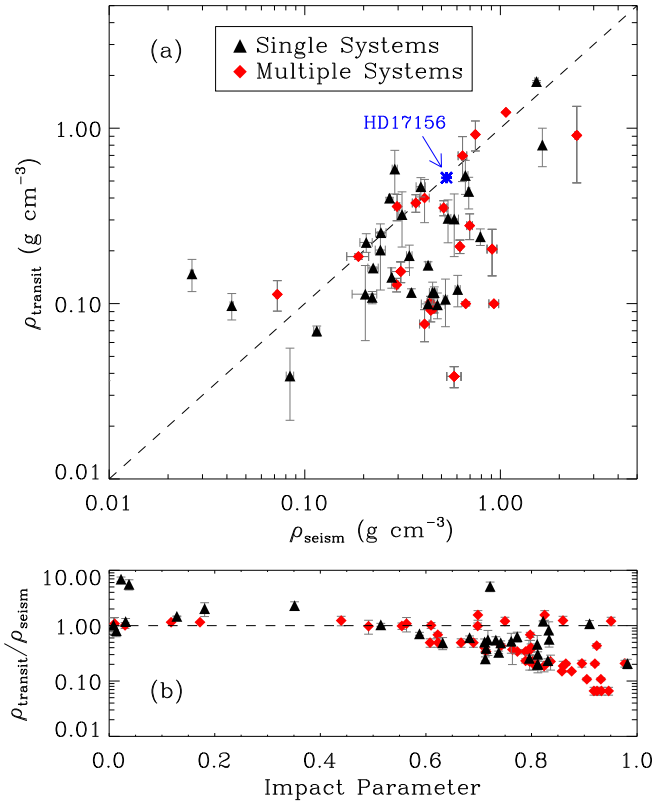}}
\caption{\textit{(a)} Mean stellar density as measured from the transit assuming 
circular orbits ($d/R_{*} = a/R_{*}$) versus the density measured from 
asteroseismology. \textit{(b)} Fractional difference between the density measured 
from the transit and asteroseismology as a function of the modeled impact parameter. 
Note that since all planets are assumed to orbit the same star, 
panel (a) shows one datapoint for each host star, while panel (b) shows one 
datapoint for each planet candidate.}
\label{fig:rho}
\end{center}
\end{figure}

Figure \ref{fig:rho}a compares the stellar densities derived from Equation 
(\ref{equ:rho}) assuming $d/R_{*} = a/R_{*}$ in the table of \citet{batalha12} 
with our independent estimates from asteroseismology. 
We emphasize that \citet{batalha12} 
explicitly report the quantity $d/R_{*}$ to point out that it is only a valid 
measurement of stellar density in the case of zero eccentricity, for which 
$d/R_{*} = a/R_{*}$. 
We also note that the uncertainties reported by \citet{batalha12} do 
not account for correlations between transit parameters, and hence the density 
uncertainties are likely underestimated. For this first comparison, 
we have excluded all hosts with a transit density uncertainty $>50\%$.

The comparison shows differences greater than 50\% for more than half of the sample, 
with mostly underestimated stellar densities from the transit model compared to the 
seismic densities. 
To investigate the cause of this discrepancy, 
Figure \ref{fig:rho}b shows the fractional difference of
stellar densities as a function of the impact parameter (the sky-projected distance 
of the planet to center of the stellar disc, expressed in units of the stellar radius) 
for each planet candidate. 
We observe a clear correlation, with large disagreements corresponding preferentially to 
high impact parameters. 
We have tested whether this bias is due to insufficiently sampled ingress and egress 
times by repeating the transit fits for a fraction of the host stars using short-cadence 
data, and found that the agreement significantly improves if short-cadence data is used.
Further investigation showed that the larger disagreements are found for the 
shallower transits, while better agreement is found for transits with the highest S/N. 
Hence, it appears that impact parameters tend to be overestimated 
for small planets, which is compensated by underestimating the density of the star 
to match the observed transit duration. The reason for this bias may 
to be due to anomalously long ingress times for small planets caused by 
smearing due to uncorrected transit timing variations or other effects.

Additional reasons for the discrepancies between the transit and seismic densities 
include planet candidates on eccentric orbits, and false-positive planet candidates. 
For eccentric orbits the stellar density derived from the transit  
can be either over- or underestimated (depending on the orientation of the orbit to 
the observer).
However, in this case no correlation with the impact parameter would be 
expected, and the distribution of impact parameters should be uniform. 
Additionally, if eccentric orbits were responsible for the majority of the 
outliers, we would expect to detect a correlation of the fractional difference in density 
with orbital period, with planet candidates on short orbital periods 
showing preferentially good agreement due to tidal circularization. However, no such 
correlation is apparent in our data.
For false-positive scenarios (e.g. a transit around a fainter background star) the 
large dilution would lead to underestimated densities, and a correlation with the 
impact parameter would be expected. However, recent results by \citet{fressin13} 
showed that the global false-positive rate is $\lesssim10$\%.
Hence, eccentric orbits and false-positives are likely not responsible for the 
majority of the outliers.

The first comparison shown here underlines the statement in \citet{batalha12} that 
stellar properties derived from the transit fits in the planet-candidate catalog 
should be viewed with caution, with further work being needed to quantify these 
differences.
We emphasize that the comparison shown here does not imply that transits cannot be 
used to accurately infer stellar densities. To demonstrate this, Figure 
\ref{fig:rho}a also includes HD\,17156, a system with an exoplanet in 
a highly eccentric orbit ($e=0.68$) for which high S/N constraints from transits, radial 
velocities and 
asteroseismology are available \citep{gilliland11,nutzman11}. The seismic and transit 
density are in excellent agreement, demonstrating that both techniques 
yield consistent results when the eccentricity and impact parameter can be accurately 
determined. Similar tests can be 
expected in future studies, making use of \kep\ exoplanet hosts for which  
asteroseismic and radial velocity 
constraints are available \citep{batalha11,gilliland13,marcy13}. 
Additionally, the precise stellar properties presented here will enable improved 
determinations of eccentricities using high S/N transit 
light curves \citep{dawson12}, and yield improved constraints for the study of 
eccentricity distributions in the \kep\ planet sample compared to planets detected with
radial velocities \citep[see, e.g.,][]{wang11,moorhead11,kane12,plavchan12}

\section{Exoplanet -- Host Star Correlations}

Accurate stellar properties of exoplanet hosts, as presented in this 
study, are valuable for testing theories of planet formation, many of which are 
related to host-star properties. While both the limited sample size and the 
difficult characterization of detection biases push a comprehensive 
investigation of exoplanet -- host star correlations beyond the 
scope of this paper, we present a first qualitative comparison here.
Note that in the following we have omitted planet candidates that have been identified 
as false positives (see Section 3.1) or for which unpublished 
radial-velocity follow-up has indicated a low-mass stellar companion (see Table 3).

\subsection{Background}

The favored theoretical scenario 
for the formation of terrestrial planets and the cores of gas giants 
is the core-accretion model, a slow process involving the collision 
of planetesimals \citep{safronov69}, with giant planets growing massive enough to gravitationally 
trap light gases \citep{mizuno80,pollack96,lissauer09,movshovitz10}. 
The efficiency of this process is predicted to be correlated with the disk properties and 
therefore the characteristics of the host star, such as stellar mass \citep{thommes08}.

Observationally, Doppler velocity surveys have yielded 
two important correlations: gas giant planets
occur more frequently around stars of high metallicity
\citep{gonzalez97,santos04,fischer05} and around more massive stars
\citep{laws03,johnson07,lovis07,johnson10}.
Early results using \kep\ planet candidates showed that single planets appear to be
more common around hotter stars, while multiple 
planetary system are preferentially found around cooler stars \citep{latham11}. 
Both results are in-line with 
\citet{howard11}, who found that small planets are more common around cool 
stars. More recently, \citet{steffen12c} showed evidence 
that hot Jupiters indeed tend to be found in single planet systems, while 
\citet{fressin13} found that the occurrence of small planets appears to be 
independent of the host-star spectral type. 
While most of these findings have been 
interpreted in favor of the core-accretion scenario, many results have so 
far relied on uncertain or indirect estimates of stellar mass (such as \teff).
The sample of host stars presented in this study allow 
us to test these results using accurate radii and masses from asteroseismology.

\subsection{Planet Radius versus Stellar Mass}

Following \citet{howard11}, we attempted to account for detection biases by estimating 
the smallest detectable planet for a 
given planet-candidate host in our sample as:

\begin{equation}
R_{\rm min} = R_{\star} (\rm{SNR} \, \sigma_{\rm CDPP})^{0.5} \left(\frac{n_{\rm tr} t_{\rm dur}}{6 \rm{hr}}\right)^{0.25} \: .
\end{equation}

\noindent
Here, $R_{\star}$ is the host-star radius, SNR is the required signal-to-noise ratio, 
$\sigma_{\rm CDPP}$ is the 6\,hr combined differential photometric precision 
\citep{christiansen12}, $n_{\rm tr}$ is 
the number of transits observed and $t_{\rm dur}$ is the duration of the transits which, 
for the simplified case of circular orbits and a central transit (impact parameter 
$b=0$), is given by:

\begin{equation}
t_{\rm dur} = \frac{R_{*} P}{\pi a} \: .
\end{equation}

\begin{figure}
\begin{center}
\resizebox{\hsize}{!}{\includegraphics{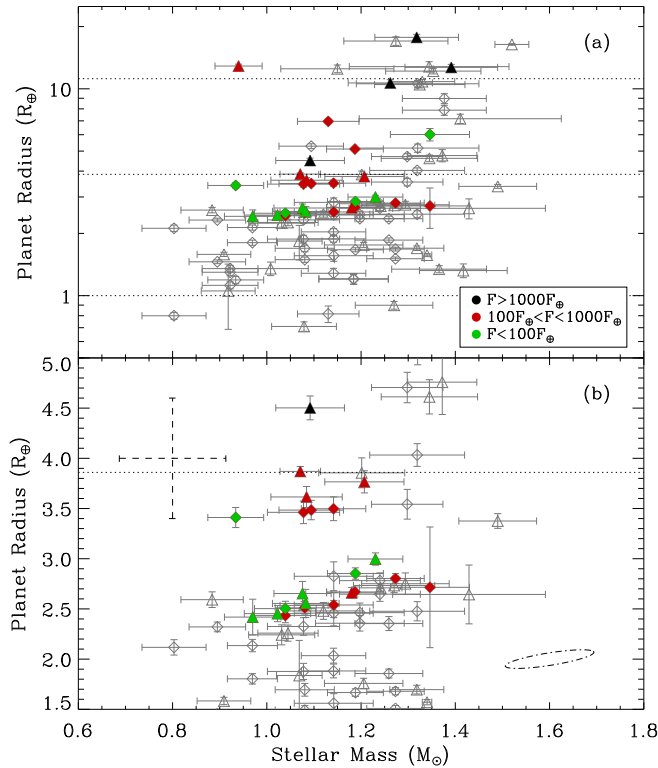}}
\caption{\textit{(a)} Planet radius versus host-star mass for all planet-candidate hosts 
in our sample. 
Planet candidates with radii $>2.4\,\rearth$ and $R_{\rm min}<2.4\,\rearth$ are 
shown as filled symbols, while other candidates are shown as open grey symbols. 
Triangles and diamonds denote candidates in single and multiple systems, respectively. 
Colors denote the incident stellar flux, as indicated in the legend.
Horizontal dotted lines show the sizes of Earth, Neptune and Jupiter. 
\textit{(b)} Same as panel (a) but only showing candidates with radii between 1.5 and 5\,\rearth. 
The dashed error bar shows typical $1-\sigma$ uncertainties without asteroseismic 
constraints. The dashed-dotted line shows a typical $1-\sigma$ error ellipse, 
illustrating the correlation of uncertainties between stellar 
mass and planet radius in the asteroseismic sample.}
\label{fig:comp_mass}
\end{center}
\end{figure}

\noindent
Here, $P$ is the orbital period and $a$ is the semi-major axis of the orbit. 
For each candidate, we estimate $R_{\rm min}$ by
calculating $t_{\rm dur}$, adopting the median 6\,hr $\sigma_{\rm CDPP}$ from quarters 1--6 
for each star and setting a signal-to-noise threshold of 25 \citep{ciardi13}.
To de-bias our sample, we calculated for a range of 
planet sizes $R_{x}$ the number of planet candidates that are larger than $R_{x}$ and 
for which $R_{\rm min}<R_{x}$. The maximum number of 
planet candidates fulfilling these criteria was found for a value of 
$R_{x}=2.4\rearth$. In the 
following, our debiased sample consists only of planet candidates with 
$R>2.4\rearth$ and with $R_{\rm min}<2.4\rearth$.
Figure \ref{fig:comp_mass}a shows planet radii versus host-star mass 
for all planet candidates in our sample (grey symbols), with the de-biased sample 
shown as filled symbols. Planet candidates in multi systems are shown as 
diamonds, while single systems are shown as triangles. For all candidates in the 
debiased sample, symbols are additionally color-coded according 
to the incident flux as a multiple of the flux incident on Earth.

While the size of our de-biased sample precludes definite conclusions, we see that
our observations are consistent with previous studies that found 
gas-giant planets to be less common 
around low-mass stars ($<1\msun$) than around more massive stars.
A notable exception is KOI-1 (TReS-2), a 0.9\,\msun\ 
K-dwarf hosting a hot Jupiter in a 2.5-day orbit \citep{donovan06,holman07,kipping11,barclay12}. 
However, this observation is only marginally significant:
considering the full sample of stars hosting planet candidates with $R>4\rearth$, 
the probability of observing one host with $M<1\msun$ by chance is 
$\sim 1\%$, corresponding to a $\sim 2.5\sigma$ 
significance that the mass 
distribution is different.
Using our unbiased sample, no statistically significant difference is found.

For sub-Neptune sized objects, planet 
candidates are detected around stars with masses ranging over the full span of our 
sample ($\sim 0.8-1.6\,\msun$). This is illustrated in Figure \ref{fig:comp_mass}b, 
showing a close-up of the 
region of sub-Neptune planets on a linear scale. 
To demonstrate the improvement of the uncertainties in our sample, 
the dashed lines in Figure \ref{fig:comp_mass}b shows an error bar
for a typical best-case 
scenario when host-star properties are based on spectroscopy alone
\citep[10\% in stellar mass and 15\% in stellar radius,][]{basu12}, neglecting 
uncertainties arising from the measurement of the transit depth.
We note that, strictly speaking, the uncertainties in stellar mass and planet radius 
are not independent since asteroseismology constrains mostly the mean stellar density. 
The dashed-dotted line in Figure \ref{fig:comp_mass}b illustrates this by showing an  
error ellipse calculated from Monte-Carlo simulations 
for a typical host-planet pair in our sample.

The data in Figure \ref{fig:comp_mass}b show a tentative lack of planets with sizes close 
to 3\rearth, which does not seem to be related 
to detection bias in the sample.
While this gap is intriguing, it is
not compatible with previous observations of planets with radii between $3-3.5\rearth$,  
such as in the Kepler-11 system \citep{lissauer11}.
Further observations will be needed to determine 
whether the apparent gap in Figure \ref{fig:comp_mass}b is real or simply a consequence 
of the small size of the available sample.

\begin{figure}
\begin{center}
\resizebox{\hsize}{!}{\includegraphics{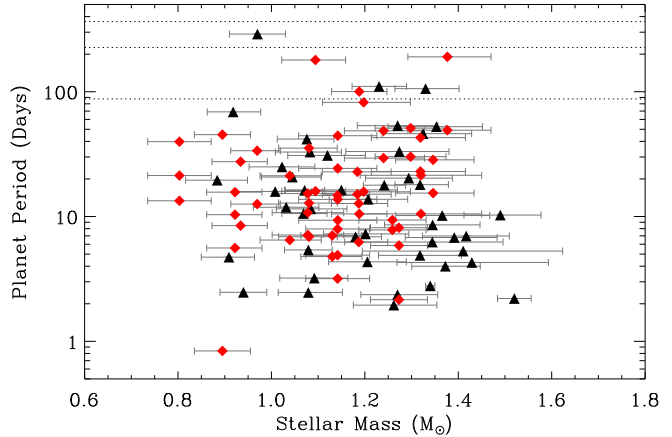}}
\caption{Orbital period versus stellar mass for all planet candidates in our sample. 
Black triangles are candidates in single systems and red diamonds are 
candidates in multiple systems. The 
orbital periods of Mercury, Venus and Earth are shown as horizontal dotted lines.}
\label{fig:hist}
\end{center}
\end{figure}

\subsection{Planet Period and Multiplicity versus Stellar Mass}

The second main observable that can be tested for correlations with the host-star mass is 
the orbital period of the planet candidates. Here, we do not expect a detection bias 
to be correlated with the host-star properties, and hence we consider the full sample. 
Figure \ref{fig:hist} compares the orbital period of the planet candidates as 
a function of host-star mass. 
There does not appear to be an overall trend, although 
there is a tendency for more single-planet candidates (black triangles) in 
close orbits ($<10$\,days) around higher mass ($\gtrsim 1.3\msun$) stars. 
This observation would qualitatively be
consistent with 
previous findings that hot-Jupiters are rare in multiple planet systems
and are more frequently found around higher mass stars.
However, we note that roughly half of the planet candidates with periods 
less than 10 days around stars with $M>1.3\msun$ have radii smaller than Neptune, 
and two have radii smaller than 2\,\rearth (see also Figure \ref{fig:comp_mass}).
Additionally, a K-S test yields only a marginal statistical difference ($\sim 2.4\sigma$) 
between host-star masses of 
single and multiple planet systems for periods $<10$\,days.

\section{Conclusions}

We have presented an asteroseismic study of \kep\ 
planet-candidate host stars in the
catalog by \citet{batalha12}. Our analysis yields new asteroseismic radii and masses 
for 66 host stars
with typical uncertainties of 3\% and 7\%, respectively, raising the total number 
of \kep\ host stars with asteroseismic solutions to 77.
Our main findings can be summarized as 
follows:

\begin{enumerate}

\item Surface gravities for subgiant and giant host stars in \citet{batalha12}
based on high-resolution spectroscopy are systematically overestimated, 
yielding underestimated 
stellar radii (and hence planet-candidate radii) by up to a factor of 1.5.
While properties for unevolved stars based on spectroscopy are in 
good agreement and
show greatly improved results compared to the KIC, 
the identified systematics illustrate 
the importance of combining spectroscopy with asteroseismic constraints 
to derive accurate and precise host-star properties.

\item We have demonstrated that asteroseismology is an efficient method to identify 
giants using \kep\ data. Our analysis yielded no detection of oscillations in host stars 
classified as M dwarfs, confirming that 
the fraction of misclassified giants in the cool planet-candidate host-star 
sample is small. 
An extension of this analysis to the complete \kep\ target sample 
is planned, and will support completeness studies of \kep\ planet detections and hence 
the determination of the 
frequency of Earth-sized planets in the habitable zone.

\item A comparison of mean stellar densities from asteroseismology and from transit 
models in \citet{batalha12}, assuming zero eccentricity, showed significant differences 
for at least 50\% of the 
sample. Preliminary investigations imply that these differences are 
mostly due to systematics in the modeled transit parameters, while some differences may
due to planet candidates in eccentric orbits. The 
independent asteroseismic densities presented here will be valuable for more detailed 
studies of the 
intrinsic eccentricity distribution of planets in this sample and for testing 
densities inferred from transits for planet-candidate host stars with available 
radial-velocity data.

\item We presented re-derived radii and semi-major axes for the \npl\ planet candidates in 
our sample based on the revised host-star properties.
We identified KOI-1230.01 and KOI-2481.01 as astrophysical false-positives, with revised 
companion radii of $64\pm2$\,\rearth\ ($0.58\pm0.02$\,\rsun), $31\pm2$\,\rearth\ ($0.29\pm0.02$\,\rsun), respectively, 
while KOI-2640.01 is a potential false-positive with a radius of $17.0\pm0.8$\,\rearth. 
On the other hand, the radius of KOI-1894.01 
decreases from the brown-dwarf/stellar regime to a
sub-Jupiter size ($7.2\pm0.4\,\rearth$). Our sample also includes accurate 
asteroseismic radii and masses for four hosts 
with confirmed planets: Kepler-4, Kepler-14, Kepler-23 and Kepler-25. 

\item We investigated correlations between host-star masses and planet-candidate 
properties, and find that our observations are consistent with previous studies 
showing that gas giants are less common around lower-mass ($\lesssim 1\msun$) stars. 
Sub-Neptune sized planets, on the other hand, 
appear to be found over the full range of host masses considered in 
this study ($\sim 0.8-1.6\,\msun$). We also observe a potential 
preference for close-in planets around higher mass 
stars to be in single systems. Due to the small sample size, 
however, these findings are tentative only and 
will have to await confirmation using 
larger samples with precise host-star properties.

\end{enumerate}

The results presented here illustrate the powerful synergy between 
asteroseismology and exoplanet studies. 
As the \kep\ mission progresses, asteroseismology will continue to 
play an important role in characterizing new \kep\ planet candidates, particularly for 
potential long-period planets in the habitable zones of F-K dwarfs. 
An important future step will also be to extend the sample of planets with determined 
masses through radial velocity follow-up \citep[see, e.g.,][]{latham10,cochran11} or 
transit-timing variations \citep[see, e.g.,][]{fabrycky12,ford12,steffen12}
for hosts for which asteroseismic constraints 
are available. This will enable precise constraints on planet densities. 
Additionally, the analysis of individual frequencies for planet-candidate hosts 
with high S/N detections will allow the precise determination of stellar ages, which can 
be used to investigate the chronology of their planetary systems. 

\section*{Acknowledgments}
We thank Willie Torres, Josh Winn and our anonymous referee for helpful comments and 
discussions. We furthermore gratefully acknowledge the entire \kep\ team and everyone 
involved in the \kep\ mission for making this paper possible. 
Funding for the \kep\ Mission is provided by NASA's Science Mission Directorate. 
DH is supported by an appointment to the NASA Postdoctoral Program at Ames Research Center, 
administered by Oak Ridge Associated Universities through a contract with NASA. 
SB acknowledges NSF grant AST-1105930. SH acknowledges financial support from the 
Netherlands Organisation for Scientific Research (NWO). 
TSM acknowledges NASA grant NNX13AE91G.
Funding for the Stellar Astrophysics Centre is provided by The Danish  
National Research Foundation (Grant DNRF106). The research is supported by the  
ASTERISK project (ASTERoseismic Investigations with SONG and Kepler)  
funded by the European Research Council (Grant agreement no.: 267864).

\bibliographystyle{apj}
\bibliography{/Users/daniel/science/codes/latex/references}

\begin{thebibliography}{155}
\expandafter\ifx\csname natexlab\endcsname\relax\def\natexlab#1{#1}\fi

\bibitem[{{Aerts} {et~al.}(2010){Aerts}, {Christensen-Dalsgaard}, \&
  {Kurtz}}]{aerts10}
{Aerts}, C., {Christensen-Dalsgaard}, J., \& {Kurtz}, D.~W. 2010,
  {Asteroseismology} (Springer: Dodrecht)

\bibitem[{{Barclay} {et~al.}(2012){Barclay}, {Huber}, {Rowe}, {Fortney},
  {Morley}, {Quintana}, {Fabrycky}, {Barentsen}, {Bloemen}, {Christiansen},
  {Demory}, {Fulton}, {Jenkins}, {Mullally}, {Ragozzine}, {Seader}, {Shporer},
  {Tenenbaum}, \& {Thompson}}]{barclay12}
{Barclay}, T., {et~al.} 2012, \apj, 761, 53

\bibitem[{{Barclay} {et~al.}(2013){Barclay}, {Rowe}, {Lissauer}, {Huber},
  {Fressin}, {Howell}, {Bryson}, {Chaplin}, {D{\'e}sert}, {Lopez}, {Marcy},
  {Mullally}, {Ragozzine}, {Torres}, {Adams}, {Agol}, {Barrado}, {Basu},
  {Bedding}, {Buchhave}, {Charbonneau}, {Christiansen},
  {Christensen-Dalsgaard}, {Ciardi}, {Cochran}, {Dupree}, {Elsworth},
  {Everett}, {Fischer}, {Ford}, {Fortney}, {Geary}, {Haas}, {Handberg},
  {Hekker}, {Henze}, {Horch}, {Howard}, {Hunter}, {Isaacson}, {Jenkins},
  {Karoff}, {Kawaler}, {Kjeldsen}, {Klaus}, {Latham}, {Li}, {Lillo-Box},
  {Lund}, {Lundkvist}, {Metcalfe}, {Miglio}, {Morris}, {Quintana}, {Stello},
  {Smith}, {Still}, \& {Thompson}}]{barclay12b}
---. 2013, \nat, 494, 452

\bibitem[{{Basu} {et~al.}(2010){Basu}, {Chaplin}, \& {Elsworth}}]{basu10}
{Basu}, S., {Chaplin}, W.~J., \& {Elsworth}, Y. 2010, \apj, 710, 1596

\bibitem[{{Basu} {et~al.}(2012){Basu}, {Verner}, {Chaplin}, \&
  {Elsworth}}]{basu12}
{Basu}, S., {Verner}, G.~A., {Chaplin}, W.~J., \& {Elsworth}, Y. 2012, \apj,
  746, 76

\bibitem[{{Basu} {et~al.}(2011){Basu}, {Grundahl}, {Stello}, {Kallinger},
  {Hekker}, {Mosser}, {Garc{\'{\i}}a}, {Mathur}, {Brogaard}, {Bruntt},
  {Chaplin}, {Gai}, {Elsworth}, {Esch}, {Ballot}, {Bedding}, {Gruberbauer},
  {Huber}, {Miglio}, {Yildiz}, {Kjeldsen}, {Christensen-Dalsgaard},
  {Gilliland}, {Fanelli}, {Ibrahim}, \& {Smith}}]{basu11}
{Basu}, S., {et~al.} 2011, \apjl, 729, L10

\bibitem[{{Batalha} {et~al.}(2011){Batalha}, {Borucki}, {Bryson}, {Buchhave},
  {Caldwell}, {Christensen-Dalsgaard}, {Ciardi}, {Dunham}, {Fressin},
  {Gautier}, {Gilliland}, {Haas}, {Howell}, {Jenkins}, {Kjeldsen}, {Koch},
  {Latham}, {Lissauer}, {Marcy}, {Rowe}, {Sasselov}, {Seager}, {Steffen},
  {Torres}, {Basri}, {Brown}, {Charbonneau}, {Christiansen}, {Clarke},
  {Cochran}, {Dupree}, {Fabrycky}, {Fischer}, {Ford}, {Fortney}, {Girouard},
  {Holman}, {Johnson}, {Isaacson}, {Klaus}, {Machalek}, {Moorehead},
  {Morehead}, {Ragozzine}, {Tenenbaum}, {Twicken}, {Quinn}, {VanCleve},
  {Walkowicz}, {Welsh}, {Devore}, \& {Gould}}]{batalha11}
{Batalha}, N.~M., {et~al.} 2011, \apj, 729, 27

\bibitem[{{Batalha} {et~al.}(2013){Batalha}, {Rowe}, {Bryson}, {Barclay},
  {Burke}, {Caldwell}, {Christiansen}, {Mullally}, {Thompson}, {Brown},
  {Dupree}, {Fabrycky}, {Ford}, {Fortney}, {Gilliland}, {Isaacson}, {Latham},
  {Marcy}, {Quinn}, {Ragozzine}, {Shporer}, {Borucki}, {Ciardi}, {Gautier},
  {Haas}, {Jenkins}, {Koch}, {Lissauer}, {Rapin}, {Basri}, {Boss}, {Buchhave},
  {Carter}, {Charbonneau}, {Christensen-Dalsgaard}, {Clarke}, {Cochran},
  {Demory}, {Desert}, {Devore}, {Doyle}, {Esquerdo}, {Everett}, {Fressin},
  {Geary}, {Girouard}, {Gould}, {Hall}, {Holman}, {Howard}, {Howell},
  {Ibrahim}, {Kinemuchi}, {Kjeldsen}, {Klaus}, {Li}, {Lucas}, {Meibom},
  {Morris}, {Pr{\v s}a}, {Quintana}, {Sanderfer}, {Sasselov}, {Seader},
  {Smith}, {Steffen}, {Still}, {Stumpe}, {Tarter}, {Tenenbaum}, {Torres},
  {Twicken}, {Uddin}, {Van Cleve}, {Walkowicz}, \& {Welsh}}]{batalha12}
---. 2013, \apjs, 204, 24

\bibitem[{{Bedding}(2011)}]{bedding11b}
{Bedding}, T.~R. 2011, ArXiv e-prints \rm (arXiv:1107.1723)

\bibitem[{{Bedding} {et~al.}(2010){Bedding}, {Huber}, {Stello}, {Elsworth},
  {Hekker}, {Kallinger}, {Mathur}, {Mosser}, {Preston}, {Ballot}, {Barban},
  {Broomhall}, {Buzasi}, {Chaplin}, {Garc{\'{\i}}a}, {Gruberbauer}, {Hale}, {De
  Ridder}, {Frandsen}, {Borucki}, {Brown}, {Christensen-Dalsgaard},
  {Gilliland}, {Jenkins}, {Kjeldsen}, {Koch}, {Belkacem}, {Bildsten}, {Bruntt},
  {Campante}, {Deheuvels}, {Derekas}, {Dupret}, {Goupil}, {Hatzes}, {Houdek},
  {Ireland}, {Jiang}, {Karoff}, {Kiss}, {Lebreton}, {Miglio}, {Montalb{\'a}n},
  {Noels}, {Roxburgh}, {Sangaralingam}, {Stevens}, {Suran}, {Tarrant}, \&
  {Weiss}}]{bedding10b}
{Bedding}, T.~R., {et~al.} 2010, \apjl, 713, L176

\bibitem[{{Belkacem}(2012)}]{belkacem12}
{Belkacem}, K. 2012, in SF2A-2012: Proceedings of the Annual meeting of the
  French Society of Astronomy and Astrophysics, ed. S.~{Boissier}, P.~{de
  Laverny}, N.~{Nardetto}, R.~{Samadi}, D.~{Valls-Gabaud}, \& H.~{Wozniak},
  173--188

\bibitem[{{Belkacem} {et~al.}(2011){Belkacem}, {Goupil}, {Dupret}, {Samadi},
  {Baudin}, {Noels}, \& {Mosser}}]{belkacem11}
{Belkacem}, K., {Goupil}, M.~J., {Dupret}, M.~A., {Samadi}, R., {Baudin}, F.,
  {Noels}, A., \& {Mosser}, B. 2011, \aap, 530, A142

\bibitem[{{Borkovits} {et~al.}(2013){Borkovits}, {Derekas}, {Kiss},
  {Kir{\'a}ly}, {Forg{\'a}cs-Dajka}, {B{\'{\i}}r{\'o}}, {Bedding}, {Bryson},
  {Huber}, \& {Szab{\'o}}}]{borkovits12}
{Borkovits}, T., {et~al.} 2013, \mnras, 428, 1656

\bibitem[{{Borucki} {et~al.}(2010{\natexlab{a}}){Borucki}, {Koch}, {Brown},
  {Basri}, {Batalha}, {Caldwell}, {Cochran}, {Dunham}, {Gautier}, {Geary},
  {Gilliland}, {Howell}, {Jenkins}, {Latham}, {Lissauer}, {Marcy}, {Monet},
  {Rowe}, \& {Sasselov}}]{borucki10b}
{Borucki}, W.~J., {et~al.} 2010{\natexlab{a}}, \apjl, 713, L126

\bibitem[{{Borucki} {et~al.}(2010{\natexlab{b}}){Borucki}, {Koch}, {Basri},
  {Batalha}, {Brown}, {Caldwell}, {Caldwell}, {Christensen-Dalsgaard},
  {Cochran}, {DeVore}, {Dunham}, {Dupree}, {Gautier}, {Geary}, {Gilliland},
  {Gould}, {Howell}, {Jenkins}, {Kondo}, {Latham}, {Marcy}, {Meibom},
  {Kjeldsen}, {Lissauer}, {Monet}, {Morrison}, {Sasselov}, {Tarter}, {Boss},
  {Brownlee}, {Owen}, {Buzasi}, {Charbonneau}, {Doyle}, {Fortney}, {Ford},
  {Holman}, {Seager}, {Steffen}, {Welsh}, {Rowe}, {Anderson}, {Buchhave},
  {Ciardi}, {Walkowicz}, {Sherry}, {Horch}, {Isaacson}, {Everett}, {Fischer},
  {Torres}, {Johnson}, {Endl}, {MacQueen}, {Bryson}, {Dotson}, {Haas},
  {Kolodziejczak}, {Van Cleve}, {Chandrasekaran}, {Twicken}, {Quintana},
  {Clarke}, {Allen}, {Li}, {Wu}, {Tenenbaum}, {Verner}, {Bruhweiler}, {Barnes},
  \& {Prsa}}]{borucki10}
---. 2010{\natexlab{b}}, Science, 327, 977

\bibitem[{{Borucki} {et~al.}(2011{\natexlab{a}}){Borucki}, {Koch}, {Basri},
  {Batalha}, {Boss}, {Brown}, {Caldwell}, {Christensen-Dalsgaard}, {Cochran},
  {DeVore}, {Dunham}, {Dupree}, {Gautier}, {Geary}, {Gilliland}, {Gould},
  {Howell}, {Jenkins}, {Kjeldsen}, {Latham}, {Lissauer}, {Marcy}, {Monet},
  {Sasselov}, {Tarter}, {Charbonneau}, {Doyle}, {Ford}, {Fortney}, {Holman},
  {Seager}, {Steffen}, {Welsh}, {Allen}, {Bryson}, {Buchhave},
  {Chandrasekaran}, {Christiansen}, {Ciardi}, {Clarke}, {Dotson}, {Endl},
  {Fischer}, {Fressin}, {Haas}, {Horch}, {Howard}, {Isaacson}, {Kolodziejczak},
  {Li}, {MacQueen}, {Meibom}, {Prsa}, {Quintana}, {Rowe}, {Sherry},
  {Tenenbaum}, {Torres}, {Twicken}, {Van Cleve}, {Walkowicz}, \&
  {Wu}}]{borucki11b}
---. 2011{\natexlab{a}}, \apj, 728, 117

\bibitem[{{Borucki} {et~al.}(2011{\natexlab{b}}){Borucki}, {Koch}, {Basri},
  {Batalha}, {Brown}, {Bryson}, {Caldwell}, {Christensen-Dalsgaard}, {Cochran},
  {DeVore}, {Dunham}, {Gautier}, {Geary}, {Gilliland}, {Gould}, {Howell},
  {Jenkins}, {Latham}, {Lissauer}, {Marcy}, {Rowe}, {Sasselov}, {Boss},
  {Charbonneau}, {Ciardi}, {Doyle}, {Dupree}, {Ford}, {Fortney}, {Holman},
  {Seager}, {Steffen}, {Tarter}, {Welsh}, {Allen}, {Buchhave}, {Christiansen},
  {Clarke}, {Das}, {D{\'e}sert}, {Endl}, {Fabrycky}, {Fressin}, {Haas},
  {Horch}, {Howard}, {Isaacson}, {Kjeldsen}, {Kolodziejczak}, {Kulesa}, {Li},
  {Lucas}, {Machalek}, {McCarthy}, {MacQueen}, {Meibom}, {Miquel}, {Prsa},
  {Quinn}, {Quintana}, {Ragozzine}, {Sherry}, {Shporer}, {Tenenbaum}, {Torres},
  {Twicken}, {Van Cleve}, {Walkowicz}, {Witteborn}, \& {Still}}]{borucki11}
---. 2011{\natexlab{b}}, \apj, 736, 19

\bibitem[{{Borucki} {et~al.}(2012){Borucki}, {Koch}, {Batalha}, {Bryson},
  {Rowe}, {Fressin}, {Torres}, {Caldwell}, {Christensen-Dalsgaard}, {Cochran},
  {DeVore}, {Gautier}, {Geary}, {Gilliland}, {Gould}, {Howell}, {Jenkins},
  {Latham}, {Lissauer}, {Marcy}, {Sasselov}, {Boss}, {Charbonneau}, {Ciardi},
  {Kaltenegger}, {Doyle}, {Dupree}, {Ford}, {Fortney}, {Holman}, {Steffen},
  {Mullally}, {Still}, {Tarter}, {Ballard}, {Buchhave}, {Carter},
  {Christiansen}, {Demory}, {D{\'e}sert}, {Dressing}, {Endl}, {Fabrycky},
  {Fischer}, {Haas}, {Henze}, {Horch}, {Howard}, {Isaacson}, {Kjeldsen},
  {Johnson}, {Klaus}, {Kolodziejczak}, {Barclay}, {Li}, {Meibom}, {Prsa},
  {Quinn}, {Quintana}, {Robertson}, {Sherry}, {Shporer}, {Tenenbaum},
  {Thompson}, {Twicken}, {Van Cleve}, {Welsh}, {Basu}, {Chaplin}, {Miglio},
  {Kawaler}, {Arentoft}, {Stello}, {Metcalfe}, {Verner}, {Karoff}, {Lundkvist},
  {Lund}, {Handberg}, {Elsworth}, {Hekker}, {Huber}, {Bedding}, \&
  {Rapin}}]{borucki12}
---. 2012, \apj, 745, 120

\bibitem[{{Bouchy} {et~al.}(2005){Bouchy}, {Bazot}, {Santos}, {Vauclair}, \&
  {Sosnowska}}]{bouchy05}
{Bouchy}, F., {Bazot}, M., {Santos}, N.~C., {Vauclair}, S., \& {Sosnowska}, D.
  2005, \aap, 440, 609

\bibitem[{{Brogaard} {et~al.}(2012){Brogaard}, {VandenBerg}, {Bruntt},
  {Grundahl}, {Frandsen}, {Bedin}, {Milone}, {Dotter}, {Feiden}, {Stetson},
  {Sandquist}, {Miglio}, {Stello}, \& {Jessen-Hansen}}]{brogaard12}
{Brogaard}, K., {et~al.} 2012, \aap, 543, A106

\bibitem[{Brown \& Gilliland(1994)}]{brown94}
Brown, T.~M., \& Gilliland, R.~L. 1994, \araa, 32, 37

\bibitem[{{Brown} {et~al.}(1991){Brown}, {Gilliland}, {Noyes}, \&
  {Ramsey}}]{brown91}
{Brown}, T.~M., {Gilliland}, R.~L., {Noyes}, R.~W., \& {Ramsey}, L.~W. 1991,
  \apj, 368, 599

\bibitem[{{Brown} {et~al.}(2011){Brown}, {Latham}, {Everett}, \&
  {Esquerdo}}]{brown11}
{Brown}, T.~M., {Latham}, D.~W., {Everett}, M.~E., \& {Esquerdo}, G.~A. 2011,
  \aj, 142, 112

\bibitem[{{Bruntt} {et~al.}(2012){Bruntt}, {Basu}, {Smalley}, {Chaplin},
  {Verner}, {Bedding}, {Catala}, {Gazzano}, {Molenda-{\.Z}akowicz}, {Thygesen},
  {Uytterhoeven}, {Hekker}, {Huber}, {Karoff}, {Mathur}, {Mosser},
  {Appourchaux}, {Campante}, {Elsworth}, {Garc{\'{\i}}a}, {Handberg},
  {Metcalfe}, {Quirion}, {R{\'e}gulo}, {Roxburgh}, {Stello},
  {Christensen-Dalsgaard}, {Kawaler}, {Kjeldsen}, {Morris}, {Quintana}, \&
  {Sanderfer}}]{bruntt12}
{Bruntt}, H., {et~al.} 2012, \mnras, 423, 122

\bibitem[{{Buchhave} {et~al.}(2011){Buchhave}, {Latham}, {Carter},
  {D{\'e}sert}, {Torres}, {Adams}, {Bryson}, {Charbonneau}, {Ciardi}, {Kulesa},
  {Dupree}, {Fischer}, {Fressin}, {Gautier}, {Gilliland}, {Howell}, {Isaacson},
  {Jenkins}, {Marcy}, {McCarthy}, {Rowe}, {Batalha}, {Borucki}, {Brown},
  {Caldwell}, {Christiansen}, {Cochran}, {Deming}, {Dunham}, {Everett}, {Ford},
  {Fortney}, {Geary}, {Girouard}, {Haas}, {Holman}, {Horch}, {Klaus},
  {Knutson}, {Koch}, {Kolodziejczak}, {Lissauer}, {Machalek}, {Mullally},
  {Still}, {Quinn}, {Seager}, {Thompson}, \& {Van Cleve}}]{buchhave11}
{Buchhave}, L.~A., {et~al.} 2011, \apjs, 197, 3

\bibitem[{{Buchhave} {et~al.}(2012){Buchhave}, {Latham}, {Johansen},
  {Bizzarro}, {Torres}, {Rowe}, {Batalha}, {Borucki}, {Brugamyer}, {Caldwell},
  {Bryson}, {Ciardi}, {Cochran}, {Endl}, {Esquerdo}, {Ford}, {Geary},
  {Gilliland}, {Hansen}, {Isaacson}, {Laird}, {Lucas}, {Marcy}, {Morse},
  {Robertson}, {Shporer}, {Stefanik}, {Still}, \& {Quinn}}]{buchhave12}
---. 2012, \nat, 486, 375

\bibitem[{{Carter} {et~al.}(2012){Carter}, {Agol}, {Chaplin}, {Basu},
  {Bedding}, {Buchhave}, {Christensen-Dalsgaard}, {Deck}, {Elsworth},
  {Fabrycky}, {Ford}, {Fortney}, {Hale}, {Handberg}, {Hekker}, {Holman},
  {Huber}, {Karoff}, {Kawaler}, {Kjeldsen}, {Lissauer}, {Lopez}, {Lund},
  {Lundkvist}, {Metcalfe}, {Miglio}, {Rogers}, {Stello}, {Borucki}, {Bryson},
  {Christiansen}, {Cochran}, {Geary}, {Gilliland}, {Haas}, {Hall}, {Howard},
  {Jenkins}, {Klaus}, {Koch}, {Latham}, {MacQueen}, {Sasselov}, {Steffen},
  {Twicken}, \& {Winn}}]{carter12}
{Carter}, J.~A., {et~al.} 2012, Science, 337, 556

\bibitem[{{Casagrande} {et~al.}(2010){Casagrande}, {Ram{\'{\i}}rez},
  {Mel{\'e}ndez}, {Bessell}, \& {Asplund}}]{casagrande10}
{Casagrande}, L., {Ram{\'{\i}}rez}, I., {Mel{\'e}ndez}, J., {Bessell}, M., \&
  {Asplund}, M. 2010, \aap, 512, A54

\bibitem[{{Chaplin} {et~al.}(2011{\natexlab{a}}){Chaplin}, {Kjeldsen},
  {Christensen-Dalsgaard}, {Basu}, {Miglio}, {Appourchaux}, {Bedding},
  {Elsworth}, {Garc{\'{\i}}a}, {Gilliland}, {Girardi}, {Houdek}, {Karoff},
  {Kawaler}, {Metcalfe}, {Molenda-{\.Z}akowicz}, {Monteiro}, {Thompson},
  {Verner}, {Ballot}, {Bonanno}, {Brand{\~a}o}, {Broomhall}, {Bruntt},
  {Campante}, {Corsaro}, {Creevey}, {Do{\u g}an}, {Esch}, {Gai}, {Gaulme},
  {Hale}, {Handberg}, {Hekker}, {Huber}, {Jim{\'e}nez}, {Mathur}, {Mazumdar},
  {Mosser}, {New}, {Pinsonneault}, {Pricopi}, {Quirion}, {R{\'e}gulo},
  {Salabert}, {Serenelli}, {Silva Aguirre}, {Sousa}, {Stello}, {Stevens},
  {Suran}, {Uytterhoeven}, {White}, {Borucki}, {Brown}, {Jenkins}, {Kinemuchi},
  {Van Cleve}, \& {Klaus}}]{chaplin11a}
{Chaplin}, W.~J., {et~al.} 2011{\natexlab{a}}, Science, 332, 213

\bibitem[{{Chaplin} {et~al.}(2011{\natexlab{b}}){Chaplin}, {Kjeldsen},
  {Bedding}, {Christensen-Dalsgaard}, {Gilliland}, {Kawaler}, {Appourchaux},
  {Elsworth}, {Garc{\'{\i}}a}, {Houdek}, {Karoff}, {Metcalfe},
  {Molenda-{\.Z}akowicz}, {Monteiro}, {Thompson}, {Verner}, {Batalha},
  {Borucki}, {Brown}, {Bryson}, {Christiansen}, {Clarke}, {Jenkins}, {Klaus},
  {Koch}, {An}, {Ballot}, {Basu}, {Benomar}, {Bonanno}, {Broomhall},
  {Campante}, {Corsaro}, {Creevey}, {Esch}, {Gai}, {Gaulme}, {Hale},
  {Handberg}, {Hekker}, {Huber}, {Mathur}, {Mosser}, {New}, {Pinsonneault},
  {Pricopi}, {Quirion}, {R{\'e}gulo}, {Roxburgh}, {Salabert}, {Stello}, \&
  {Suran}}]{chaplin11b}
---. 2011{\natexlab{b}}, \apj, 732, 54

\bibitem[{{Chaplin} {et~al.}(2013){Chaplin}, {Sanchis-Ojeda}, {Campante},
  {Handberg}, {Stello}, {Winn}, {Basu}, {Christensen-Dalsgaard}, {Davies},
  {Metcalfe}, {Buchhave}, {Fischer}, {Bedding}, {Cochran}, {Elsworth},
  {Gilliland}, {Hekker}, {Huber}, {Isaacson}, {Karoff}, {Kawaler}, {Kjeldsen},
  {Latham}, {Lund}, {Lundkvist}, {Marcy}, {Miglio}, {Barclay}, \&
  {Lissauer}}]{chaplin12}
---. 2013, \apj, 766, 101

\bibitem[{Christensen-Dalsgaard(2004)}]{CD04}
Christensen-Dalsgaard, J. 2004, \solphys, 220, 137

\bibitem[{{Christensen-Dalsgaard}(2008)}]{CD08a}
{Christensen-Dalsgaard}, J. 2008, \apss, 316, 13

\bibitem[{{Christensen-Dalsgaard} {et~al.}(2010){Christensen-Dalsgaard},
  {Kjeldsen}, {Brown}, {Gilliland}, {Arentoft}, {Frandsen}, {Quirion},
  {Borucki}, {Koch}, \& {Jenkins}}]{cd10}
{Christensen-Dalsgaard}, J., {et~al.} 2010, \apjl, 713, L164

\bibitem[{{Christiansen} {et~al.}(2012){Christiansen}, {Jenkins}, {Caldwell},
  {Burke}, {Tenenbaum}, {Seader}, {Thompson}, {Barclay}, {Clarke}, {Li},
  {Smith}, {Stumpe}, {Twicken}, \& {Van Cleve}}]{christiansen12}
{Christiansen}, J.~L., {et~al.} 2012, \pasp, 124, 1279

\bibitem[{{Ciardi} {et~al.}(2013){Ciardi}, {Fabrycky}, {Ford}, {Gautier},
  {Howell}, {Lissauer}, {Ragozzine}, \& {Rowe}}]{ciardi13}
{Ciardi}, D.~R., {Fabrycky}, D.~C., {Ford}, E.~B., {Gautier}, III, T.~N.,
  {Howell}, S.~B., {Lissauer}, J.~J., {Ragozzine}, D., \& {Rowe}, J.~F. 2013,
  \apj, 763, 41

\bibitem[{{Cochran} {et~al.}(2011){Cochran}, {Fabrycky}, {Torres}, {Fressin},
  {D{\'e}sert}, {Ragozzine}, {Sasselov}, {Fortney}, {Rowe}, {Brugamyer},
  {Bryson}, {Carter}, {Ciardi}, {Howell}, {Steffen}, {Borucki}, {Koch}, {Winn},
  {Welsh}, {Uddin}, {Tenenbaum}, {Still}, {Seager}, {Quinn}, {Mullally},
  {Miller}, {Marcy}, {MacQueen}, {Lucas}, {Lissauer}, {Latham}, {Knutson},
  {Kinemuchi}, {Johnson}, {Jenkins}, {Isaacson}, {Howard}, {Horch}, {Holman},
  {Henze}, {Haas}, {Gilliland}, {Gautier}, {Ford}, {Fischer}, {Everett},
  {Endl}, {Demory}, {Deming}, {Charbonneau}, {Caldwell}, {Buchhave}, {Brown},
  \& {Batalha}}]{cochran11}
{Cochran}, W.~D., {et~al.} 2011, \apjs, 197, 7

\bibitem[{{Creevey} {et~al.}(2012){Creevey}, {Do{\v g}an}, {Frasca},
  {Thygesen}, {Basu}, {Bhattacharya}, {Biazzo}, {Brand{\~a}o}, {Bruntt},
  {Mazumdar}, {Niemczura}, {Shrotriya}, {Sousa}, {Stello}, {Subramaniam},
  {Campante}, {Handberg}, {Mathur}, {Bedding}, {Garc{\'{\i}}a}, {R{\'e}gulo},
  {Salabert}, {Molenda-{\.Z}akowicz}, {Quirion}, {White}, {Bonanno}, {Chaplin},
  {Christensen-Dalsgaard}, {Christiansen}, {Elsworth}, {Fanelli}, {Karoff},
  {Kinemuchi}, {Kjeldsen}, {Gai}, {Monteiro}, \& {Su{\'a}rez}}]{creevey12}
{Creevey}, O.~L., {et~al.} 2012, \aap, 537, A111

\bibitem[{{da Silva} {et~al.}(2006){da Silva}, {Girardi}, {Pasquini},
  {Setiawan}, {von der L{\"u}he}, {de Medeiros}, {Hatzes}, {D{\"o}llinger}, \&
  {Weiss}}]{dasilva06}
{da Silva}, L., {et~al.} 2006, \aap, 458, 609

\bibitem[{{Dawson} \& {Johnson}(2012)}]{dawson12}
{Dawson}, R.~I., \& {Johnson}, J.~A. 2012, \apj, 756, 122

\bibitem[{{Demarque} {et~al.}(2008){Demarque}, {Guenther}, {Li}, {Mazumdar}, \&
  {Straka}}]{demarque08}
{Demarque}, P., {Guenther}, D.~B., {Li}, L.~H., {Mazumdar}, A., \& {Straka},
  C.~W. 2008, \apss, 316, 31

\bibitem[{{Demarque} {et~al.}(2004){Demarque}, {Woo}, {Kim}, \&
  {Yi}}]{demarque04}
{Demarque}, P., {Woo}, J.-H., {Kim}, Y.-C., \& {Yi}, S.~K. 2004, \apjs, 155,
  667

\bibitem[{{Derekas} {et~al.}(2011){Derekas}, {Kiss}, {Borkovits}, {Huber},
  {Lehmann}, {Southworth}, {Bedding}, {Balam}, {Hartmann}, {Hrudkova},
  {Ireland}, {Kov{\'a}cs}, {Mez{\H o}}, {Mo{\'o}r}, {Niemczura}, {Sarty},
  {Szab{\'o}}, {Szab{\'o}}, {Telting}, {Tkachenko}, {Uytterhoeven}, {Benk{\H
  o}}, {Bryson}, {Maestro}, {Simon}, {Stello}, {Schaefer}, {Aerts}, {ten
  Brummelaar}, {De Cat}, {McAlister}, {Maceroni}, {M{\'e}rand}, {Still},
  {Sturmann}, {Sturmann}, {Turner}, {Tuthill}, {Christensen-Dalsgaard},
  {Gilliland}, {Kjeldsen}, {Quintana}, {Tenenbaum}, \& {Twicken}}]{derekas11}
{Derekas}, A., {et~al.} 2011, Science, 332, 216

\bibitem[{{di Mauro} {et~al.}(2011){di Mauro}, {Cardini}, {Catanzaro},
  {Ventura}, {Barban}, {Bedding}, {Christensen-Dalsgaard}, {De Ridder},
  {Hekker}, {Huber}, {Kallinger}, {Miglio}, {Montalban}, {Mosser}, {Stello},
  {Uytterhoeven}, {Kinemuchi}, {Kjeldsen}, {Mullally}, \& {Still}}]{dimauro11}
{di Mauro}, M.~P., {et~al.} 2011, \mnras, 415, 3783

\bibitem[{{Djupvik} \& {Andersen}(2010)}]{djupvik10}
{Djupvik}, A.~A., \& {Andersen}, J. 2010, in Highlights of Spanish Astrophysics
  V, ed. J.~M. {Diego}, L.~J. {Goicoechea}, J.~I. {Gonz{\'a}lez-Serrano}, \&
  J.~{Gorgas}, 211

\bibitem[{{Dotter} {et~al.}(2008){Dotter}, {Chaboyer}, {Jevremovi{\'c}},
  {Kostov}, {Baron}, \& {Ferguson}}]{dotter08}
{Dotter}, A., {Chaboyer}, B., {Jevremovi{\'c}}, D., {Kostov}, V., {Baron}, E.,
  \& {Ferguson}, J.~W. 2008, \apjs, 178, 89

\bibitem[{{Do{\u g}an} {et~al.}(2010){Do{\u g}an}, {Brand{\~a}o}, {Bedding},
  {Christensen-Dalsgaard}, {Cunha}, \& {Kjeldsen}}]{dogan10}
{Do{\u g}an}, G., {Brand{\~a}o}, I.~M., {Bedding}, T.~R.,
  {Christensen-Dalsgaard}, J., {Cunha}, M.~S., \& {Kjeldsen}, H. 2010, \apss,
  328, 101

\bibitem[{{Fabrycky} {et~al.}(2012){Fabrycky}, {Ford}, {Steffen}, {Rowe},
  {Carter}, {Moorhead}, {Batalha}, {Borucki}, {Bryson}, {Buchhave},
  {Christiansen}, {Ciardi}, {Cochran}, {Endl}, {Fanelli}, {Fischer}, {Fressin},
  {Geary}, {Haas}, {Hall}, {Holman}, {Jenkins}, {Koch}, {Latham}, {Li},
  {Lissauer}, {Lucas}, {Marcy}, {Mazeh}, {McCauliff}, {Quinn}, {Ragozzine},
  {Sasselov}, \& {Shporer}}]{fabrycky12}
{Fabrycky}, D.~C., {et~al.} 2012, \apj, 750, 114

\bibitem[{{Fischer} \& {Valenti}(2005)}]{fischer05}
{Fischer}, D.~A., \& {Valenti}, J. 2005, \apj, 622, 1102

\bibitem[{{Ford} {et~al.}(2012){Ford}, {Fabrycky}, {Steffen}, {Carter},
  {Fressin}, {Holman}, {Lissauer}, {Moorhead}, {Morehead}, {Ragozzine}, {Rowe},
  {Welsh}, {Allen}, {Batalha}, {Borucki}, {Bryson}, {Buchhave}, {Burke},
  {Caldwell}, {Charbonneau}, {Clarke}, {Cochran}, {D{\'e}sert}, {Endl},
  {Everett}, {Fischer}, {Gautier}, {Gilliland}, {Jenkins}, {Haas}, {Horch},
  {Howell}, {Ibrahim}, {Isaacson}, {Koch}, {Latham}, {Li}, {Lucas}, {MacQueen},
  {Marcy}, {McCauliff}, {Mullally}, {Quinn}, {Quintana}, {Shporer}, {Still},
  {Tenenbaum}, {Thompson}, {Torres}, {Twicken}, {Wohler}, \& {the Kepler
  Science Team}}]{ford12}
{Ford}, E.~B., {et~al.} 2012, \apj, 750, 113

\bibitem[{{Fressin} {et~al.}(2013){Fressin}, {Torres}, {Charbonneau}, {Bryson},
  {Christiansen}, {Dressing}, {Jenkins}, {Walkowicz}, \& {Batalha}}]{fressin13}
{Fressin}, F., {et~al.} 2013, \apj, 766, 81

\bibitem[{{Fuller} {et~al.}(2013){Fuller}, {Derekas}, {Borkovits}, {Huber},
  {Bedding}, \& {Kiss}}]{fuller13}
{Fuller}, J., {Derekas}, A., {Borkovits}, T., {Huber}, D., {Bedding}, T.~R., \&
  {Kiss}, L.~L. 2013, \mnras, 429, 2425

\bibitem[{{F\"{u}r\'esz}(2008)}]{furesz08}
{F\"{u}r\'esz}, G. 2008, PhD thesis, University of Szeged, Szeged, Hungary

\bibitem[{{Gai} {et~al.}(2011){Gai}, {Basu}, {Chaplin}, \& {Elsworth}}]{gai11}
{Gai}, N., {Basu}, S., {Chaplin}, W.~J., \& {Elsworth}, Y. 2011, \apj, 730, 63

\bibitem[{{Gaidos} \& {Mann}(2013)}]{gaidos13}
{Gaidos}, E., \& {Mann}, A.~W. 2013, \apj, 762, 41

\bibitem[{{Gautier} {et~al.}(2010){Gautier}, {Batalha}, {Borucki}, {Cochran},
  {Dunham}, {Howell}, {Koch}, {Latham}, {Marcy}, {Buchhave}, {Ciardi}, {Endl},
  {Furesz}, {Isaacson}, {MacQueen}, {Mandushev}, \& {Walkowicz}}]{gautier10}
{Gautier}, III, T.~N., {et~al.} 2010, ArXiv e-prints (arXiv:1001.0352)

\bibitem[{{Gilliland} {et~al.}(2011){Gilliland}, {McCullough}, {Nelan},
  {Brown}, {Charbonneau}, {Nutzman}, {Christensen-Dalsgaard}, \&
  {Kjeldsen}}]{gilliland11}
{Gilliland}, R.~L., {McCullough}, P.~R., {Nelan}, E.~P., {Brown}, T.~M.,
  {Charbonneau}, D., {Nutzman}, P., {Christensen-Dalsgaard}, J., \& {Kjeldsen},
  H. 2011, \apj, 726, 2

\bibitem[{{Gilliland} {et~al.}(2010{\natexlab{a}}){Gilliland}, {Jenkins},
  {Borucki}, {Bryson}, {Caldwell}, {Clarke}, {Dotson}, {Haas}, {Hall}, {Klaus},
  {Koch}, {McCauliff}, {Quintana}, {Twicken}, \& {van Cleve}}]{gilliland10b}
{Gilliland}, R.~L., {et~al.} 2010{\natexlab{a}}, \apjl, 713, L160

\bibitem[{{Gilliland} {et~al.}(2010{\natexlab{b}}){Gilliland}, {Brown},
  {Christensen-Dalsgaard}, {Kjeldsen}, {Aerts}, {Appourchaux}, {Basu},
  {Bedding}, {Chaplin}, {Cunha}, {De Cat}, {De Ridder}, {Guzik}, {Handler},
  {Kawaler}, {Kiss}, {Kolenberg}, {Kurtz}, {Metcalfe}, {Monteiro}, {Szab{\'o}},
  {Arentoft}, {Balona}, {Debosscher}, {Elsworth}, {Quirion}, {Stello},
  {Su{\'a}rez}, {Borucki}, {Jenkins}, {Koch}, {Kondo}, {Latham}, {Rowe}, \&
  {Steffen}}]{gilliland10}
---. 2010{\natexlab{b}}, \pasp, 122, 131

\bibitem[{{Gilliland} {et~al.}(2013){Gilliland}, {Marcy}, {Rowe}, {Rogers},
  {Torres}, {Fressin}, {Lopez}, {Buchhave}, {Christensen-Dalsgaard},
  {D{\'e}sert}, {Henze}, {Isaacson}, {Jenkins}, {Lissauer}, {Chaplin}, {Basu},
  {Metcalfe}, {Elsworth}, {Handberg}, {Hekker}, {Huber}, {Karoff}, {Kjeldsen},
  {Lund}, {Lundkvist}, {Miglio}, {Charbonneau}, {Ford}, {Fortney}, {Haas},
  {Howard}, {Howell}, {Ragozzine}, \& {Thompson}}]{gilliland13}
---. 2013, \apj, 766, 40

\bibitem[{{Gonzalez}(1997)}]{gonzalez97}
{Gonzalez}, G. 1997, \mnras, 285, 403

\bibitem[{Gough(1986)}]{gough86}
Gough, D.~O. 1986, in Hydrodynamic and Magnetodynamic Problems in the Sun and
  Stars, ed. Y.~{Osaki} (Uni. of Tokyo Press), 117

\bibitem[{{Hekker} {et~al.}(2009){Hekker}, {Kallinger}, {Baudin}, {De Ridder},
  {Barban}, {Carrier}, {Hatzes}, {Weiss}, \& {Baglin}}]{hekker09}
{Hekker}, S., {et~al.} 2009, \aap, 506, 465

\bibitem[{{Hekker} {et~al.}(2010){Hekker}, {Broomhall}, {Chaplin}, {Elsworth},
  {Fletcher}, {New}, {Arentoft}, {Quirion}, \& {Kjeldsen}}]{hekker10c}
---. 2010, \mnras, 402, 2049

\bibitem[{{Hekker} {et~al.}(2011{\natexlab{a}}){Hekker}, {Gilliland},
  {Elsworth}, {Chaplin}, {De Ridder}, {Stello}, {Kallinger}, {Ibrahim},
  {Klaus}, \& {Li}}]{hekker11c}
---. 2011{\natexlab{a}}, \mnras, 414, 2594

\bibitem[{{Hekker} {et~al.}(2011{\natexlab{b}}){Hekker}, {Elsworth}, {De
  Ridder}, {Mosser}, {Garc{\'{\i}}a}, {Kallinger}, {Mathur}, {Huber}, {Buzasi},
  {Preston}, {Hale}, {Ballot}, {Chaplin}, {R{\'e}gulo}, {Bedding}, {Stello},
  {Borucki}, {Koch}, {Jenkins}, {Allen}, {Gilliland}, {Kjeldsen}, \&
  {Christensen-Dalsgaard}}]{hekker11}
---. 2011{\natexlab{b}}, \aap, 525, A131

\bibitem[{{Hekker} {et~al.}(2012){Hekker}, {Elsworth}, {Mosser}, {Kallinger},
  {Chaplin}, {De Ridder}, {Garc{\'{\i}}a}, {Stello}, {Clarke}, {Hall}, \&
  {Ibrahim}}]{hekker12b}
---. 2012, \aap, 544, A90

\bibitem[{{Holman} {et~al.}(2007){Holman}, {Winn}, {Latham}, {O'Donovan},
  {Charbonneau}, {Torres}, {Sozzetti}, {Fernandez}, \& {Everett}}]{holman07}
{Holman}, M.~J., {et~al.} 2007, \apj, 664, 1185

\bibitem[{{Houdek}(2006)}]{houdek06}
{Houdek}, G. 2006, in ESA Special Publication, Vol. 624, Proceedings of SOHO
  18/GONG 2006/HELAS I, Beyond the spherical Sun

\bibitem[{Houdek {et~al.}(1999)Houdek, {Balmforth}, {Christensen-Dalsgaard}, \&
  {Gough}}]{houdek99}
Houdek, G., {Balmforth}, N.~J., {Christensen-Dalsgaard}, J., \& {Gough}, D.~O.
  1999, \aap, 351, 582

\bibitem[{{Howard} {et~al.}(2012){Howard}, {Marcy}, {Bryson}, {Jenkins},
  {Rowe}, {Batalha}, {Borucki}, {Koch}, {Dunham}, {Gautier}, {Van Cleve},
  {Cochran}, {Latham}, {Lissauer}, {Torres}, {Brown}, {Gilliland}, {Buchhave},
  {Caldwell}, {Christensen-Dalsgaard}, {Ciardi}, {Fressin}, {Haas}, {Howell},
  {Kjeldsen}, {Seager}, {Rogers}, {Sasselov}, {Steffen}, {Basri},
  {Charbonneau}, {Christiansen}, {Clarke}, {Dupree}, {Fabrycky}, {Fischer},
  {Ford}, {Fortney}, {Tarter}, {Girouard}, {Holman}, {Johnson}, {Klaus},
  {Machalek}, {Moorhead}, {Morehead}, {Ragozzine}, {Tenenbaum}, {Twicken},
  {Quinn}, {Isaacson}, {Shporer}, {Lucas}, {Walkowicz}, {Welsh}, {Boss},
  {Devore}, {Gould}, {Smith}, {Morris}, {Prsa}, {Morton}, {Still}, {Thompson},
  {Mullally}, {Endl}, \& {MacQueen}}]{howard11}
{Howard}, A.~W., {et~al.} 2012, \apjs, 201, 15

\bibitem[{{Howell} {et~al.}(2012){Howell}, {Rowe}, {Bryson}, {Quinn}, {Marcy},
  {Isaacson}, {Ciardi}, {Chaplin}, {Metcalfe}, {Monteiro}, {Appourchaux},
  {Basu}, {Creevey}, {Gilliland}, {Quirion}, {Stello}, {Kjeldsen},
  {Christensen-Dalsgaard}, {Elsworth}, {Garc{\'{\i}}a}, {Houdek}, {Karoff},
  {Molenda-{\.Z}akowicz}, {Thompson}, {Verner}, {Torres}, {Fressin}, {Crepp},
  {Adams}, {Dupree}, {Sasselov}, {Dressing}, {Borucki}, {Koch}, {Lissauer},
  {Latham}, {Buchhave}, {Gautier}, {Everett}, {Horch}, {Batalha}, {Dunham},
  {Szkody}, {Silva}, {Mighell}, {Holberg}, {Ballot}, {Bedding}, {Bruntt},
  {Campante}, {Handberg}, {Hekker}, {Huber}, {Mathur}, {Mosser}, {R{\'e}gulo},
  {White}, {Christiansen}, {Middour}, {Haas}, {Hall}, {Jenkins}, {McCaulif},
  {Fanelli}, {Kulesa}, {McCarthy}, \& {Henze}}]{howell12}
{Howell}, S.~B., {et~al.} 2012, \apj, 746, 123

\bibitem[{{Huber} {et~al.}(2009){Huber}, {Stello}, {Bedding}, {Chaplin},
  {Arentoft}, {Quirion}, \& {Kjeldsen}}]{huber09}
{Huber}, D., {Stello}, D., {Bedding}, T.~R., {Chaplin}, W.~J., {Arentoft}, T.,
  {Quirion}, P.-O., \& {Kjeldsen}, H. 2009, Communications in Asteroseismology,
  160, 74

\bibitem[{{Huber} {et~al.}(2010){Huber}, {Bedding}, {Stello}, {Mosser},
  {Mathur}, {Kallinger}, {Hekker}, {Elsworth}, {Buzasi}, {De Ridder},
  {Gilliland}, {Kjeldsen}, {Chaplin}, {Garc{\'{\i}}a}, {Hale}, {Preston},
  {White}, {Borucki}, {Christensen-Dalsgaard}, {Clarke}, {Jenkins}, \&
  {Koch}}]{huber10}
{Huber}, D., {et~al.} 2010, \apj, 723, 1607

\bibitem[{{Huber} {et~al.}(2011){Huber}, {Bedding}, {Stello}, {Hekker},
  {Mathur}, {Mosser}, {Verner}, {Bonanno}, {Buzasi}, {Campante}, {Elsworth},
  {Hale}, {Kallinger}, {Silva Aguirre}, {Chaplin}, {De Ridder},
  {Garc{\'{\i}}a}, {Appourchaux}, {Frandsen}, {Houdek}, {Molenda-{\.Z}akowicz},
  {Monteiro}, {Christensen-Dalsgaard}, {Gilliland}, {Kawaler}, {Kjeldsen},
  {Broomhall}, {Corsaro}, {Salabert}, {Sanderfer}, {Seader}, \&
  {Smith}}]{huber11b}
---. 2011, \apj, 743, 143

\bibitem[{{Huber} {et~al.}(2012){Huber}, {Ireland}, {Bedding}, {Brand{\~a}o},
  {Piau}, {Maestro}, {White}, {Bruntt}, {Casagrande}, {Molenda-{\.Z}akowicz},
  {Silva Aguirre}, {Sousa}, {Barclay}, {Burke}, {Chaplin},
  {Christensen-Dalsgaard}, {Cunha}, {De Ridder}, {Farrington}, {Frasca},
  {Garc{\'{\i}}a}, {Gilliland}, {Goldfinger}, {Hekker}, {Kawaler}, {Kjeldsen},
  {McAlister}, {Metcalfe}, {Miglio}, {Monteiro}, {Pinsonneault}, {Schaefer},
  {Stello}, {Stumpe}, {Sturmann}, {Sturmann}, {ten Brummelaar}, {Thompson},
  {Turner}, \& {Uytterhoeven}}]{huber12b}
---. 2012, \apj, 760, 32

\bibitem[{{Huber} {et~al.}(2013){Huber}, {Basu}, {Smalley}, {Chaplin},
  {Verner}, {Bedding}, {Catala}, {Gazzano}, {Molenda-Zakowicz}, {Thygesen},
  {Uytterhoeven}, {Hekker}, {Huber}, {Karoff}, {Mathur}, {Mosser},
  {Appourchaux}, {Campante}, {Elsworth}, {Garcia}, {Handberg}, {Metcalfe},
  {Quirion}, {Regulo}, {Roxburgh}, {Stello}, {Christensen-Dalsgaard},
  {Kawaler}, {Kjeldsen}, {Morris}, {Quintana}, \& {Sanderfer}}]{huber12d}
---. 2013, in preparation

\bibitem[{{Jenkins} {et~al.}(2010){Jenkins}, {Caldwell}, {Chandrasekaran},
  {Twicken}, {Bryson}, {Quintana}, {Clarke}, {Li}, {Allen}, {Tenenbaum}, {Wu},
  {Klaus}, {Van Cleve}, {Dotson}, {Haas}, {Gilliland}, {Koch}, \&
  {Borucki}}]{jenkins10}
{Jenkins}, J.~M., {et~al.} 2010, \apjl, 713, L120

\bibitem[{{Jiang} {et~al.}(2011){Jiang}, {Jiang}, {Christensen-Dalsgaard},
  {Bedding}, {Stello}, {Huber}, {Frandsen}, {Kjeldsen}, {Karoff}, {Mosser},
  {Demarque}, {Fanelli}, {Kinemuchi}, \& {Mullally}}]{jiang11}
{Jiang}, C., {et~al.} 2011, \apj, 742, 120

\bibitem[{{Johnson} {et~al.}(2010){Johnson}, {Aller}, {Howard}, \&
  {Crepp}}]{johnson10}
{Johnson}, J.~A., {Aller}, K.~M., {Howard}, A.~W., \& {Crepp}, J.~R. 2010,
  \pasp, 122, 905

\bibitem[{{Johnson} {et~al.}(2007){Johnson}, {Butler}, {Marcy}, {Fischer},
  {Vogt}, {Wright}, \& {Peek}}]{johnson07}
{Johnson}, J.~A., {Butler}, R.~P., {Marcy}, G.~W., {Fischer}, D.~A., {Vogt},
  S.~S., {Wright}, J.~T., \& {Peek}, K.~M.~G. 2007, \apj, 670, 833

\bibitem[{{Johnson} {et~al.}(2012){Johnson}, {Gazak}, {Apps}, {Muirhead},
  {Crepp}, {Crossfield}, {Boyajian}, {von Braun}, {Rojas-Ayala}, {Howard},
  {Covey}, {Schlawin}, {Hamren}, {Morton}, {Marcy}, \& {Lloyd}}]{johnson12}
{Johnson}, J.~A., {et~al.} 2012, \aj, 143, 111

\bibitem[{{Kallinger} {et~al.}(2010{\natexlab{a}}){Kallinger}, {Mosser},
  {Hekker}, {Huber}, {Stello}, {Mathur}, {Basu}, {Bedding}, {Chaplin}, {De
  Ridder}, {Elsworth}, {Frandsen}, {Garc{\'{\i}}a}, {Gruberbauer}, {Matthews},
  {Borucki}, {Bruntt}, {Christensen-Dalsgaard}, {Gilliland}, {Kjeldsen}, \&
  {Koch}}]{kallinger10}
{Kallinger}, T., {et~al.} 2010{\natexlab{a}}, \aap, 522, A1

\bibitem[{{Kallinger} {et~al.}(2010{\natexlab{b}}){Kallinger}, {Weiss},
  {Barban}, {Baudin}, {Cameron}, {Carrier}, {De Ridder}, {Goupil},
  {Gruberbauer}, {Hatzes}, {Hekker}, {Samadi}, \& {Deleuil}}]{kallinger10c}
---. 2010{\natexlab{b}}, \aap, 509, A77

\bibitem[{{Kane} {et~al.}(2012){Kane}, {Ciardi}, {Gelino}, \& {von
  Braun}}]{kane12}
{Kane}, S.~R., {Ciardi}, D.~R., {Gelino}, D.~M., \& {von Braun}, K. 2012,
  \mnras, 425, 757

\bibitem[{{Karoff} {et~al.}(2010){Karoff}, {Campante}, \& {Chaplin}}]{karoff10}
{Karoff}, C., {Campante}, T.~L., \& {Chaplin}, W.~J. 2010, ArXiv e-prints
  (arXiv:1003.4167)

\bibitem[{{Kipping} \& {Bakos}(2011)}]{kipping11}
{Kipping}, D., \& {Bakos}, G. 2011, \apj, 733, 36

\bibitem[{{Kjeldsen} \& {Bedding}(1995)}]{KB95}
{Kjeldsen}, H., \& {Bedding}, T.~R. 1995, \aap, 293, 87

\bibitem[{{Koch} {et~al.}(2010){Koch}, {Borucki}, {Basri}, {Batalha}, {Brown},
  {Caldwell}, {Christensen-Dalsgaard}, {Cochran}, {DeVore}, {Dunham},
  {Gautier}, {Geary}, {Gilliland}, {Gould}, {Jenkins}, {Kondo}, {Latham},
  {Lissauer}, {Marcy}, {Monet}, {Sasselov}, {Boss}, {Brownlee}, {Caldwell},
  {Dupree}, {Howell}, {Kjeldsen}, {Meibom}, {Morrison}, {Owen}, {Reitsema},
  {Tarter}, {Bryson}, {Dotson}, {Gazis}, {Haas}, {Kolodziejczak}, {Rowe}, {Van
  Cleve}, {Allen}, {Chandrasekaran}, {Clarke}, {Li}, {Quintana}, {Tenenbaum},
  {Twicken}, \& {Wu}}]{koch10b}
{Koch}, D.~G., {et~al.} 2010, \apjl, 713, L79

\bibitem[{{Latham} {et~al.}(2010){Latham}, {Borucki}, {Koch}, {Brown},
  {Buchhave}, {Basri}, {Batalha}, {Caldwell}, {Cochran}, {Dunham}, {F{\H
  u}r{\'e}sz}, {Gautier}, {Geary}, {Gilliland}, {Howell}, {Jenkins},
  {Lissauer}, {Marcy}, {Monet}, {Rowe}, \& {Sasselov}}]{latham10}
{Latham}, D.~W., {et~al.} 2010, \apjl, 713, L140

\bibitem[{{Latham} {et~al.}(2011){Latham}, {Rowe}, {Quinn}, {Batalha},
  {Borucki}, {Brown}, {Bryson}, {Buchhave}, {Caldwell}, {Carter},
  {Christiansen}, {Ciardi}, {Cochran}, {Dunham}, {Fabrycky}, {Ford}, {Gautier},
  {Gilliland}, {Holman}, {Howell}, {Ibrahim}, {Isaacson}, {Jenkins}, {Koch},
  {Lissauer}, {Marcy}, {Quintana}, {Ragozzine}, {Sasselov}, {Shporer},
  {Steffen}, {Welsh}, \& {Wohler}}]{latham11}
---. 2011, \apjl, 732, L24

\bibitem[{{Laws} {et~al.}(2003){Laws}, {Gonzalez}, {Walker}, {Tyagi},
  {Dodsworth}, {Snider}, \& {Suntzeff}}]{laws03}
{Laws}, C., {Gonzalez}, G., {Walker}, K.~M., {Tyagi}, S., {Dodsworth}, J.,
  {Snider}, K., \& {Suntzeff}, N.~B. 2003, \aj, 125, 2664

\bibitem[{{Lissauer} {et~al.}(2009){Lissauer}, {Hubickyj}, {D'Angelo}, \&
  {Bodenheimer}}]{lissauer09}
{Lissauer}, J.~J., {Hubickyj}, O., {D'Angelo}, G., \& {Bodenheimer}, P. 2009,
  \icarus, 199, 338

\bibitem[{{Lissauer} {et~al.}(2011){Lissauer}, {Fabrycky}, {Ford}, {Borucki},
  {Fressin}, {Marcy}, {Orosz}, {Rowe}, {Torres}, {Welsh}, {Batalha}, {Bryson},
  {Buchhave}, {Caldwell}, {Carter}, {Charbonneau}, {Christiansen}, {Cochran},
  {Desert}, {Dunham}, {Fanelli}, {Fortney}, {Gautier}, {Geary}, {Gilliland},
  {Haas}, {Hall}, {Holman}, {Koch}, {Latham}, {Lopez}, {McCauliff}, {Miller},
  {Morehead}, {Quintana}, {Ragozzine}, {Sasselov}, {Short}, \&
  {Steffen}}]{lissauer11}
{Lissauer}, J.~J., {et~al.} 2011, \nat, 470, 53

\bibitem[{{Lovis} \& {Mayor}(2007)}]{lovis07}
{Lovis}, C., \& {Mayor}, M. 2007, \aap, 472, 657

\bibitem[{{Lund} {et~al.}(2012){Lund}, {Chaplin}, \& {Kjeldsen}}]{lund12}
{Lund}, M.~N., {Chaplin}, W.~J., \& {Kjeldsen}, H. 2012, \mnras, 427, 1784

\bibitem[{{Mann} {et~al.}(2012){Mann}, {Gaidos}, {L{\'e}pine}, \&
  {Hilton}}]{mann12}
{Mann}, A.~W., {Gaidos}, E., {L{\'e}pine}, S., \& {Hilton}, E.~J. 2012, \apj,
  753, 90

\bibitem[{{Marcy} {et~al.}(2013){Marcy}, {Ireland}, {Bedding}, {Howell},
  {Maestro}, {M{\'e}rand}, {Tuthill}, {White}, {Farrington}, {Goldfinger},
  {McAlister}, {Schaefer}, {Sturmann}, {Sturmann}, {ten Brummelaar}, \&
  {Turner}}]{marcy13}
{Marcy}, G.~W., {et~al.} 2013, in preparation

\bibitem[{{Marigo} {et~al.}(2008){Marigo}, {Girardi}, {Bressan}, {Groenewegen},
  {Silva}, \& {Granato}}]{marigo08}
{Marigo}, P., {Girardi}, L., {Bressan}, A., {Groenewegen}, M.~A.~T., {Silva},
  L., \& {Granato}, G.~L. 2008, \aap, 482, 883

\bibitem[{{Mathur} {et~al.}(2012){Mathur}, {Metcalfe}, {Woitaszek}, {Bruntt},
  {Verner}, {Christensen-Dalsgaard}, {Creevey}, {Do{\u g}an}, {Basu}, {Karoff},
  {Stello}, {Appourchaux}, {Campante}, {Chaplin}, {Garc{\'{\i}}a}, {Bedding},
  {Benomar}, {Bonanno}, {Deheuvels}, {Elsworth}, {Gaulme}, {Guzik}, {Handberg},
  {Hekker}, {Herzberg}, {Monteiro}, {Piau}, {Quirion}, {R{\'e}gulo}, {Roth},
  {Salabert}, {Serenelli}, {Thompson}, {Trampedach}, {White}, {Ballot},
  {Brand{\~a}o}, {Molenda-{\.Z}akowicz}, {Kjeldsen}, {Twicken}, {Uddin}, \&
  {Wohler}}]{mathur12}
{Mathur}, S., {et~al.} 2012, \apj, 749, 152

\bibitem[{{Metcalfe} {et~al.}(2010){Metcalfe}, {Monteiro}, {Thompson},
  {Molenda-{\.Z}akowicz}, {Appourchaux}, {Chaplin}, {Do{\u g}an},
  {Eggenberger}, {Bedding}, {Bruntt}, {Creevey}, {Quirion}, {Stello},
  {Bonanno}, {Silva Aguirre}, {Basu}, {Esch}, {Gai}, {Di Mauro}, {Kosovichev},
  {Kitiashvili}, {Su{\'a}rez}, {Moya}, {Piau}, {Garc{\'{\i}}a}, {Marques},
  {Frasca}, {Biazzo}, {Sousa}, {Dreizler}, {Bazot}, {Karoff}, {Frandsen},
  {Wilson}, {Brown}, {Christensen-Dalsgaard}, {Gilliland}, {Kjeldsen},
  {Campante}, {Fletcher}, {Handberg}, {R{\'e}gulo}, {Salabert}, {Schou},
  {Verner}, {Ballot}, {Broomhall}, {Elsworth}, {Hekker}, {Huber}, {Mathur},
  {New}, {Roxburgh}, {Sato}, {White}, {Borucki}, {Koch}, \&
  {Jenkins}}]{metcalfe10}
{Metcalfe}, T.~S., {et~al.} 2010, \apj, 723, 1583

\bibitem[{{Metcalfe} {et~al.}(2012){Metcalfe}, {Chaplin}, {Appourchaux},
  {Garc{\'{\i}}a}, {Basu}, {Brand{\~a}o}, {Creevey}, {Deheuvels}, {Do{\u g}an},
  {Eggenberger}, {Karoff}, {Miglio}, {Stello}, {Y{\i}ld{\i}z}, {{\c C}elik},
  {Antia}, {Benomar}, {Howe}, {R{\'e}gulo}, {Salabert}, {Stahn}, {Bedding},
  {Davies}, {Elsworth}, {Gizon}, {Hekker}, {Mathur}, {Mosser}, {Bryson},
  {Still}, {Christensen-Dalsgaard}, {Gilliland}, {Kawaler}, {Kjeldsen},
  {Ibrahim}, {Klaus}, \& {Li}}]{metcalfe12}
---. 2012, \apjl, 748, L10

\bibitem[{{Miglio}(2012)}]{miglio11}
{Miglio}, A. 2012, in Red Giants as Probes of the Structure and Evolution of
  the Milky Way, ed. A.~Miglio, J.~Montalb{\'a}n, \& A.~Noels, ApSS Proceedings

\bibitem[{{Miglio} {et~al.}(2012{\natexlab{a}}){Miglio}, {Morel}, {Barbieri},
  {Mosser}, {Girardi}, {Montalb{\'a}n}, \& {Valentini}}]{miglio12c}
{Miglio}, A., {Morel}, T., {Barbieri}, M., {Mosser}, B., {Girardi}, L.,
  {Montalb{\'a}n}, J., \& {Valentini}, M. 2012{\natexlab{a}}, in European
  Physical Journal Web of Conferences, Vol.~19, European Physical Journal Web
  of Conferences, 5012

\bibitem[{{Miglio} {et~al.}(2009){Miglio}, {Montalb{\'a}n}, {Baudin},
  {Eggenberger}, {Noels}, {Hekker}, {De Ridder}, {Weiss}, \&
  {Baglin}}]{miglio09}
{Miglio}, A., {et~al.} 2009, \aap, 503, L21

\bibitem[{{Miglio} {et~al.}(2012{\natexlab{b}}){Miglio}, {Brogaard}, {Stello},
  {Chaplin}, {D'Antona}, {Montalb{\'a}n}, {Basu}, {Bressan}, {Grundahl},
  {Pinsonneault}, {Serenelli}, {Elsworth}, {Hekker}, {Kallinger}, {Mosser},
  {Ventura}, {Bonanno}, {Noels}, {Silva Aguirre}, {Szabo}, {Li}, {McCauliff},
  {Middour}, \& {Kjeldsen}}]{miglio12b}
---. 2012{\natexlab{b}}, \mnras, 419, 2077

\bibitem[{{Miglio} {et~al.}(2013{\natexlab{a}}){Miglio}, {Chiappini}, {Morel},
  {Barbieri}, {Chaplin}, {Girardi}, {Montalban}, {Noels}, {Valentini},
  {Mosser}, {Baudin}, {Casagrande}, {Fossati}, {Silva Aguirre}, \&
  {Baglin}}]{miglio13}
---. 2013{\natexlab{a}}, ArXiv e-prints (arXiv:1301.1515)

\bibitem[{{Miglio} {et~al.}(2013{\natexlab{b}}){Miglio}, {Chiappini}, {Morel},
  {Barbieri}, {Chaplin}, {Girardi}, {Montalb{\'a}n}, {Valentini}, {Mosser},
  {Baudin}, {Casagrande}, {Fossati}, {Aguirre}, \& {Baglin}}]{miglio12}
---. 2013{\natexlab{b}}, \mnras, 429, 423

\bibitem[{{Mizuno}(1980)}]{mizuno80}
{Mizuno}, H. 1980, Progress of Theoretical Physics, 64, 544

\bibitem[{{Molenda-{\.Z}akowicz} {et~al.}(2011){Molenda-{\.Z}akowicz},
  {Latham}, {Catanzaro}, {Frasca}, \& {Quinn}}]{molenda11}
{Molenda-{\.Z}akowicz}, J., {Latham}, D.~W., {Catanzaro}, G., {Frasca}, A., \&
  {Quinn}, S.~N. 2011, \mnras, 412, 1210

\bibitem[{{Moorhead} {et~al.}(2011){Moorhead}, {Ford}, {Morehead}, {Rowe},
  {Borucki}, {Batalha}, {Bryson}, {Caldwell}, {Fabrycky}, {Gautier}, {Koch},
  {Holman}, {Jenkins}, {Li}, {Lissauer}, {Lucas}, {Marcy}, {Quinn}, {Quintana},
  {Ragozzine}, {Shporer}, {Still}, \& {Torres}}]{moorhead11}
{Moorhead}, A.~V., {et~al.} 2011, \apjs, 197, 1

\bibitem[{{Morel} \& {Miglio}(2012)}]{morel12}
{Morel}, T., \& {Miglio}, A. 2012, \mnras, 419, L34

\bibitem[{{Mosser} {et~al.}(2010){Mosser}, {Belkacem}, {Goupil}, {Miglio},
  {Morel}, {Barban}, {Baudin}, {Hekker}, {Samadi}, {De Ridder}, {Weiss},
  {Auvergne}, \& {Baglin}}]{mosser10}
{Mosser}, B., {et~al.} 2010, \aap, 517, A22

\bibitem[{{Mosser} {et~al.}(2012){Mosser}, {Elsworth}, {Hekker}, {Huber},
  {Kallinger}, {Mathur}, {Belkacem}, {Goupil}, {Samadi}, {Barban}, {Bedding},
  {Chaplin}, {Garc{\'{\i}}a}, {Stello}, {De Ridder}, {Middour}, {Morris}, \&
  {Quintana}}]{mosser11c}
---. 2012, \aap, 537, A30

\bibitem[{{Mosser} {et~al.}(2013){Mosser}, {Michel}, {Belkacem}, {Goupil},
  {Baglin}, {Barban}, {Provost}, {Samadi}, {Auvergne}, \& {Catala}}]{mosser13}
---. 2013, \aap, 550, A126

\bibitem[{{Movshovitz} {et~al.}(2010){Movshovitz}, {Bodenheimer}, {Podolak}, \&
  {Lissauer}}]{movshovitz10}
{Movshovitz}, N., {Bodenheimer}, P., {Podolak}, M., \& {Lissauer}, J.~J. 2010,
  \icarus, 209, 616

\bibitem[{{Muirhead} {et~al.}(2012{\natexlab{a}}){Muirhead}, {Hamren},
  {Schlawin}, {Rojas-Ayala}, {Covey}, \& {Lloyd}}]{muirhead12}
{Muirhead}, P.~S., {Hamren}, K., {Schlawin}, E., {Rojas-Ayala}, B., {Covey},
  K.~R., \& {Lloyd}, J.~P. 2012{\natexlab{a}}, \apjl, 750, L37

\bibitem[{{Muirhead} {et~al.}(2012{\natexlab{b}}){Muirhead}, {Johnson}, {Apps},
  {Carter}, {Morton}, {Fabrycky}, {Pineda}, {Bottom}, {Rojas-Ayala},
  {Schlawin}, {Hamren}, {Covey}, {Crepp}, {Stassun}, {Pepper}, {Hebb}, {Kirby},
  {Howard}, {Isaacson}, {Marcy}, {Levitan}, {Diaz-Santos}, {Armus}, \&
  {Lloyd}}]{muirhead12b}
{Muirhead}, P.~S., {et~al.} 2012{\natexlab{b}}, \apj, 747, 144

\bibitem[{{Nutzman} {et~al.}(2011){Nutzman}, {Gilliland}, {McCullough},
  {Charbonneau}, {Christensen-Dalsgaard}, {Kjeldsen}, {Nelan}, {Brown}, \&
  {Holman}}]{nutzman11}
{Nutzman}, P., {et~al.} 2011, \apj, 726, 3

\bibitem[{{O'Donovan} {et~al.}(2006){O'Donovan}, {Charbonneau}, {Mandushev},
  {Dunham}, {Latham}, {Torres}, {Sozzetti}, {Brown}, {Trauger}, {Belmonte},
  {Rabus}, {Almenara}, {Alonso}, {Deeg}, {Esquerdo}, {Falco}, {Hillenbrand},
  {Roussanova}, {Stefanik}, \& {Winn}}]{donovan06}
{O'Donovan}, F.~T., {et~al.} 2006, \apjl, 651, L61

\bibitem[{{Pietrinferni} {et~al.}(2004){Pietrinferni}, {Cassisi}, {Salaris}, \&
  {Castelli}}]{basti}
{Pietrinferni}, A., {Cassisi}, S., {Salaris}, M., \& {Castelli}, F. 2004, \apj,
  612, 168

\bibitem[{{Pinsonneault} {et~al.}(2012){Pinsonneault}, {An},
  {Molenda-{\.Z}akowicz}, {Chaplin}, {Metcalfe}, \& {Bruntt}}]{pinsonneault11}
{Pinsonneault}, M.~H., {An}, D., {Molenda-{\.Z}akowicz}, J., {Chaplin}, W.~J.,
  {Metcalfe}, T.~S., \& {Bruntt}, H. 2012, \apjs, 199, 30

\bibitem[{{Plavchan} {et~al.}(2012){Plavchan}, {Bilinski}, \&
  {Currie}}]{plavchan12}
{Plavchan}, P., {Bilinski}, C., \& {Currie}, T. 2012, ArXiv e-prints
  (arXiv:1203.1887)

\bibitem[{{Pollack} {et~al.}(1996){Pollack}, {Hubickyj}, {Bodenheimer},
  {Lissauer}, {Podolak}, \& {Greenzweig}}]{pollack96}
{Pollack}, J.~B., {Hubickyj}, O., {Bodenheimer}, P., {Lissauer}, J.~J.,
  {Podolak}, M., \& {Greenzweig}, Y. 1996, \icarus, 124, 62

\bibitem[{{Safronov} \& {Zvjagina}(1969)}]{safronov69}
{Safronov}, V.~S., \& {Zvjagina}, E.~V. 1969, \icarus, 10, 109

\bibitem[{{Samadi} {et~al.}(2007){Samadi}, {Georgobiani}, {Trampedach R.},
  {Goupil}, {Stein}, \& {Nordlund}}]{samadi07}
{Samadi}, R., {Georgobiani}, D., {Trampedach R.}, R., {Goupil}, M.~J., {Stein},
  R.~F., \& {Nordlund}, A. 2007, \aap, 463, 297

\bibitem[{{Santos} {et~al.}(2004){Santos}, {Israelian}, \& {Mayor}}]{santos04}
{Santos}, N.~C., {Israelian}, G., \& {Mayor}, M. 2004, \aap, 415, 1153

\bibitem[{{Savitzky} \& {Golay}(1964)}]{savitzky64}
{Savitzky}, A., \& {Golay}, M.~J.~E. 1964, Analytical Chemistry, 36, 1627

\bibitem[{{Seager} \& {Mall{\'e}n-Ornelas}(2003)}]{seager03}
{Seager}, S., \& {Mall{\'e}n-Ornelas}, G. 2003, \apj, 585, 1038

\bibitem[{{Silva Aguirre} {et~al.}(2011){Silva Aguirre}, {Chaplin}, {Ballot},
  {Basu}, {Bedding}, {Serenelli}, {Verner}, {Miglio}, {Monteiro}, {Weiss},
  {Appourchaux}, {Bonanno}, {Broomhall}, {Bruntt}, {Campante}, {Casagrande},
  {Corsaro}, {Elsworth}, {Garc{\'{\i}}a}, {Gaulme}, {Handberg}, {Hekker},
  {Huber}, {Karoff}, {Mathur}, {Mosser}, {Salabert}, {Sch{\"o}nrich}, {Sousa},
  {Stello}, {White}, {Christensen-Dalsgaard}, {Gilliland}, {Kawaler},
  {Kjeldsen}, {Houdek}, {Metcalfe}, {Molenda-{\.Z}akowicz}, {Thompson},
  {Caldwell}, {Christiansen}, \& {Wohler}}]{silvaaguirre11}
{Silva Aguirre}, V., {et~al.} 2011, \apjl, 740, L2

\bibitem[{{Silva Aguirre} {et~al.}(2012){Silva Aguirre}, {Casagrande}, {Basu},
  {Campante}, {Chaplin}, {Huber}, {Miglio}, {Serenelli}, {Ballot}, {Bedding},
  {Christensen-Dalsgaard}, {Creevey}, {Elsworth}, {Garc{\'{\i}}a}, {Gilliland},
  {Hekker}, {Kjeldsen}, {Mathur}, {Metcalfe}, {Monteiro}, {Mosser},
  {Pinsonneault}, {Stello}, {Weiss}, {Tenenbaum}, {Twicken}, \&
  {Uddin}}]{silva12}
---. 2012, \apj, 757, 99

\bibitem[{{Silva Aguirre} {et~al.}(2013){Silva Aguirre}, {Casagrande}, {Basu},
  {Campante}, {Chaplin}, {Huber}, {Miglio}, {Serenelli}, {Ballot}, {Bedding},
  {Christensen-Dalsgaard}, {Creevey}, {Elsworth}, {Garc{\'{\i}}a}, {Gilliland},
  {Hekker}, {Kjeldsen}, {Mathur}, {Metcalfe}, {Monteiro}, {Mosser},
  {Pinsonneault}, {Stello}, {Weiss}, {Tenenbaum}, {Twicken}, \&
  {Uddin}}]{silva13}
---. 2013, \apj, submitted

\bibitem[{{Southworth}(2011)}]{southworth11}
{Southworth}, J. 2011, \mnras, 417, 2166

\bibitem[{{Southworth}(2012)}]{southworth12}
---. 2012, \mnras, 426, 1291

\bibitem[{{Sozzetti} {et~al.}(2007){Sozzetti}, {Torres}, {Charbonneau},
  {Latham}, {Holman}, {Winn}, {Laird}, \& {O'Donovan}}]{sozzetti07}
{Sozzetti}, A., {Torres}, G., {Charbonneau}, D., {Latham}, D.~W., {Holman},
  M.~J., {Winn}, J.~N., {Laird}, J.~B., \& {O'Donovan}, F.~T. 2007, \apj, 664,
  1190

\bibitem[{{Steffen} {et~al.}(2012{\natexlab{a}}){Steffen}, {Ragozzine},
  {Fabrycky}, {Carter}, {Ford}, {Holman}, {Rowe}, {Welsh}, {Borucki}, {Boss},
  {Ciardi}, \& {Quinn}}]{steffen12c}
{Steffen}, J.~H., {et~al.} 2012{\natexlab{a}}, Proceedings of the National
  Academy of Science, 109, 7982

\bibitem[{{Steffen} {et~al.}(2012{\natexlab{b}}){Steffen}, {Fabrycky}, {Ford},
  {Carter}, {D{\'e}sert}, {Fressin}, {Holman}, {Lissauer}, {Moorhead}, {Rowe},
  {Ragozzine}, {Welsh}, {Batalha}, {Borucki}, {Buchhave}, {Bryson}, {Caldwell},
  {Charbonneau}, {Ciardi}, {Cochran}, {Endl}, {Everett}, {Gautier},
  {Gilliland}, {Girouard}, {Jenkins}, {Horch}, {Howell}, {Isaacson}, {Klaus},
  {Koch}, {Latham}, {Li}, {Lucas}, {MacQueen}, {Marcy}, {McCauliff}, {Middour},
  {Morris}, {Mullally}, {Quinn}, {Quintana}, {Shporer}, {Still}, {Tenenbaum},
  {Thompson}, {Twicken}, \& {Van Cleve}}]{steffen12}
---. 2012{\natexlab{b}}, \mnras, 421, 2342

\bibitem[{{Stello} {et~al.}(2008){Stello}, {Bruntt}, {Preston}, \&
  {Buzasi}}]{stello08}
{Stello}, D., {Bruntt}, H., {Preston}, H., \& {Buzasi}, D. 2008, \apjl, 674,
  L53

\bibitem[{{Stello} {et~al.}(2009){Stello}, {Chaplin}, {Bruntt}, {Creevey},
  {Garc{\'{\i}}a-Hern{\'a}ndez}, {Monteiro}, {Moya}, {Quirion}, {Sousa},
  {Su{\'a}rez}, {Appourchaux}, {Arentoft}, {Ballot}, {Bedding},
  {Christensen-Dalsgaard}, {Elsworth}, {Fletcher}, {Garc{\'{\i}}a}, {Houdek},
  {Jim{\'e}nez-Reyes}, {Kjeldsen}, {New}, {R{\'e}gulo}, {Salabert}, \&
  {Toutain}}]{stello09}
{Stello}, D., {et~al.} 2009, \apj, 700, 1589

\bibitem[{Tassoul(1980)}]{tassoul80}
Tassoul, M. 1980, \apjs, 43, 469

\bibitem[{{Thommes} {et~al.}(2008){Thommes}, {Matsumura}, \&
  {Rasio}}]{thommes08}
{Thommes}, E.~W., {Matsumura}, S., \& {Rasio}, F.~A. 2008, Science, 321, 814

\bibitem[{{Thygesen} {et~al.}(2012){Thygesen}, {Frandsen}, {Bruntt},
  {Kallinger}, {Andersen}, {Elsworth}, {Hekker}, {Karoff}, {Stello},
  {Brogaard}, {Burke}, {Caldwell}, \& {Christiansen}}]{thygesen12}
{Thygesen}, A.~O., {et~al.} 2012, \aap, 543, A160

\bibitem[{{Torres} {et~al.}(2012){Torres}, {Fischer}, {Sozzetti}, {Buchhave},
  {Winn}, {Holman}, \& {Carter}}]{torres12}
{Torres}, G., {Fischer}, D.~A., {Sozzetti}, A., {Buchhave}, L.~A., {Winn},
  J.~N., {Holman}, M.~J., \& {Carter}, J.~A. 2012, \apj, 757, 161

\bibitem[{{Ulrich}(1986)}]{ulrich86}
{Ulrich}, R.~K. 1986, \apjl, 306, L37

\bibitem[{{Valenti} \& {Piskunov}(1996)}]{valenti96}
{Valenti}, J.~A., \& {Piskunov}, N. 1996, \aaps, 118, 595

\bibitem[{{Vandakurov}(1968)}]{vandakurov68}
{Vandakurov}, Y.~V. 1968, \sovast, 11, 630

\bibitem[{{Vauclair} {et~al.}(2008){Vauclair}, {Laymand}, {Bouchy}, {Vauclair},
  {Hui Bon Hoa}, {Charpinet}, \& {Bazot}}]{vauclair08}
{Vauclair}, S., {Laymand}, M., {Bouchy}, F., {Vauclair}, G., {Hui Bon Hoa}, A.,
  {Charpinet}, S., \& {Bazot}, M. 2008, \aap, 482, L5

\bibitem[{{Verner} \& {Roxburgh}(2011)}]{verner11c}
{Verner}, G.~A., \& {Roxburgh}, I.~W. 2011, ArXiv e-prints \rm
  (arXiv:1104.0631)

\bibitem[{{Verner} {et~al.}(2011{\natexlab{a}}){Verner}, {Elsworth}, {Chaplin},
  {Campante}, {Corsaro}, {Gaulme}, {Hekker}, {Huber}, {Karoff}, {Mathur},
  {Mosser}, {Appourchaux}, {Ballot}, {Bedding}, {Bonanno}, {Broomhall},
  {Garc{\'{\i}}a}, {Handberg}, {New}, {Stello}, {R{\'e}gulo}, {Roxburgh},
  {Salabert}, {White}, {Caldwell}, {Christiansen}, \& {Fanelli}}]{verner11}
{Verner}, G.~A., {et~al.} 2011{\natexlab{a}}, \mnras, 415, 3539

\bibitem[{{Verner} {et~al.}(2011{\natexlab{b}}){Verner}, {Chaplin}, {Basu},
  {Brown}, {Hekker}, {Huber}, {Karoff}, {Mathur}, {Metcalfe}, {Mosser},
  {Quirion}, {Appourchaux}, {Bedding}, {Bruntt}, {Campante}, {Elsworth},
  {Garc{\'{\i}}a}, {Handberg}, {R{\'e}gulo}, {Roxburgh}, {Stello},
  {Christensen-Dalsgaard}, {Gilliland}, {Kawaler}, {Kjeldsen}, {Allen},
  {Clarke}, \& {Girouard}}]{verner11b}
---. 2011{\natexlab{b}}, \apjl, 738, L28

\bibitem[{{Vogt} {et~al.}(1994){Vogt}, {Allen}, {Bigelow}, {Bresee}, {Brown},
  {Cantrall}, {Conrad}, {Couture}, {Delaney}, {Epps}, {Hilyard}, {Hilyard},
  {Horn}, {Jern}, {Kanto}, {Keane}, {Kibrick}, {Lewis}, {Osborne},
  {Pardeilhan}, {Pfister}, {Ricketts}, {Robinson}, {Stover}, {Tucker}, {Ward},
  \& {Wei}}]{vogt94}
{Vogt}, S.~S., {et~al.} 1994, in Society of Photo-Optical Instrumentation
  Engineers (SPIE) Conference Series, Vol. 2198, Society of Photo-Optical
  Instrumentation Engineers (SPIE) Conference Series, ed. D.~L. {Crawford} \&
  E.~R. {Craine}, 362

\bibitem[{{Wang} \& {Ford}(2011)}]{wang11}
{Wang}, J., \& {Ford}, E.~B. 2011, \mnras, 418, 1822

\bibitem[{{White} {et~al.}(2011){White}, {Bedding}, {Stello},
  {Christensen-Dalsgaard}, {Huber}, \& {Kjeldsen}}]{white11}
{White}, T.~R., {Bedding}, T.~R., {Stello}, D., {Christensen-Dalsgaard}, J.,
  {Huber}, D., \& {Kjeldsen}, H. 2011, \apj, 743, 161

\bibitem[{{Winn}(2010)}]{winn10b}
{Winn}, J.~N. 2010, ArXiv e-prints (arXiv:1001.2010)

\bibitem[{{Wright} {et~al.}(2011){Wright}, {Chen{\'e}}, {De Cat}, {Marois},
  {Mathias}, {Macintosh}, {Isaacs}, {Lehmann}, \& {Hartmann}}]{wright11}
{Wright}, D.~J., {et~al.} 2011, \apjl, 728, L20

\end{thebibliography}

\newpage

\begin{footnotesize}
\begin{center}
\renewcommand\thetable{1}      
\begin{longtable*}{l l c | c c c c c | c c c c c | c}  
\caption{Asteroseismic and Spectroscopic Observations of
\nkoi\ \kep\ Planet-Candidate Hosts.}
\\
\hline
KOI & KIC & Kp & \multicolumn{5}{c|}{Asteroseismology} & \multicolumn{5}{c|}{Spectroscopy}  & Notes \\
    &     &    & \numax\ (\muHz) & \Dnu\ (\muHz) & HBR & M & Cad & $v\sin(i)$ & S/N & CCF & Sp & Obs  &  \\
\hline         
    1   &   11446443   &   11.34  &             --  & $   141.0  \pm   1.4 $ &   --    &  25   &   SC  &             --   &  --    &      --    &  --   &   --    &       Kepler-1$^{\rm a}$    \\
    2   &   10666592   &   10.46  &             --  & $   59.22  \pm  0.59 $ &   --    &   1   &   SC  &             --   &  --    &      --    &  --   &   --    &       Kepler-2$^{\rm b}$    \\
    5   &    8554498   &   11.66  & $ 1153  \pm  76$ & $   61.98  \pm  0.96 $ & 1.079   &  30   &   SC  & $10.3  \pm 0.5 $ &  167   &    0.978   &   3   &    HM   &             --              \\
    7   &   11853905   &   12.21  & $ 1436  \pm  42$ & $    74.4  \pm   1.1 $ & 1.080   &  25   &   SC  & $ 2.3  \pm 0.5 $ &  168   &    0.971   &   5   &    FH   &              Kepler-4       \\
   41   &    6521045   &   11.20  & $ 1502  \pm  31$ & $    77.0  \pm   1.1 $ & 1.215   &  27   &   SC  & $ 2.9  \pm 0.5 $ &  300   &    0.987   &   5   &    HM   &             --              \\
   42   &    8866102   &    9.36  & $ 2014  \pm  32$ & $   94.50  \pm  0.60 $ & 1.348   &  28   &   SC  & $15.0  \pm 0.5 $ &  176   &    0.968   &   2   &    HM   &             --              \\
   64   &    7051180   &   13.14  & $  681  \pm  19$ & $   40.05  \pm  0.54 $ & 1.100   &  27   &   SC  & $ 2.4  \pm 0.5 $ &  131   &    0.985   &   3   &     H   &             --              \\
   69   &    3544595   &    9.93  & $ 3366  \pm  81$ & $  145.77  \pm  0.45 $ & 1.036   &  28   &   SC  & $ 2.0  \pm 0.5 $ &  175   &    0.986   &  10   &   FHM   &             --              \\
   72   &   11904151   &   10.96  &             --  & $  118.20  \pm  0.20 $ &   --    &   5   &   SC  &             --   &  --    &      --    &  --   &   --    &       Kepler-10$^{\rm c}$   \\
   75   &    7199397   &   10.78  & $  643  \pm  17$ & $   38.63  \pm  0.68 $ & 1.859   &  28   &   SC  & $ 5.6  \pm 0.5 $ &   91   &    0.964   &   5   &   HMT   &             --              \\
   85   &    5866724   &   11.02  & $ 1880  \pm  60$ & $   89.56  \pm  0.48 $ &   --    &  27   &   SC  &             --   &  --    &      --    &  --   &   --    &       Kepler-50$^{\rm d}$   \\
   87   &   10593626   &   11.66  &             --  & $   137.5  \pm   1.4 $ &   --    &  18   &   SC  &             --   &  --    &      --    &  --   &   --    &       Kepler-22$^{\rm e}$   \\
   97   &    5780885   &   12.88  &              --  & $    56.4  \pm   1.7 $ & 1.050   &  18   &   SC  & $ 3.3  \pm 0.5 $ &   73   &    0.975   &  10   &    FM   &             --              \\
   98   &   10264660   &   12.13  &              --  & $    53.9  \pm   1.6 $ & 1.078   &  27   &   SC  & $10.8  \pm 0.5 $ &   89   &    0.970   &  10   &    FH   &              Kepler-14      \\
  107   &   11250587   &   12.70  &              --  & $    74.4  \pm   2.8 $ & 1.035   &  18   &   SC  & $ 3.3  \pm 0.7 $ &   39   &    0.889   &   2   &     F   &             --              \\
  108   &    4914423   &   12.29  & $ 1663  \pm  56$ & $    81.5  \pm   1.6 $ & 1.062   &  27   &   SC  & $ 3.8  \pm 0.7 $ &  108   &    0.945   &   5   &    HM   &             --              \\
  113   &    2306756   &   12.39  & $ 1412  \pm  50$ & $    70.4  \pm   2.2 $ & 1.054   &  12   &   SC  & $ 2.5  \pm 0.5 $ &   25   &    0.915   &   3   &   FMT   &             --              \\
  117   &   10875245   &   12.49  & $ 1711  \pm 107$ & $    86.7  \pm   3.7 $ & 1.043   &  21   &   SC  & $ 3.5  \pm 0.5 $ &   39   &    0.978   &   2   &     M   &             --              \\
  118   &    3531558   &   12.38  &              --  & $    86.8  \pm   2.2 $ & 1.042   &  18   &   SC  & $ 2.2  \pm 0.5 $ &   25   &    0.894   &   1   &     F   &             --              \\
  119   &    9471974   &   12.65  &              --  & $    49.4  \pm   3.2 $ & 1.052   &  12   &   SC  & $ 5.3  \pm 0.6 $ &   27   &    0.871   &   3   &     F   &             --              \\
  122   &    8349582   &   12.35  & $ 1677  \pm  90$ & $    83.6  \pm   1.4 $ & 1.061   &  27   &   SC  & $ 2.3  \pm 0.5 $ &  107   &    0.973   &   5   &    HM   &             --              \\
  123   &    5094751   &   12.37  & $ 1745  \pm 117$ & $    91.1  \pm   2.3 $ & 1.046   &  27   &   SC  & $ 2.8  \pm 0.5 $ &  117   &    0.963   &   6   &    HM   &             --              \\
  168   &   11512246   &   13.44  &              --  & $    72.9  \pm   2.1 $ & 1.026   &  27   &   SC  & $ 2.9  \pm 0.5 $ &   24   &    0.850   &   2   &    FM   &              Kepler-23      \\
  244   &    4349452   &   10.73  & $ 2106  \pm  50$ & $   98.27  \pm  0.57 $ & 1.082   &  22   &   SC  & $11.1  \pm 0.5 $ &  123   &    0.939   &   5   &   HMT   &              Kepler-25      \\
  245   &    8478994   &    9.71  &             --  & $   178.7  \pm   1.4 $ &   --    &  15   &   SC  &             --   &  --    &      --    &  --   &   --    &       Kepler-37$^{\rm f}$   \\
  246   &   11295426   &   10.00  & $ 2154  \pm  13$ & $  101.57  \pm  0.10 $ &   --    &  22   &   SC  &             --   &  --    &      --    &  --   &   --    &       Kepler-68$^{\rm g}$   \\
  257   &    5514383   &   10.87  &              --  & $   113.3  \pm   2.0 $ & 1.081   &  16   &   SC  & $ 9.4  \pm 0.5 $ &   41   &    0.909   &   4   &    MT   &             --              \\
  260   &    8292840   &   10.50  & $ 1983  \pm  37$ & $   92.85  \pm  0.35 $ & 1.219   &  21   &   SC  & $10.4  \pm 0.5 $ &   47   &    0.868   &   3   &    MT   &             --              \\
  262   &   11807274   &   10.42  & $ 1496  \pm  56$ & $   75.71  \pm  0.31 $ &   --    &  18   &   SC  &             --   &  --    &      --    &  --   &   --    &       Kepler-65$^{\rm d}$   \\
  263   &   10514430   &   10.82  & $ 1303  \pm  30$ & $    70.0  \pm   1.0 $ & 1.077   &  22   &   SC  & $ 2.7  \pm 0.6 $ &  101   &    0.876   &   4   &    HT   &             --              \\
  268   &    3425851   &   10.56  & $ 2038  \pm  60$ & $    92.6  \pm   1.5 $ & 1.125   &  10   &   SC  & $ 9.5  \pm 0.5 $ &   47   &    0.900   &   2   &    FT   &             --              \\
  269   &    7670943   &   10.93  & $ 1895  \pm  73$ & $    88.6  \pm   1.3 $ & 1.143   &  19   &   SC  & $13.5  \pm 0.6 $ &   53   &    0.827   &   2   &     T   &             --              \\
  270   &    6528464   &   11.41  & $ 1380  \pm  56$ & $    75.4  \pm   1.4 $ & 1.144   &   9   &   SC  & $ 3.5  \pm 0.5 $ &   34   &    0.855   &   2   &     T   &             --              \\
  271   &    9451706   &   11.48  & $ 1988  \pm  86$ & $    95.0  \pm   1.6 $ & 1.067   &  19   &   SC  & $ 8.1  \pm 0.6 $ &   37   &    0.837   &   3   &     T   &             --              \\
  273   &    3102384   &   11.46  &              --  & $   124.3  \pm   1.3 $ & 1.026   &  19   &   SC  & $ 2.4  \pm 0.5 $ &   99   &    0.951   &   4   &   HMT   &             --              \\
  274   &    8077137   &   11.39  & $ 1324  \pm  39$ & $   68.80  \pm  0.64 $ & 1.186   &  18   &   SC  & $ 7.7  \pm 0.5 $ &  149   &    0.960   &   2   &    HM   &             --              \\
  275   &   10586004   &   11.70  & $ 1395  \pm  40$ & $    69.2  \pm   1.4 $ & 1.249   &   6   &   SC  & $ 3.4  \pm 0.5 $ &   37   &    0.904   &   2   &     T   &             --              \\
  276   &   11133306   &   11.85  & $ 2381  \pm  95$ & $   107.9  \pm   1.9 $ & 1.033   &  18   &   SC  & $ 2.8  \pm 0.5 $ &   33   &    0.901   &   1   &     T   &             --              \\
  277   &   11401755   &   11.87  & $ 1250  \pm  44$ & $    67.9  \pm   1.2 $ &   --    &  15   &   SC  &             --   &  --    &      --    &  --   &   --    &       Kepler-36$^{\rm h}$   \\
  279   &   12314973   &   11.68  &              --  & $    78.7  \pm   5.3 $ & 1.078   &  19   &   SC  & $15.2  \pm 0.5 $ &   27   &    0.880   &   1   &     M   &             --              \\
  280   &    4141376   &   11.07  & $ 2928  \pm  97$ & $   128.8  \pm   1.3 $ & 1.037   &  20   &   SC  & $ 3.5  \pm 0.5 $ &   56   &    0.876   &   1   &     T   &             --              \\
  281   &    4143755   &   11.95  & $ 1458  \pm  57$ & $    77.2  \pm   1.3 $ & 1.043   &  18   &   SC  & $ 2.2  \pm 0.6 $ &   34   &    0.836   &   2   &    FT   &             --              \\
  282   &    5088536   &   11.53  &              --  & $   109.6  \pm   3.1 $ & 1.052   &   4   &   SC  & $ 3.0  \pm 0.5 $ &   28   &    0.941   &   1   &     M   &             --              \\
  285   &    6196457   &   11.56  & $ 1299  \pm  53$ & $    66.6  \pm   1.1 $ & 1.118   &   9   &   SC  & $ 4.2  \pm 0.6 $ &   33   &    0.884   &   2   &    MT   &             --              \\
  288   &    9592705   &   11.02  & $ 1008  \pm  21$ & $   53.54  \pm  0.32 $ & 1.317   &  13   &   SC  & $ 9.4  \pm 0.5 $ &   41   &    0.870   &   3   &    MT   &             --              \\
  319   &    8684730   &   12.71  & $  962  \pm  39$ & $    51.7  \pm   1.9 $ & 1.201   &   9   &   SC  & $ 5.9  \pm 0.5 $ &   32   &    0.885   &   2   &     T   &             --              \\
  370   &    8494142   &   11.93  & $ 1133  \pm  81$ & $   61.80  \pm  0.76 $ & 1.166   &  15   &   SC  & $ 8.9  \pm 0.5 $ &   29   &    0.835   &   2   &     T   &             --              \\
  371   &    5652983   &   12.19  &              --  & $   29.27  \pm  0.36 $ & 1.196   &  15   &   SC  & $ 3.1  \pm 0.6 $ &   22   &    0.866   &   2   &     T   &             --              \\
  623   &   12068975   &   11.81  & $ 2298  \pm 105$ & $   108.4  \pm   3.1 $ & 1.048   &  15   &   SC  & $ 2.6  \pm 0.5 $ &   33   &    0.847   &   3   &     T   &             --              \\
  674   &    7277317   &   13.78  & $  533  \pm  15$ & $   33.00  \pm  0.72 $ & 1.254   &   6   &   SC  & $ 1.6  \pm 0.5 $ &   54   &    0.972   &   1   &     H   &             --              \\
  974   &    9414417   &    9.58  & $ 1115  \pm  32$ & $   60.05  \pm  0.27 $ & 1.636   &  16   &   SC  & $ 9.3  \pm 0.5 $ &   55   &    0.953   &   2   &     T   &             --              \\
  975   &    3632418   &    8.22  & $ 1153  \pm  32$ & $   60.86  \pm  0.55 $ &   --    &   1   &   SC  &             --   &  --    &      --    &  --   &   --    &       Kepler-21$^{\rm i}$   \\
  981   &    8607720   &   10.73  & $159.0  \pm 2.4$ & $   12.86  \pm  0.06 $ & 2.864   &  31   &   LC  & $ 2.0  \pm 0.5 $ &   54   &    0.964   &   5   &    FT   &             --              \\
 1019   &    8179973   &   10.27  & $341.8  \pm 6.4$ & $   22.93  \pm  0.13 $ & 4.575   &  16   &   SC  & $ 2.2  \pm 0.5 $ &   93   &    0.979   &   4   &   HMT   &             --              \\
 1054   &    6032981   &   11.90  & $ 35.1  \pm 0.6 $ &                  --   & 2.697   &  31   &   LC  & $ 7.3  \pm 0.5 $ &   24   &    0.881   &   3   &     M   &             --              \\
 1221   &    3640905   &   11.58  & $500.7  \pm 7.0$ & $   30.63  \pm  0.20 $ & 3.201   &   9   &   SC  & $ 2.5  \pm 0.5 $ &   32   &    0.945   &   2   &     T   &             --              \\
 1222   &    4060815   &   12.20  & $  333  \pm  11$ & $   22.33  \pm  0.56 $ & 4.676   &   1   &   SC  & $ 2.6  \pm 0.5 $ &   25   &    0.923   &   2   &     T   &             --              \\
 1230   &    6470149   &   12.26  & $118.2  \pm 3.7$ & $    9.59  \pm  0.04 $ & 1.591   &  31   &   LC  & $ 3.3  \pm 0.5 $ &   22   &    0.903   &   1   &     T   &             --              \\
 1241   &    6448890   &   12.44  & $244.3  \pm 3.4$ & $   17.40  \pm  0.24 $ &   --    &  31   &   LC  &             --   &  --    &      --    &  --   &   --    &       Kepler-56$^{\rm j}$   \\
 1282   &    8822366   &   12.55  &              --  & $    71.3  \pm   1.9 $ & 1.069   &   9   &   SC  & $ 7.2  \pm 0.5 $ &   29   &    0.873   &   2   &     T   &             --              \\
 1299   &   10864656   &   12.18  & $259.5  \pm 4.3$ & $   18.53  \pm  0.17 $ & 3.696   &   9   &   SC  & $ 2.7  \pm 0.5 $ &   32   &    0.940   &  10   &    FT   &             --              \\
 1314   &   10585852   &   13.24  & $352.1  \pm 6.9$ & $   23.42  \pm  0.23 $ & 1.868   &   9   &   SC  & $ 2.2  \pm 0.5 $ &   25   &    0.935   &   2   &    MT   &             --              \\
 1537   &    9872292   &   11.74  &              --  & $    63.8  \pm   1.2 $ & 1.135   &   9   &   SC  & $ 9.6  \pm 0.7 $ &   23   &    0.810   &   2   &     M   &             --              \\
 1612   &   10963065   &    8.77  & $ 2193  \pm  48$ & $  103.20  \pm  0.63 $ & 1.571   &  16   &   SC  & $ 3.4  \pm 0.5 $ &  180   &    0.990   &   4   &    HT   &             --              \\
 1613   &    6268648   &   11.05  &              --  & $    88.9  \pm   2.0 $ & 1.105   &   3   &   SC  & $10.6  \pm 0.7 $ &   28   &    0.811   &   2   &     T   &             --              \\
 1618   &    7215603   &   11.60  &              --  & $    82.4  \pm   1.1 $ & 1.107   &   7   &   SC  & $11.7  \pm 0.5 $ &   28   &    0.871   &   2   &     T   &             --              \\
 1621   &    5561278   &   11.86  & $ 1023  \pm  36$ & $    56.2  \pm   1.7 $ & 1.177   &   6   &   SC  & $ 6.4  \pm 0.5 $ &   39   &    0.936   &   1   &     T   &             --              \\
 1890   &    7449136   &   11.70  &              --  & $    76.4  \pm   1.4 $ & 1.104   &   4   &   SC  & $ 7.8  \pm 0.5 $ &   41   &    0.936   &   2   &    FM   &             --              \\
 1894   &   11673802   &   13.43  &              --  & $   21.76  \pm  0.38 $ & 1.338   &  31   &   LC  & $ 2.8  \pm 0.5 $ &   22   &    0.942   &   1   &     T   &             --              \\
 1924   &    5108214   &    7.84  & $  703  \pm  35$ & $   41.34  \pm  0.67 $ & 3.125   &   3   &   SC  & $ 4.8  \pm 0.5 $ &  187   &    0.990   &   6   &   FHM   &             --              \\
 1925   &    9955598   &    9.44  & $ 3546  \pm 119$ & $  153.18  \pm  0.14 $ & 1.058   &  22   &   SC  & $ 1.6  \pm 0.5 $ &  208   &    0.985   &   4   &    FH   &             --              \\
 1930   &    5511081   &   12.12  &              --  & $    63.3  \pm   3.4 $ & 1.121   &   3   &   SC  & $ 4.3  \pm 0.5 $ &   32   &    0.932   &   2   &    FM   &             --              \\
 1962   &    5513648   &   10.77  &              --  & $    78.6  \pm   3.7 $ & 1.106   &   3   &   SC  & $ 3.9  \pm 0.5 $ &   41   &    0.926   &   3   &   FMT   &             --              \\
 2133   &    8219268   &   12.49  & $108.9  \pm 3.0$ & $    9.39  \pm  0.22 $ & 5.393   &  31   &   LC  & $ 3.2  \pm 0.5 $ &   21   &    0.918   &   3   &     T   &             --              \\
 2481   &    4476423   &   13.61  & $ 51.2  \pm 1.6$ & $    5.09  \pm  0.12 $ & 4.626   &  13   &   LC  & $ 5.9  \pm 1.3 $ &   13   &    0.707   &   1   &     M   &             --              \\
 2545   &    9696358   &   11.75  &              --  & $    51.4  \pm   3.7 $ & 1.388   &   1   &   SC  & $ 8.9  \pm 0.5 $ &   29   &    0.941   &   1   &     M   &             --              \\
 2640   &    9088780   &   13.23  & $ 76.7  \pm 1.3$ & $    7.46  \pm  0.09 $ & 5.433   &  16   &   LC  & $ 2.0  \pm 0.5 $ &   56   &    0.981   &   1   &     H   &             --              \\
\hline
\end{longtable*} 
\flushleft The solar reference values for the asteroseismic observations are 
$\nu_{\rm max,\sun}=3090\pm30\,\muHz$ and $\Delta\nu_{\sun}=135.1\pm0.1\,\muHz$ \citep{huber11b}.
``HBR'' denotes the height-to-background ratio of the power excess 
\citep[a measure of signal-to-noise, see e.g.][]{kallinger10}, ``Cad'' the type 
of \kep\ data used for the detection (SC = short-cadence, LC = long-cadence), and 
``M'' the number of months of \kep\ data used for the analysis. 
For spectroscopic observations, ``CCF'' denotes the cross-correlation function 
\citep[a measure of the quality of the fit compared to the spectral template, see][]{buchhave12}, 
``Sp'' the number of spectra used in the analysis, and ``Obs'' the spectrographs 
used for the observations (F = FIES, H = HIRES, M = McDonald, T = TRES).
References to solutions published in separate 
papers: a - \citet{barclay12}, 
b - \citet{cd10}, c - \citet{batalha11}, d - \citet{chaplin12}, e - \citet{borucki12},
f - \citet{barclay12b}, g - \citet{gilliland13}, 
h - \citet{carter12}, i - \citet{howell12}, j - \citet{huber12d}. 
Note that stars with solutions published in separate papers were not re-analyzed in this 
study, and hence the columns ``HBR'' as well as spectroscopic information are not available 
for these host stars.
\label{tab:results}
\end{center}
\end{footnotesize}

\begin{footnotesize}
\begin{center}
\renewcommand\thetable{2}      
\begin{longtable*}{l l c c c c c l}  
\caption{Fundamental Properties of \nkoi\ \kep\ Planet-Candidate Hosts.}
\\
\hline
KOI & KIC & $T_{\rm eff}$ (K) & [Fe/H] & $\rho$ (g cm$^{-3}$) & $ R (\rsun)$ & $M (\msun)$ & Notes  \\
\hline         
      1       &   11446443& $ 5850 \pm 50$ & $-0.15\pm0.10$ & $     1.530\pm     0.030$ & $ 0.950\pm0.020$ & $0.940\pm0.050 $ & Kepler-1$^{\rm a}$  \\
      2       &   10666592& $ 6350 \pm 80$ & $+0.26\pm0.08$ & $    0.2712\pm    0.0032$ & $ 1.991\pm0.018$ & $1.520\pm0.036 $ & Kepler-2$^{\rm b}$  \\
      5       &    8554498& $ 5753 \pm 75$ & $+0.05\pm0.10$ & $    0.2965\pm    0.0092$ & $ 1.747\pm0.042$ & $1.130\pm0.065 $ &                  -- \\
      7       &   11853905& $ 5781 \pm 76$ & $+0.09\pm0.10$ & $     0.427\pm     0.013$ & $ 1.533\pm0.040$ & $1.092\pm0.073 $ &            Kepler-4 \\
     41       &    6521045& $ 5825 \pm 75$ & $+0.02\pm0.10$ & $     0.457\pm     0.013$ & $ 1.490\pm0.035$ & $1.080\pm0.063 $ &                  -- \\
     42       &    8866102& $ 6325 \pm 75$ & $+0.01\pm0.10$ & $    0.6892\pm    0.0088$ & $ 1.361\pm0.018$ & $1.242\pm0.045 $ &                  -- \\
     64       &    7051180& $ 5302 \pm 75$ & $-0.00\pm0.10$ & $    0.1238\pm    0.0033$ & $ 2.437\pm0.072$ & $1.262\pm0.089 $ &                  -- \\
     69       &    3544595& $ 5669 \pm 75$ & $-0.18\pm0.10$ & $     1.640\pm     0.010$ & $ 0.921\pm0.020$ & $0.909\pm0.057 $ &                  -- \\
     72       &   11904151& $ 5627 \pm 44$ & $-0.15\pm0.04$ & $    1.0680\pm    0.0080$ & $ 1.056\pm0.021$ & $0.895\pm0.060 $ & Kepler-10$^{\rm c}$ \\
     75       &    7199397& $ 5896 \pm 75$ & $-0.17\pm0.10$ & $    0.1152\pm    0.0040$ & $ 2.527\pm0.059$ & $1.330\pm0.069 $ &                  -- \\
     85       &    5866724& $ 6169 \pm 50$ & $+0.09\pm0.08$ & $     0.621\pm     0.011$ & $ 1.424\pm0.024$ & $1.273\pm0.061 $ & Kepler-50$^{\rm d}$ \\
     87       &   10593626& $ 5642 \pm 50$ & $-0.27\pm0.08$ & $     1.458\pm     0.030$ & $ 0.979\pm0.020$ & $0.970\pm0.060 $ & Kepler-22$^{\rm e}$ \\
     97       &    5780885& $ 6027 \pm 75$ & $+0.10\pm0.10$ & $     0.245\pm     0.015$ & $ 1.962\pm0.066$ & $1.318\pm0.089 $ &                  -- \\
     98       &   10264660& $ 6378 \pm 75$ & $-0.02\pm0.10$ & $     0.224\pm     0.014$ & $ 2.075\pm0.070$ & $1.391\pm0.098 $ &           Kepler-14 \\
    107       &   11250587& $ 5862 \pm 97$ & $+0.27\pm0.11$ & $     0.427\pm     0.032$ & $ 1.586\pm0.061$ & $1.201\pm0.091 $ &                  -- \\
    108       &    4914423& $ 5845 \pm 88$ & $+0.07\pm0.11$ & $     0.513\pm     0.020$ & $ 1.436\pm0.039$ & $1.094\pm0.068 $ &                  -- \\
    113       &    2306756& $ 5543 \pm 79$ & $+0.44\pm0.10$ & $     0.382\pm     0.024$ & $ 1.580\pm0.064$ & $1.103\pm0.097 $ &                  -- \\
    117       &   10875245& $ 5851 \pm 75$ & $+0.27\pm0.10$ & $     0.581\pm     0.049$ & $ 1.411\pm0.047$ & $1.142\pm0.068 $ &                  -- \\
    118       &    3531558& $ 5747 \pm 85$ & $+0.03\pm0.10$ & $     0.581\pm     0.030$ & $ 1.357\pm0.040$ & $1.023\pm0.070 $ &                  -- \\
    119       &    9471974& $ 5854 \pm 92$ & $+0.31\pm0.11$ & $     0.188\pm     0.024$ & $ 2.192\pm0.121$ & $1.377\pm0.089 $ &                  -- \\
    122       &    8349582& $ 5699 \pm 74$ & $+0.30\pm0.10$ & $     0.540\pm     0.019$ & $ 1.415\pm0.039$ & $1.084\pm0.076 $ &                  -- \\
    123       &    5094751& $ 5952 \pm 75$ & $-0.08\pm0.10$ & $     0.641\pm     0.032$ & $ 1.323\pm0.037$ & $1.039\pm0.065 $ &                  -- \\
    168       &   11512246& $ 5828 \pm100$ & $-0.05\pm0.10$ & $     0.410\pm     0.023$ & $ 1.548\pm0.048$ & $1.078\pm0.077 $ &           Kepler-23 \\
    244       &    4349452& $ 6270 \pm 79$ & $-0.04\pm0.10$ & $    0.7453\pm    0.0086$ & $ 1.309\pm0.023$ & $1.187\pm0.060 $ &           Kepler-25 \\
    245       &    8478994& $ 5417 \pm 75$ & $-0.32\pm0.07$ & $     2.458\pm     0.046$ & $ 0.772\pm0.026$ & $0.803\pm0.068 $ & Kepler-37$^{\rm f}$ \\
    246       &   11295426& $ 5793 \pm 74$ & $+0.12\pm0.07$ & $    0.7903\pm    0.0054$ & $ 1.243\pm0.019$ & $1.079\pm0.051 $ & Kepler-68$^{\rm g}$ \\
    257       &    5514383& $ 6184 \pm 81$ & $+0.12\pm0.10$ & $     0.990\pm     0.034$ & $ 1.188\pm0.022$ & $1.180\pm0.053 $ &                  -- \\
    260       &    8292840& $ 6239 \pm 94$ & $-0.14\pm0.10$ & $    0.6652\pm    0.0050$ & $ 1.358\pm0.024$ & $1.188\pm0.059 $ &                  -- \\
    262       &   11807274& $ 6225 \pm 75$ & $-0.00\pm0.08$ & $    0.4410\pm    0.0040$ & $ 1.584\pm0.031$ & $1.259\pm0.072 $ & Kepler-65$^{\rm d}$ \\
    263       &   10514430& $ 5784 \pm 98$ & $-0.11\pm0.11$ & $     0.378\pm     0.011$ & $ 1.574\pm0.039$ & $1.045\pm0.064 $ &                  -- \\
    268       &    3425851& $ 6343 \pm 85$ & $-0.04\pm0.10$ & $     0.662\pm     0.021$ & $ 1.366\pm0.026$ & $1.230\pm0.058 $ &                  -- \\
    269       &    7670943& $ 6463 \pm110$ & $+0.09\pm0.11$ & $     0.605\pm     0.018$ & $ 1.447\pm0.026$ & $1.318\pm0.057 $ &                  -- \\
    270       &    6528464& $ 5588 \pm 99$ & $-0.10\pm0.10$ & $     0.439\pm     0.016$ & $ 1.467\pm0.033$ & $0.969\pm0.053 $ &                  -- \\
    271       &    9451706& $ 6106 \pm106$ & $+0.33\pm0.10$ & $     0.697\pm     0.023$ & $ 1.359\pm0.035$ & $1.240\pm0.086 $ &                  -- \\
    273       &    3102384& $ 5739 \pm 75$ & $+0.35\pm0.10$ & $     1.193\pm     0.025$ & $ 1.081\pm0.019$ & $1.069\pm0.048 $ &                  -- \\
    274       &    8077137& $ 6072 \pm 75$ & $-0.09\pm0.10$ & $    0.3653\pm    0.0068$ & $ 1.659\pm0.038$ & $1.184\pm0.074 $ &                  -- \\
    275       &   10586004& $ 5770 \pm 83$ & $+0.29\pm0.10$ & $     0.370\pm     0.015$ & $ 1.641\pm0.051$ & $1.197\pm0.094 $ &                  -- \\
    276       &   11133306& $ 5982 \pm 82$ & $-0.02\pm0.10$ & $     0.898\pm     0.032$ & $ 1.185\pm0.026$ & $1.076\pm0.061 $ &                  -- \\
    277       &   11401755& $ 5911 \pm 66$ & $-0.20\pm0.06$ & $    0.3508\pm    0.0056$ & $ 1.626\pm0.019$ & $1.071\pm0.043 $ & Kepler-36$^{\rm h}$ \\
    279       &   12314973& $ 6215 \pm 89$ & $+0.28\pm0.10$ & $     0.478\pm     0.064$ & $ 1.570\pm0.085$ & $1.346\pm0.084 $ &                  -- \\
    280       &    4141376& $ 6134 \pm 91$ & $-0.24\pm0.10$ & $     1.281\pm     0.027$ & $ 1.042\pm0.026$ & $1.032\pm0.070 $ &                  -- \\
    281       &    4143755& $ 5622 \pm106$ & $-0.40\pm0.11$ & $     0.459\pm     0.015$ & $ 1.406\pm0.041$ & $0.884\pm0.066 $ &                  -- \\
    282       &    5088536& $ 5884 \pm 75$ & $-0.22\pm0.10$ & $     0.927\pm     0.053$ & $ 1.127\pm0.033$ & $0.934\pm0.059 $ &                  -- \\
    285       &    6196457& $ 5871 \pm 94$ & $+0.17\pm0.11$ & $     0.342\pm     0.011$ & $ 1.703\pm0.048$ & $1.207\pm0.084 $ &                  -- \\
    288       &    9592705& $ 6174 \pm 92$ & $+0.22\pm0.10$ & $    0.2212\pm    0.0027$ & $ 2.114\pm0.042$ & $1.490\pm0.082 $ &                  -- \\
    319       &    8684730& $ 5882 \pm 87$ & $+0.16\pm0.10$ & $     0.206\pm     0.015$ & $ 2.064\pm0.076$ & $1.325\pm0.096 $ &                  -- \\
    370       &    8494142& $ 6144 \pm106$ & $+0.13\pm0.10$ & $    0.2948\pm    0.0072$ & $ 1.850\pm0.050$ & $1.319\pm0.101 $ &                  -- \\
    371       &    5652983& $ 5198 \pm 95$ & $+0.19\pm0.11$ & $    0.0661\pm    0.0016$ & $ 3.207\pm0.107$ & $1.552\pm0.154 $ &                  -- \\
    623       &   12068975& $ 6004 \pm102$ & $-0.38\pm0.10$ & $     0.907\pm     0.052$ & $ 1.120\pm0.033$ & $0.922\pm0.059 $ &                  -- \\
    674       &    7277317& $ 4883 \pm 75$ & $+0.16\pm0.10$ & $    0.0840\pm    0.0037$ & $ 2.690\pm0.114$ & $1.150\pm0.120 $ &                  -- \\
    974       &    9414417& $ 6253 \pm 75$ & $-0.13\pm0.10$ & $    0.2783\pm    0.0025$ & $ 1.851\pm0.044$ & $1.270\pm0.086 $ &                  -- \\
    975       &    3632418& $ 6131 \pm 44$ & $-0.15\pm0.06$ & $    0.2886\pm    0.0087$ & $ 1.860\pm0.020$ & $1.340\pm0.010 $ & Kepler-21$^{\rm i}$ \\
    981       &    8607720& $ 5066 \pm 75$ & $-0.33\pm0.10$ & $   0.01275\pm   0.00011$ & $ 5.324\pm0.107$ & $1.372\pm0.073 $ &                  -- \\
   1019       &    8179973& $ 5009 \pm 75$ & $-0.02\pm0.10$ & $   0.04057\pm   0.00047$ & $ 3.585\pm0.090$ & $1.327\pm0.094 $ &                  -- \\
   1054$^{*}$ &    6032981& $ 5254 \pm 97$ & $-0.94\pm0.16 $ &         --               &      --          &     --           &                  -- \\
   1221       &    3640905& $ 4991 \pm 75$ & $+0.28\pm0.10$ & $   0.07240\pm   0.00094$ & $ 2.935\pm0.066$ & $1.298\pm0.076 $ &                  -- \\
   1222       &    4060815& $ 5055 \pm 75$ & $-0.07\pm0.10$ & $    0.0385\pm    0.0019$ & $ 3.733\pm0.176$ & $1.429\pm0.162 $ &                  -- \\
   1230       &    6470149& $ 5015 \pm 97$ & $-0.21\pm0.16$ & $  0.007091\pm  0.000067$ & $ 7.062\pm0.258$ & $1.782\pm0.193 $ &                  -- \\
   1241       &    6448890& $ 4840 \pm 97$ & $+0.20\pm0.16$ & $   0.02460\pm   0.00060$ & $ 4.230\pm0.150$ & $1.320\pm0.130 $ & Kepler-56$^{\rm j}$ \\
   1282       &    8822366& $ 6034 \pm 92$ & $-0.14\pm0.10$ & $     0.392\pm     0.021$ & $ 1.592\pm0.050$ & $1.120\pm0.083 $ &                  -- \\
   1299       &   10864656& $ 4995 \pm 78$ & $-0.07\pm0.10$ & $   0.02650\pm   0.00049$ & $ 4.160\pm0.120$ & $1.353\pm0.101 $ &                  -- \\
   1314       &   10585852& $ 5048 \pm 75$ & $-0.03\pm0.10$ & $   0.04233\pm   0.00082$ & $ 3.549\pm0.104$ & $1.345\pm0.102 $ &                  -- \\
   1537       &    9872292& $ 6260 \pm116$ & $+0.10\pm0.11$ & $     0.314\pm     0.011$ & $ 1.824\pm0.049$ & $1.366\pm0.101 $ &                  -- \\
   1612       &   10963065& $ 6104 \pm 74$ & $-0.20\pm0.10$ & $     0.822\pm     0.010$ & $ 1.225\pm0.027$ & $1.079\pm0.069 $ &                  -- \\
   1613       &    6268648& $ 6044 \pm117$ & $-0.24\pm0.11$ & $     0.609\pm     0.028$ & $ 1.327\pm0.041$ & $1.008\pm0.080 $ &                  -- \\
   1618       &    7215603& $ 6173 \pm 93$ & $+0.17\pm0.10$ & $     0.524\pm     0.013$ & $ 1.506\pm0.035$ & $1.270\pm0.082 $ &                  -- \\
   1621       &    5561278& $ 6081 \pm 75$ & $-0.03\pm0.10$ & $     0.244\pm     0.014$ & $ 1.954\pm0.064$ & $1.294\pm0.093 $ &                  -- \\
   1890       &    7449136& $ 6099 \pm 75$ & $+0.04\pm0.10$ & $     0.450\pm     0.017$ & $ 1.557\pm0.040$ & $1.206\pm0.083 $ &                  -- \\
   1894       &   11673802& $ 4992 \pm 75$ & $+0.04\pm0.10$ & $    0.0365\pm    0.0013$ & $ 3.790\pm0.190$ & $1.410\pm0.214 $ &                  -- \\
   1924       &    5108214& $ 5844 \pm 75$ & $+0.21\pm0.10$ & $    0.1319\pm    0.0043$ & $ 2.490\pm0.055$ & $1.443\pm0.080 $ &                  -- \\
   1925       &    9955598& $ 5460 \pm 75$ & $+0.08\pm0.10$ & $    1.8106\pm    0.0032$ & $ 0.893\pm0.018$ & $0.918\pm0.057 $ &                  -- \\
   1930       &    5511081& $ 5923 \pm 77$ & $-0.07\pm0.10$ & $     0.309\pm     0.034$ & $ 1.735\pm0.082$ & $1.142\pm0.084 $ &                  -- \\
   1962       &    5513648& $ 5904 \pm 85$ & $-0.07\pm0.10$ & $     0.477\pm     0.045$ & $ 1.470\pm0.060$ & $1.083\pm0.071 $ &                  -- \\
   2133       &    8219268& $ 4605 \pm 97$ & $+0.29\pm0.16$ & $   0.00681\pm   0.00032$ & $ 6.528\pm0.352$ & $1.344\pm0.169 $ &                  -- \\
   2481       &    4476423& $ 4553 \pm 97$ & $+0.42\pm0.16$ & $  0.001999\pm  0.000096$ & $10.472\pm0.696$ & $1.616\pm0.256 $ &                  -- \\
   2545       &    9696358& $ 6131 \pm 75$ & $+0.13\pm0.10$ & $     0.204\pm     0.029$ & $ 2.134\pm0.134$ & $1.417\pm0.093 $ &                  -- \\
   2640       &    9088780& $ 4854 \pm 97$ & $-0.33\pm0.16$ & $   0.00429\pm   0.00010$ & $ 7.478\pm0.246$ & $1.274\pm0.110 $ &                  -- \\
\hline
\end{longtable*} 
\flushleft Note that
$\rho$ is derived directly from scaling relations, while 
\rsun\ and \msun\ are modeled values using asteroseismic constraints.
For references to solutions published in separate 
papers, see Table 1. \\
$^{*}$ No full solution was derived for KOI-1054 due to the lack of a reliable \Dnu\ 
measurement. The asteroseismic surface gravity constrained using \numax\ is  
$\logg=2.47\pm0.01$\,dex.
\label{tab:results}
\end{center}
\end{footnotesize}

\begin{footnotesize}
\begin{center}
\renewcommand\thetable{3}            
\begin{longtable*}{l c c c c c}  
\caption{Re-derived Properties of \npl\ Planet Candidates in the Sample.}
\\
\hline
KOI & $P$ (d) & $R_{p}/R_{*}$ & $R_{p} (\rearth)$ & $a$ (AU) & $F (F_{\oplus})$  \\
\hline  
          1.01   & $ 2.470613200\pm 0.000000100 $ & $  0.124320\pm  0.000080 $ & $  12.89\pm   0.27 $ & $0.03504\pm0.00062 $ & $       773 $ \\
          2.01   & $ 2.204735400\pm 0.000000100 $ & $  0.075450\pm  0.000020 $ & $  16.39\pm   0.15 $ & $0.03812\pm0.00030 $ & $      3983 $ \\
          5.01   & $   4.7803287\pm   0.0000030 $ & $   0.03651\pm   0.00026 $ & $   6.96\pm   0.17 $ & $ 0.0579\pm 0.0011 $ & $       897 $ \\
          5.02   & $     7.05186\pm     0.00028 $ & $   0.00428\pm   0.00038 $ & $  0.816\pm  0.075 $ & $ 0.0750\pm 0.0014 $ & $       534 $ \\
          7.01   & $   3.2136641\pm   0.0000042 $ & $  0.026920\pm  0.000070 $ & $   4.50\pm   0.12 $ & $0.04389\pm0.00098 $ & $      1223 $ \\
         41.01   & $   12.815735\pm    0.000053 $ & $  0.015440\pm  0.000100 $ & $  2.511\pm  0.062 $ & $ 0.1100\pm 0.0021 $ & $       190 $ \\
         41.02   & $    6.887099\pm    0.000062 $ & $   0.00918\pm   0.00016 $ & $  1.493\pm  0.044 $ & $ 0.0727\pm 0.0014 $ & $       434 $ \\
         41.03   & $    35.33314\pm     0.00063 $ & $   0.01042\pm   0.00030 $ & $  1.694\pm  0.063 $ & $ 0.2162\pm 0.0042 $ & $        49 $ \\
         42.01   & $   17.834381\pm    0.000031 $ & $  0.018250\pm  0.000100 $ & $  2.710\pm  0.039 $ & $ 0.1436\pm 0.0017 $ & $       129 $ \\
         64.01   & $   1.9510914\pm   0.0000040 $ & $   0.04015\pm   0.00093 $ & $  10.68\pm   0.40 $ & $0.03302\pm0.00078 $ & $      3864 $ \\
         69.01   & $   4.7267482\pm   0.0000052 $ & $   0.01575\pm   0.00011 $ & $  1.583\pm  0.037 $ & $ 0.0534\pm 0.0011 $ & $       276 $ \\
         72.01   & $   0.8374903\pm   0.0000015 $ & $  0.012650\pm  0.000070 $ & $  1.458\pm  0.030 $ & $0.01676\pm0.00037 $ & $      3574 $ \\
         72.02   & $    45.29404\pm     0.00020 $ & $   0.02014\pm   0.00013 $ & $  2.321\pm  0.049 $ & $ 0.2397\pm 0.0054 $ & $        17 $ \\
         75.01   & $   105.88531\pm     0.00033 $ & $  0.039310\pm  0.000080 $ & $  10.84\pm   0.25 $ & $ 0.4817\pm 0.0083 $ & $        30 $ \\
         85.01   & $    5.859933\pm    0.000011 $ & $  0.018040\pm  0.000080 $ & $  2.803\pm  0.049 $ & $ 0.0689\pm 0.0011 $ & $       555 $ \\
         85.02   & $   2.1549189\pm   0.0000083 $ & $  0.009710\pm  0.000090 $ & $  1.509\pm  0.029 $ & $0.03539\pm0.00057 $ & $      2106 $ \\
         85.03   & $    8.131146\pm    0.000051 $ & $   0.01081\pm   0.00014 $ & $  1.680\pm  0.036 $ & $ 0.0858\pm 0.0014 $ & $       358 $ \\
         87.01   & $    289.8622\pm      0.0018 $ & $    0.0226\pm    0.0016 $ & $   2.42\pm   0.18 $ & $  0.849\pm  0.017 $ & $       1.2 $ \\
         97.01   & $  4.88548920\pm  0.00000090 $ & $  0.082840\pm  0.000030 $ & $  17.74\pm   0.59 $ & $ 0.0618\pm 0.0014 $ & $      1195 $ \\
         98.01   & $   6.7901235\pm   0.0000036 $ & $  0.056080\pm  0.000050 $ & $  12.70\pm   0.43 $ & $ 0.0783\pm 0.0018 $ & $      1042 $ \\
        107.01   & $    7.256999\pm    0.000019 $ & $   0.02228\pm   0.00011 $ & $   3.86\pm   0.15 $ & $ 0.0780\pm 0.0020 $ & $       438 $ \\
        108.01   & $   15.965349\pm    0.000047 $ & $   0.02224\pm   0.00014 $ & $  3.484\pm  0.096 $ & $ 0.1279\pm 0.0027 $ & $       132 $ \\
        108.02   & $    179.6010\pm      0.0012 $ & $   0.03371\pm   0.00017 $ & $   5.28\pm   0.14 $ & $  0.642\pm  0.013 $ & $       5.2 $ \\
        117.01   & $   14.749102\pm    0.000076 $ & $   0.02271\pm   0.00016 $ & $   3.50\pm   0.12 $ & $ 0.1230\pm 0.0025 $ & $       139 $ \\
        117.02   & $    4.901467\pm    0.000039 $ & $   0.01321\pm   0.00020 $ & $  2.034\pm  0.074 $ & $ 0.0590\pm 0.0012 $ & $       602 $ \\
        117.03   & $    3.179962\pm    0.000025 $ & $   0.01222\pm   0.00019 $ & $  1.882\pm  0.069 $ & $0.04423\pm0.00088 $ & $      1071 $ \\
        117.04   & $     7.95792\pm     0.00017 $ & $   0.00833\pm   0.00038 $ & $  1.283\pm  0.072 $ & $ 0.0815\pm 0.0016 $ & $       315 $ \\
        118.01   & $    24.99334\pm     0.00019 $ & $   0.01656\pm   0.00024 $ & $  2.453\pm  0.080 $ & $ 0.1686\pm 0.0039 $ & $        64 $ \\
        119.01   & $    49.18431\pm     0.00013 $ & $  0.037540\pm  0.000090 $ & $   8.98\pm   0.50 $ & $ 0.2923\pm 0.0063 $ & $        59 $ \\
        119.02   & $    190.3134\pm      0.0020 $ & $   0.03297\pm   0.00020 $ & $   7.89\pm   0.44 $ & $  0.720\pm  0.016 $ & $       9.8 $ \\
        122.01   & $   11.523063\pm    0.000028 $ & $   0.02342\pm   0.00020 $ & $   3.62\pm   0.10 $ & $ 0.1026\pm 0.0024 $ & $       180 $ \\
        123.01   & $    6.481674\pm    0.000018 $ & $   0.01686\pm   0.00015 $ & $  2.434\pm  0.071 $ & $ 0.0689\pm 0.0014 $ & $       415 $ \\
        123.02   & $   21.222534\pm    0.000093 $ & $   0.01734\pm   0.00016 $ & $  2.503\pm  0.073 $ & $ 0.1520\pm 0.0032 $ & $        85 $ \\
        168.01   & $   10.742491\pm    0.000063 $ & $   0.02049\pm   0.00018 $ & $   3.46\pm   0.11 $ & $ 0.0977\pm 0.0023 $ & $       260 $ \\
        168.02   & $    15.27448\pm     0.00019 $ & $   0.01376\pm   0.00029 $ & $  2.325\pm  0.088 $ & $ 0.1235\pm 0.0029 $ & $       163 $ \\
        168.03   & $     7.10690\pm     0.00012 $ & $   0.01111\pm   0.00033 $ & $  1.877\pm  0.081 $ & $ 0.0742\pm 0.0018 $ & $       451 $ \\
        244.01   & $  12.7203650\pm   0.0000070 $ & $   0.03583\pm   0.00017 $ & $  5.117\pm  0.095 $ & $ 0.1129\pm 0.0019 $ & $       186 $ \\
        244.02   & $   6.2385593\pm   0.0000075 $ & $  0.018690\pm  0.000090 $ & $  2.669\pm  0.050 $ & $ 0.0702\pm 0.0012 $ & $       482 $ \\
        245.01   & $   39.792196\pm    0.000052 $ & $   0.02513\pm   0.00033 $ & $  2.117\pm  0.077 $ & $ 0.2120\pm 0.0060 $ & $        10 $ \\
        245.02   & $    21.30185\pm     0.00013 $ & $   0.00949\pm   0.00017 $ & $  0.799\pm  0.030 $ & $ 0.1398\pm 0.0039 $ & $        24 $ \\
        245.03   & $    13.36726\pm     0.00031 $ & $   0.00422\pm   0.00027 $ & $  0.356\pm  0.026 $ & $ 0.1025\pm 0.0029 $ & $        44 $ \\
        246.01   & $   5.3987665\pm   0.0000071 $ & $  0.018690\pm  0.000080 $ & $  2.535\pm  0.040 $ & $0.06178\pm0.00097 $ & $       409 $ \\
        257.01   & $    6.883403\pm    0.000012 $ & $   0.02052\pm   0.00015 $ & $  2.661\pm  0.054 $ & $ 0.0748\pm 0.0011 $ & $       331 $ \\
        260.01   & $   10.495678\pm    0.000075 $ & $   0.01125\pm   0.00015 $ & $  1.667\pm  0.037 $ & $ 0.0994\pm 0.0017 $ & $       254 $ \\
        260.02   & $   100.28319\pm     0.00085 $ & $   0.01925\pm   0.00016 $ & $  2.852\pm  0.057 $ & $ 0.4474\pm 0.0074 $ & $        13 $ \\
        262.01   & $    7.812512\pm    0.000052 $ & $   0.01074\pm   0.00015 $ & $  1.856\pm  0.045 $ & $ 0.0832\pm 0.0016 $ & $       489 $ \\
        262.02   & $    9.376137\pm    0.000056 $ & $   0.01362\pm   0.00030 $ & $  2.354\pm  0.069 $ & $ 0.0940\pm 0.0018 $ & $       383 $ \\
        263.01   & $    20.71936\pm     0.00019 $ & $   0.01315\pm   0.00034 $ & $  2.259\pm  0.081 $ & $ 0.1498\pm 0.0031 $ & $       111 $ \\
        268.01   & $   110.37908\pm     0.00084 $ & $   0.02011\pm   0.00015 $ & $  2.997\pm  0.061 $ & $ 0.4826\pm 0.0076 $ & $        12 $ \\
        269.01   & $    18.01134\pm     0.00022 $ & $   0.01074\pm   0.00019 $ & $  1.696\pm  0.043 $ & $ 0.1475\pm 0.0021 $ & $       151 $ \\
        270.01   & $    12.58250\pm     0.00014 $ & $   0.01127\pm   0.00020 $ & $  1.804\pm  0.052 $ & $ 0.1048\pm 0.0019 $ & $       172 $ \\
        270.02   & $    33.67312\pm     0.00049 $ & $   0.01334\pm   0.00024 $ & $  2.135\pm  0.062 $ & $ 0.2020\pm 0.0037 $ & $        46 $ \\
        271.01   & $    48.63070\pm     0.00041 $ & $   0.01876\pm   0.00021 $ & $  2.782\pm  0.078 $ & $ 0.2801\pm 0.0065 $ & $        29 $ \\
        271.02   & $    29.39234\pm     0.00018 $ & $   0.01785\pm   0.00016 $ & $  2.647\pm  0.072 $ & $ 0.2003\pm 0.0046 $ & $        57 $ \\
        273.01   & $   10.573769\pm    0.000026 $ & $    0.0156\pm    0.0029 $ & $   1.84\pm   0.35 $ & $ 0.0964\pm 0.0015 $ & $       122 $ \\
        274.01   & $    15.09205\pm     0.00035 $ & $   0.00663\pm   0.00033 $ & $  1.200\pm  0.066 $ & $ 0.1264\pm 0.0026 $ & $       210 $ \\
        274.02   & $    22.79519\pm     0.00063 $ & $   0.00667\pm   0.00036 $ & $  1.208\pm  0.071 $ & $ 0.1665\pm 0.0035 $ & $       121 $ \\
        275.01   & $    15.79186\pm     0.00014 $ & $   0.01316\pm   0.00015 $ & $  2.357\pm  0.078 $ & $ 0.1308\pm 0.0034 $ & $       157 $ \\
        275.02   & $     82.1997\pm      0.0020 $ & $   0.01375\pm   0.00032 $ & $  2.463\pm  0.096 $ & $  0.393\pm  0.010 $ & $        17 $ \\
        276.01   & $    41.74591\pm     0.00018 $ & $   0.02052\pm   0.00082 $ & $   2.65\pm   0.12 $ & $ 0.2413\pm 0.0045 $ & $        28 $ \\
        277.01   & $   16.231204\pm    0.000054 $ & $  0.021810\pm  0.000100 $ & $  3.870\pm  0.049 $ & $ 0.1284\pm 0.0017 $ & $       176 $ \\
        279.01   & $   28.455113\pm    0.000053 $ & $    0.0351\pm    0.0015 $ & $   6.01\pm   0.41 $ & $ 0.2014\pm 0.0042 $ & $        81 $ \\
        279.02   & $    15.41298\pm     0.00012 $ & $    0.0158\pm    0.0034 $ & $   2.72\pm   0.60 $ & $ 0.1338\pm 0.0028 $ & $       184 $ \\
        280.01   & $   11.872914\pm    0.000023 $ & $   0.01972\pm   0.00073 $ & $  2.242\pm  0.100 $ & $ 0.1029\pm 0.0023 $ & $       130 $ \\
        281.01   & $    19.55663\pm     0.00011 $ & $   0.01689\pm   0.00012 $ & $  2.592\pm  0.078 $ & $ 0.1363\pm 0.0034 $ & $        95 $ \\
        282.01   & $   27.508733\pm    0.000086 $ & $   0.02774\pm   0.00013 $ & $   3.41\pm   0.10 $ & $ 0.1743\pm 0.0037 $ & $        45 $ \\
        282.02   & $    8.457489\pm    0.000097 $ & $   0.00969\pm   0.00028 $ & $  1.191\pm  0.049 $ & $ 0.0794\pm 0.0017 $ & $       217 $ \\
        285.01   & $   13.748761\pm    0.000072 $ & $   0.02026\pm   0.00017 $ & $   3.77\pm   0.11 $ & $ 0.1196\pm 0.0028 $ & $       216 $ \\
        288.01   & $   10.275394\pm    0.000058 $ & $   0.01463\pm   0.00013 $ & $  3.376\pm  0.073 $ & $ 0.1057\pm 0.0019 $ & $       522 $ \\
        319.01   & $    46.15159\pm     0.00011 $ & $   0.04655\pm   0.00024 $ & $  10.49\pm   0.39 $ & $ 0.2765\pm 0.0067 $ & $        60 $ \\
        370.01   & $    42.88255\pm     0.00031 $ & $   0.01998\pm   0.00015 $ & $   4.03\pm   0.11 $ & $ 0.2630\pm 0.0067 $ & $        63 $ \\
        370.02   & $    22.95036\pm     0.00032 $ & $   0.01227\pm   0.00034 $ & $  2.477\pm  0.096 $ & $ 0.1733\pm 0.0044 $ & $       146 $ \\
        623.01   & $    10.34971\pm     0.00012 $ & $   0.01062\pm   0.00027 $ & $  1.298\pm  0.050 $ & $ 0.0905\pm 0.0019 $ & $       179 $ \\
        623.02   & $    15.67749\pm     0.00020 $ & $   0.01106\pm   0.00026 $ & $  1.351\pm  0.051 $ & $ 0.1193\pm 0.0026 $ & $       103 $ \\
        623.03   & $    5.599359\pm    0.000062 $ & $   0.00918\pm   0.00023 $ & $  1.122\pm  0.043 $ & $ 0.0601\pm 0.0013 $ & $       405 $ \\
        674.01   & $   16.338952\pm    0.000071 $ & $   0.04259\pm   0.00022 $ & $  12.50\pm   0.54 $ & $ 0.1320\pm 0.0046 $ & $       212 $ \\
        974.01   & $    53.50607\pm     0.00061 $ & $   0.01353\pm   0.00014 $ & $  2.733\pm  0.071 $ & $ 0.3010\pm 0.0068 $ & $        52 $ \\
        975.01   & $    2.785819\pm    0.000017 $ & $  0.007740\pm  0.000100 $ & $  1.571\pm  0.026 $ & $0.04272\pm0.00011 $ & $      2405 $ \\
        981.01   & $     3.99780\pm     0.00012 $ & $   0.00819\pm   0.00053 $ & $   4.76\pm   0.32 $ & $0.05478\pm0.00098 $ & $      5586 $ \\
 1019.01$^{*}$   & $    2.497052\pm    0.000063 $ & $   0.00689\pm   0.00041 $ & $   2.70\pm   0.17 $ & $0.03958\pm0.00094 $ & $      4637 $ \\
       1221.01   & $    30.16012\pm     0.00052 $ & $   0.01469\pm   0.00034 $ & $   4.71\pm   0.15 $ & $ 0.2069\pm 0.0040 $ & $       112 $ \\
       1221.02   & $     51.0802\pm      0.0018 $ & $   0.01106\pm   0.00039 $ & $   3.54\pm   0.15 $ & $ 0.2939\pm 0.0057 $ & $        56 $ \\
       1222.01   & $     4.28553\pm     0.00014 $ & $   0.00649\pm   0.00065 $ & $   2.64\pm   0.29 $ & $ 0.0582\pm 0.0022 $ & $      2416 $ \\
 1230.01$^{+}$   & $   165.72106\pm     0.00077 $ & $   0.08259\pm   0.00018 $ & $  63.64\pm   2.33 $ & $  0.716\pm  0.026 $ & $        55 $ \\
       1241.01   & $    21.40505\pm     0.00036 $ & $   0.02292\pm   0.00033 $ & $  10.58\pm   0.40 $ & $ 0.1655\pm 0.0054 $ & $       322 $ \\
       1241.02   & $    10.50343\pm     0.00025 $ & $   0.01120\pm   0.00033 $ & $   5.17\pm   0.24 $ & $ 0.1030\pm 0.0034 $ & $       832 $ \\
       1282.01   & $    30.86392\pm     0.00031 $ & $   0.01428\pm   0.00018 $ & $  2.480\pm  0.084 $ & $ 0.2000\pm 0.0049 $ & $        75 $ \\
       1299.01   & $    52.50128\pm     0.00067 $ & $   0.02683\pm   0.00030 $ & $  12.18\pm   0.38 $ & $ 0.3035\pm 0.0076 $ & $       105 $ \\
       1314.01   & $     8.57507\pm     0.00018 $ & $   0.01191\pm   0.00027 $ & $   4.61\pm   0.17 $ & $ 0.0905\pm 0.0023 $ & $       896 $ \\
       1537.01   & $    10.19144\pm     0.00022 $ & $   0.00675\pm   0.00020 $ & $  1.343\pm  0.054 $ & $ 0.1021\pm 0.0025 $ & $       440 $ \\
       1612.01   & $    2.464999\pm    0.000020 $ & $   0.00531\pm   0.00025 $ & $  0.710\pm  0.037 $ & $0.03663\pm0.00078 $ & $      1394 $ \\
       1613.01   & $    15.86621\pm     0.00020 $ & $   0.00933\pm   0.00064 $ & $   1.35\pm   0.10 $ & $ 0.1239\pm 0.0033 $ & $       137 $ \\
       1618.01   & $    2.364320\pm    0.000036 $ & $   0.00547\pm   0.00020 $ & $  0.899\pm  0.039 $ & $0.03761\pm0.00081 $ & $      2089 $ \\
       1621.01   & $    20.31035\pm     0.00024 $ & $   0.01290\pm   0.00027 $ & $   2.75\pm   0.11 $ & $ 0.1588\pm 0.0038 $ & $       186 $ \\
       1890.01   & $    4.336491\pm    0.000029 $ & $   0.01035\pm   0.00014 $ & $  1.758\pm  0.051 $ & $ 0.0554\pm 0.0013 $ & $       981 $ \\
       1894.01   & $    5.288016\pm    0.000045 $ & $   0.01733\pm   0.00019 $ & $   7.17\pm   0.37 $ & $ 0.0666\pm 0.0034 $ & $      1805 $ \\
 1924.01$^{*}$   & $    2.119128\pm    0.000038 $ & $   0.00645\pm   0.00022 $ & $  1.753\pm  0.071 $ & $0.03649\pm0.00067 $ & $      4878 $ \\
       1925.01   & $    68.95800\pm     0.00090 $ & $    0.0108\pm    0.0038 $ & $   1.05\pm   0.37 $ & $ 0.3199\pm 0.0066 $ & $       6.2 $ \\
       1930.01   & $    13.72686\pm     0.00014 $ & $   0.01341\pm   0.00020 $ & $   2.54\pm   0.13 $ & $ 0.1173\pm 0.0029 $ & $       242 $ \\
       1930.02   & $    24.31058\pm     0.00036 $ & $   0.01298\pm   0.00026 $ & $   2.46\pm   0.13 $ & $ 0.1717\pm 0.0042 $ & $       113 $ \\
       1930.03   & $    44.43150\pm     0.00076 $ & $   0.01492\pm   0.00027 $ & $   2.83\pm   0.14 $ & $ 0.2566\pm 0.0063 $ & $        51 $ \\
       1930.04   & $     9.34131\pm     0.00021 $ & $   0.00824\pm   0.00032 $ & $  1.560\pm  0.096 $ & $ 0.0907\pm 0.0022 $ & $       404 $ \\
       1962.01   & $    32.85833\pm     0.00033 $ & $   0.01593\pm   0.00058 $ & $   2.55\pm   0.14 $ & $ 0.2062\pm 0.0045 $ & $        55 $ \\
       2133.01   & $    6.246580\pm    0.000082 $ & $   0.01798\pm   0.00026 $ & $  12.81\pm   0.72 $ & $ 0.0733\pm 0.0031 $ & $      3205 $ \\
 2481.01$^{+}$   & $    33.84760\pm     0.00091 $ & $   0.02750\pm   0.00072 $ & $  31.42\pm   2.24 $ & $  0.240\pm  0.013 $ & $       733 $ \\
       2545.01   & $     6.98158\pm     0.00019 $ & $   0.00568\pm   0.00025 $ & $   1.32\pm   0.10 $ & $ 0.0803\pm 0.0018 $ & $       896 $ \\
       2640.01   & $     33.1809\pm      0.0014 $ & $   0.02086\pm   0.00072 $ & $  17.02\pm   0.81 $ & $ 0.2191\pm 0.0063 $ & $       581 $ \\
\hline
\end{longtable*} 
\flushleft Planet period and planet-star size ratio have been adopted from \citet{batalha12}. 
The incident flux $F (F_{\oplus})$ has been 
estimated using the planet-star separation given in \citet{batalha12} and 
assuming circular orbits ($d/R_{*} = a/R_{*}$). 
Note that KOI-113.01, KOI-245.04, KOI-371.01 and KOI-1054.01 have been omitted either 
due to large uncertainties in the 
transit parameters given in \citet{batalha12} or due to evidence that the 
transit events are false-positives. \\
$^{+}$ Asteroseismic false-positive.\\
$^{*}$ Low-mass stellar companion detected by follow-up radial-velocity observations.
\label{tab:results2}
\end{center}
\end{footnotesize}

\end{document}